\documentclass[3p,12pt]{elsarticle}
\usepackage{lineno,hyperref}
\usepackage{amsthm}
\newtheorem{remark}{Remark}
\usepackage{bm}
\usepackage{graphicx}
\usepackage{graphics}
\usepackage[fleqn]{amsmath}
\usepackage{cases}
\usepackage{amssymb}
\usepackage{booktabs}
\usepackage{multirow}
\usepackage{color}
\usepackage{xcolor}
\usepackage[normalem]{ulem}
\usepackage{tabularx}
\usepackage{siunitx}
\usepackage[caption=false]{subfig}
\usepackage{comment}
\usepackage{grffile}
\usepackage{rotating}

\journal{Journal of Computational Physics}

\newcommand{\bn}{\bm n}

\newcommand{\bU}{\bm U}

\newcommand{\mbD}{\mathbf D}
\newcommand{\mbL}{\mathbf L}
\newcommand{\mbU}{\mathbf U}

\newcommand{\dvvv}{ \mathrm{d}^3 \vec \xi}
\newcommand{\dvv}{ \mathrm{d}^2 \vec \xi}
\newcommand{\dCCC}{ \mathrm{d}^3 \vec C}
\newcommand{\sstar}{\,*}

\newcommand{\pp}[2]{\frac{\partial #1}{\partial #2}}

\newcommand{\HoT}{\text{HoT}}
\newcommand{\la}{\langle}
\newcommand{\ra}{\rangle}

\title{General synthetic iteration scheme for nonlinear gas kinetic simulation of multi-scale rarefied gas flows}

 \author[strath]{Lianhua Zhu\fnref{fn1}} 
 \author[cadrc]{Xingcai Pi\fnref{fn1}}
 \author[strath]{Wei Su}
 \author[cadrc,cfdlab]{Zhi-Hui Li}
 \author[strath]{Yonghao Zhang}
 \author[sustech]{Lei Wu\corref{cor}}

 \address[strath]{James Weir Fluids Laboratory, Department of Mechanical and Aerospace Engineering, University of Strathclyde, Glasgow G1 1XJ, UK}
 \address[cadrc]{Hypervelocity Aerodynamics Institute, China Aerodynamics Research and Development Center, Mianyang, 621000, China}
 \address[cfdlab]{National Laboratory of Computational Fluid Dynamics, Beijing, 100191, China}
 \address[sustech]{Department of Mechanics and Aerospace Engineering, Southern University of Science and Technology, Shenzhen, 518055, China}
\cortext[cor]{Corresponding author:~wul@sustech.edu.cn}
\fntext[fn1]{Both authors contributed equally.}

\begin{document} 

\begin{abstract}
The general synthetic iteration scheme (GSIS) is extended to find the steady-state solution of nonlinear gas kinetic equation, removing the long-standing problems of slow convergence and requirement of ultra-fine grids in  near-continuum flows. The key ingredients of GSIS are that  the gas kinetic equation and macroscopic synthetic equations are tightly coupled, and the constitutive relations in macroscopic synthetic equations explicitly contain Newton's law of shear stress and Fourier's law of heat conduction. The higher-order constitutive relations describing rarefaction effects are calculated from the velocity distribution function, however, their constructions are simpler than our previous work (Su \textit{et al.} Journal of Computational Physics 407 (2020) 109245) for linearized gas kinetic equations. On the other hand, solutions of macroscopic synthetic equations are used to inform the evolution of gas kinetic equation at the next iteration step. A rigorous linear Fourier stability analysis in  periodic system shows that the error decay rate of GSIS can be smaller than 0.5, which means that the deviation to steady-state solution can be reduced by 3 orders of magnitude in 10 iterations. Other important advantages of the GSIS are (i) it does not rely on the specific form of Boltzmann collision operator and (ii) it can be solved by sophisticated techniques in computational fluid dynamics, making it amenable to large scale engineering applications. In this paper, the efficiency and accuracy of GSIS is demonstrated by a number of canonical test cases in rarefied gas dynamics.
\end{abstract}

\maketitle

\section{Introduction}

Multi-scale rarefied gas flow exists in many engineering applications, from the aerodynamics of re-entering vehicles in the sky to the shale gas transport in the underground. Due to the significant variation of gas density or characteristic length scale, these flows can span several regimes, e.g. the continuum, transition, and free molecular flow regimes, which are usually categorized by the Knudsen number (Kn, the ratio between the mean free path of gas molecules and the characteristic flow length). Gas flow in the continuum regime (Kn $<$ 0.001) can be accurately modeled by the Navier-Stokes-Fourier (NSF) equations. But for rarefied flows (Kn $>$ 0.001), NSF equations are inaccurate due to the linear constitutive relations given by Newton's law of shear stress and Fourier's law of heat conduction. To model rarefied gas flows that deviate far away from thermodynamic equilibrium, the Boltzmann equation, which is an integral-differential equation describing the evolution of one-particle velocity distribution function (VDF) at the mesoscopic scale, should be used~\cite{chapmanMathematicalTheoryNonuniform1970}. Although various higher-order macroscopic equations have been derived from the Boltzmann equation, either by the Chapman-Enskog expansion or the Grad's moment method~\cite{gradKineticTheoryRarefied1949,struchtrupMacroscopicTransportEquations2006,guHighorderMomentApproach2009}, none of them are valid in highly rarefied gas flows.

The common numerical methods for rarefied gas flow simulations are the discrete velocity method (DVM)~\cite{aristovDirectMethodsSolving2001} and the direct simulation Monte Carlo (DSMC) method~\cite{birdMolecularGasDynamics1994}. In DVM, the Boltzmann equation is first discretized in both the velocity and spatial spaces, and then solved deterministically by the 
computational fluid dynamics (CFD). In DSMC, simulation particles are used to mimic the streaming and collision of real gas molecules. It has been proven that DSMC solves the Boltzmann equation for monatomic gas~\cite{wagnerConvergenceProofBird1992}. The kinetic nature of the Boltzmann equation means that it is much more expensive to be solved than the NSF equations. For example, DVM requires appropriate discretization of the velocity space, and DSMC needs a large number of repeated samples, which lead to expensive computational cost.  Particularly, for low-Kn flows, DSMC becomes prohibitive due to the requirement that the cell size and time step should be respectively smaller than the mean free path and mean collision time, in order to keep the numerical dissipation small~\cite{birdMolecularGasDynamics1994}. 
The conventional DVM also suffers from the same problem due to the decoupled treatment of molecular collision and streaming~\cite{wangComparativeStudyDiscrete2018,chenComparativeStudyAsymptotic2015}.

The failure of NSF equations for rarefied gas flows and the difficulty of solving the Boltzmann equation for continuum flows make the multiscale simulation challenging. Numerous efforts have been devoted to bridging the gap of macroscopic and mesoscopic methods. The popular approach is to couple the macroscopic and mesoscopic models under the domain-decomposition framework. For example, in the method of CFD-DSMC coupling~\cite{schwartzentruberModularParticleContinuum2007,boydHybridParticleContinuumNumerical2011,alaiaHybridMethodHydrodynamickinetic2012,liApplicationHybridNS2012,darbandiHybridDSMCNavierStokes2013}, macroscopic and mesoscopic models are applied in the continuum and rarefied flow regions, respectively. The implementation of such hybrid approaches usually involves a buffer region where both macroscopic and mesoscopic models are solved and assumed to be valid. In reality, however, these methods face the dilemma of ensuring the validity of NSF equations and the efficiency of mesoscopic methods in the coupling region. Recently, a hybrid approach applying the regularized 26-moment equations rather than the NSF equations in macroscopic regions is proposed, which significantly moves the buffer zone towards high-Kn regions and hence reduces the iteration number for gas kinetic equation~\cite{yangHybridApproachCouple2020}.

An alternative approach is to solve the gas kinetic methods in the whole computational domain and use appropriate numerical schemes to remove the restrictions on cell size and time step. By coupling the collision and streaming, the unified gas kinetic scheme (UGKS) and its variants~\cite{xuUnifiedGaskineticScheme2010,xuImprovedUnifiedGaskinetic2011,guoDiscreteUnifiedGas2013,guoDiscreteUnifiedGas2015,zhuDiscreteUnifiedGas2016} are able to obtain accurate results when the numerical cell size is much larger than the mean free path $\lambda$: in the near-continuum flow regime, the cell size can be at the order of $\sqrt{\lambda}$~\cite{guoUnifiedPreservingProperties2020}. The implicit version of UGKS further reduces the number of iteration steps~\cite{zhuImplicitUnifiedGaskinetic2016,zhuUnifiedGaskineticScheme2017}.

The recently-developed general synthetic iteration scheme (GSIS) is also one of these promising multiscale methods~\cite{suCanWeFind2020}. It is a generalization of the synthetic iterative scheme that is originally developed for solving radiation transport equation in the optical thick regions~\cite{adamsFastIterativeMethods2002} and extended to some special linear rarefied gas flows~\cite{valougeorgisAccelerationSchemesDiscrete2003,lihnaropoulosFormulationStabilityAnalysis2007,szalmasFastIterativeModel2010,szalmasAcceleratedDiscreteVelocity2013,szalmasAcceleratedDiscreteVelocity2016,wuFastIterativeScheme2017,suAccurateEfficientComputation2018}. The GSIS extends the synthetic iterative scheme to general rarefied gas flows, and it is not limited to simple flows where the velocity must be perpendicular to the computational domain. The efficiency and accuracy of GSIS is demonstrated in solving two-dimensional (2D) linearized gas kinetic equation in the whole flow regime~\cite{suCanWeFind2020,suGSISEfficientAccurate2020}, where the linearized gas kinetic equation and  macroscopic synthetic equations are solved on the same grid alternately, and converged solutions are found within a few dozens of iteration steps. In each iteration of gas kinetic equation, the latest macroscopic quantities from the previous solution of macroscopic synthetic equations are used to evaluate the equilibrium distribution function. While in macroscopic synthetic equations, expressions of shear stress and heat flux explicitly include the constitutive laws at the first-order of Kn, i.e. the Newton law and the Fourier law; higher-order contributions are directly calculated by taking the velocity moments of VDF~\cite{suCanWeFind2020}. Compared with other multiscale methods~\cite{zhuImplicitUnifiedGaskinetic2016,panImplicitDiscreteUnified2019}, GSIS does not rely on specific forms of the Boltzmann collision operator~\cite{suCanWeFind2020}. In addition, sophisticated CFD techniques can be directly used to solve the gas kinetic equation and macroscopic synthetic equations. For example, in the linearized GSIS, the DVM is solved by the upwind method, while the SIMPLE algorithm or discontinuous Galerkin method is used to solve the linearized NSF equations with high-order constitutive relations treated as source terms~\cite{suCanWeFind2020}.

It is the aim of this paper to extend the GSIS for solving nonlinear gas kinetic equations and demonstrate its potential for practical applications. The overall framework of the linear GSIS will remain unchanged, i.e., we solve the macroscopic synthetic  equations and nonlinear gas kinetic equation alternately in the whole computational domain. We will propose a new way to construct the nonlinear macroscopic synthetic equations, which will be solved by compressible CFD techniques. In this paper we will use the Shakhov model equation~\cite{shakhovGeneralizationKrookKinetic1968} as example, but the method can be used to solve the full BE and other model equations straightforwardly, just as we have achieved in linear GSIS~\cite{suCanWeFind2020}.

The remainder of this paper is organized as follows. In Section~\ref{sec:gkeme}, we introduce the Shakhov model equation, the convention iterative scheme (CIS) to find the steady-state solutions and its the convergence rate.  In Section~\ref{sec:gsis},  the GSIS for nonlinear gas kinetic equation is constructed, and its convergence rate is rigorously calculated based on the Fourier stability analysis. In Section~\ref{sec4}, the numerical schemes for solving both  gas kinetic and macroscopic equations will be presented. In Section~\ref{sec:tests}, several canonical cases are carried out to assess the accuracy and efficiency of the nonlinear GSIS. Section~\ref{Conclusion&outlook} concludes with final comments and outlook.

\section{Gas kinetic equation, CIS and its convergence rate}\label{sec:gkeme}

\subsection{Gas kinetic equation, }

In gas kinetic theory, the gas dynamics is described by the one-particle VDF $f(t, \vec r, \vec \xi)$, which depends on the time $t$, the spatial location $\vec r = (x, y, z)$, and the molecular velocity $\vec \xi = (\xi_x, \xi_y, \xi_z)$. Evolution of the VDF is governed by the Boltzmann equation:
\begin{equation}\label{be}
\pp{f}{t} + \vec \xi \cdot \vec \nabla f = \mathcal{Q}(f),
\end{equation}
where $\vec \nabla$ is the spatial gradient operator and  $\mathcal{Q}(f)$ is the collision operator; they  describe the change of VDF due to the free streaming and binary collision of gas molecules, respectively. Since the Boltzmann collision operator is a complicated five-fold integral, it is usually simplified  by the  Shakhov model~\cite{shakhovGeneralizationKrookKinetic1968}:
\begin{equation}\label{s-model-dim}
\mathcal{Q}^s(f)= \frac{f^s(t,\vec r, \vec \xi)-f(t,\vec r, \vec \xi)}{\tau(t,\vec r)}, 
\end{equation}
where $\tau = \mu/p$ is the mean collision time, with $\mu$ being the shear viscosity and $p$ the pressure of gas.  In this paper, we assume the viscosity varies with the temperature $T$ by the power law: $\mu(T) = \mu_0 (T/T_0)^\omega$, where $\mu_0$ is the reference viscosity at the reference temperature $T_0$, and $\omega$ is a viscosity index. The reference VDF $f^s$ takes the following form:
\begin{equation}\label{Shakhov_nonlinear}
 f^s(t,\vec r, \vec \xi) =f^m\left[ 1 + (1-\Pr) \frac{\vec q \cdot \vec C}{5pRT} \left( \frac{C^2}{RT} - 5 \right) \right], \quad
 f^m=\frac{\rho}{(2\pi RT)^{3/2}}\exp\left ( -\frac{ C^2}{2RT}\right), 
\end{equation}
where $\rho$ is the mass density, $\vec U$ is the macroscopic flow velocity, $\vec q$ is the heat flux, $\vec C \equiv \vec \xi - \vec U$ the peculiar velocity, $R$ is the specific gas constant, and $\Pr$ is the Prandtl number. For ideal gas, the equation of state is $p = \rho R T$. The macroscopic variables including the stress tensor $\sigma_{ij}$ can be calculated by the taking moments of the VDF:
\begin{align}\label{macros}
\begin{split}
& \rho(t,\vec r) = \int f(t,\vec r, \vec \xi\,) \dvvv,~~\rho \vec U(t,\vec r) = \int f(t,\vec r, \vec \xi\,) \vec \xi \dvvv,\\
& \sigma_{ij}(t,\vec r) = \int f(t,\vec r, \vec \xi\,) C_{\la i}C_{j \ra} \dvvv,\quad
 p(t,\vec r) = \frac{1}{3} \int f(t,\vec r, \vec \xi\,) C^2 \dvvv,\\
 &\vec q(t,\vec r) = \frac{1}{2} \int f(t,\vec r, \vec \xi\,) \vec CC^2 \dvvv,
\end{split}
\end{align}
where the angle brackets $\la i, j \ra $ representing the trace-less part of a tensor, e.g. $a_{\la i} b_{j\ra} \equiv a_i b_j - (a_kb_k/3) \delta_{ij}$ with $\delta_{ij}$ being the Kronecker delta function.

\subsection{The conventional iterative scheme and its efficiency}\label{sec:dvm1}
\label{sec:dvm_cartesian}

It is noted that the turbulence is often absent in rarefied gas flows, since the Reynolds is inversely proportional to the Knudsen number. Therefore, steady-state solutions of the gas kinetic equation are of particular interest, which can be obtained in CIS by solving the following equation iteratively: 
\begin{equation}\label{bgkfd}
\vec \xi \cdot \vec \nabla {f^{k+1}} = \frac{1}{\tau^k} [f^{\mathrm{s},k} - f^{k+1}], 
\end{equation}
where $k$ is the step of iteration. Note that in order to avoid solving the nonlinear equation, the reference VDF $f_s$ is calculated from the macroscopic variables of the $k$-th iteration step, while the VDF $f$ is obtained at the $(k+1)$-th iteration. The spatial gradient operator can be approximated by the finite difference or discontinuous Galerkin schemes~\cite{suCanWeFind2020, suImplicitDiscontinuousGalerkin2020}, and the whole system can be easily solved by sweeping procedures~\cite{hoMultilevelParallelSolver2019,zhuGPUAccelerationIterative2019,suCanWeFind2020}.

We use the Fourier stability analysis to investigate the efficiency of CIS, that is, to see how fast the error decays during iterations. Since the Fourier stability analysis relies on linear systems, we rewrite the collision operator~\eqref{Shakhov_nonlinear} in the following linearized one:
\begin{equation}\label{Shakhov}
f^s=\left[\varrho+2\vec{U}\cdot\vec \xi+T\left(\xi^2-\frac{3}{2}\right)
+\frac{4}{15}\vec{q}\cdot\vec{\xi}\left(\xi^2-\frac{5}{2}\right)\right]f_{eq},
\end{equation}
where the Prandtl number is chosen as $2/3$, $f_{eq} = \exp \left( -{\xi^2}\right)/\pi^{1.5}$ is the global equilibrium VDF,
and the macroscopic quantities deviated from their corresponding equilibrium values are:
\begin{equation}\label{macroscopic}
\begin{aligned}
&\varrho=\int{f}\mathrm{d}^3\vec{\xi}, \quad \vec{U}=\int{\vec{\xi}f}\mathrm{d}^3\vec{\xi}, \quad
T=\int{\left(\frac{2}{3}\xi^2-1\right)f}\mathrm{d}^3\vec{\xi}, \\
& \sigma_{ij}  = 2\int \xi_{\la i}\xi_{j \ra} f \mathrm{d}^3\vec{\xi}, \quad
\vec{q}=\int{\vec{\xi}\left(\xi^2-\frac{5}{2}\right)f}\mathrm{d}^3\vec{\xi}.
\end{aligned}
\end{equation}
Note that after linearization the mean collision time $\tau$ in Eq.~\eqref{s-model-dim} is a constant, which has the meaning of Knudsen number. More details can be found in Ref.~\cite{suCanWeFind2020}.

We define the error functions between VDFs at two consecutive iterations as:
\begin{equation}\label{Diff_Y}
Y^{k+1}(\vec{r},\vec{\xi}\,)=f^{k+1}(\vec{r},\vec{\xi}\,)-f^{k}(\vec{r},\vec{\xi}\,), 
\end{equation}
and the error functions for macroscopic quantities $M_0=[\varrho,\vec{U}, T,\vec{q}\,]$  between two consecutive iteration steps: 
\begin{equation}\label{Macro_difference}
\begin{aligned}
\Phi^{k+1}(\vec{r}\,)=&M^{k+1}(\vec{r}\,)-M^{k}(\vec{r}\,)\\
=&\int{Y^{k+1}(\vec{r},\vec{\xi}\,)\phi(\vec{\xi}\,)}\mathrm{d}^3\vec{\xi},
\end{aligned}
\end{equation}
where
\begin{equation}
\phi(\vec{\xi}\,)=\left[1,\xi_x,\xi_y,\frac{2}{3}\xi^2-1,\xi_x\left(\xi^2-\frac{5}{2}\right),\xi_y\left(\xi^2-\frac{5}{2}\right)\right].
\end{equation}

To determine the error decay rate $e$ we perform the Fourier stability analysis by seeking the eigenfunctions $\bar{Y}(\vec{\xi}\,)$ and $\alpha=[\alpha_\varrho,\vec\alpha_{U},  \alpha_{T},\vec\alpha_{q}]$ of the following forms:
\begin{equation}\label{ansatz}
\begin{aligned}[c]
Y^{k+1}(\vec{r},\vec{\xi}\,)=e^{k}\bar{Y}(\vec{\xi}\,)\exp(i\vec{\theta}\cdot{\vec{r}}\,),\\
\Phi^{k+1}(\vec{r}\,)=e^{k+1}\alpha\exp(i\vec{\theta}\cdot{\vec{r}}\,),
\end{aligned}
\end{equation}
where $i$ is the imaginary unit and $\vec{\theta}=(\theta_x,\theta_y,\theta_z)$ is the wave vector of perturbance satisfying $|\vec\theta|=1$. The slow convergence occurs when the error decay rate $|e|$ approaches one, where the error is nearly the same when compared to that in the previous iteration, while the fast convergence is realized when $|e|<1$, especially when $|e|$ approaches zero.

\begin{figure}[t]
	\centering
	\includegraphics[scale=0.6,viewport=0 0 680 390,clip=true]{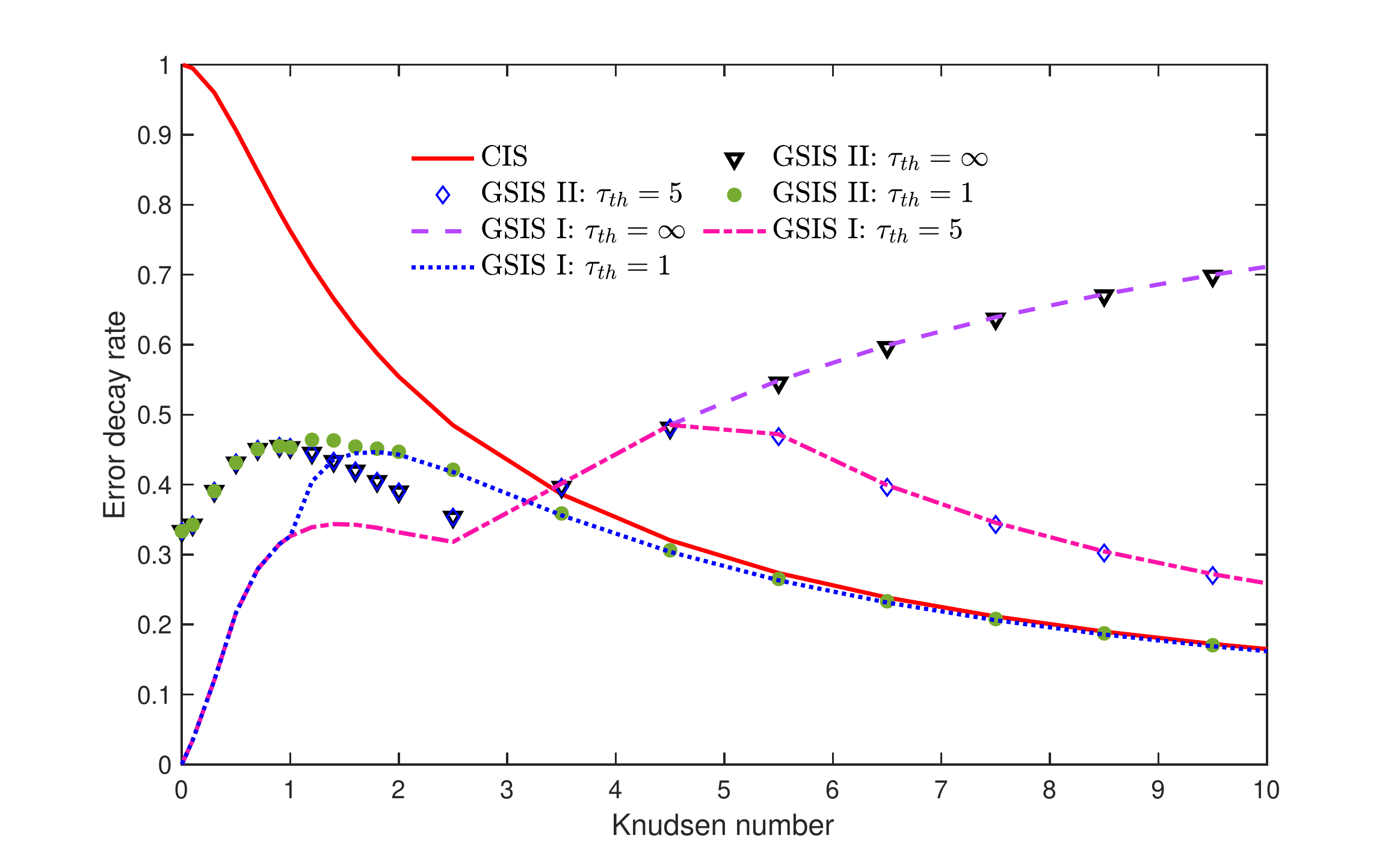}
	\caption{ The error decay rate as a function of the Knudsen number $\tau$ in both CIS and GSIS, calculated from the linearized Shakhov model based on the Fourier stability analysis. }
	\label{fig:SR}
\end{figure}

The streaming operator in Eq.~\eqref{bgkfd} is kept intact when calculating the error decay rate; the convergence rate of the discretized version of gas kinetic equation will be shown in numerical simulations in Section~\ref{sec:tests}.
Obviously, from Eqs.~\eqref{Macro_difference} and~\eqref{ansatz} we have 
\begin{equation}\label{relation}
e\alpha=\int \bar{Y}(\vec{\xi}\,)\phi(\vec{\xi}\,)\mathrm{d}^3\vec{\xi},
\end{equation}
and from  Eqs.~\eqref{bgkfd},  \eqref{Shakhov}, \eqref{Diff_Y}, and \eqref{ansatz}, we obtain the following expressions for $\bar{Y}(\vec{\xi}\,)$:
\begin{equation}\label{y0_solution_CIS}
\begin{aligned}[c]
 \bar{Y}(\vec{\xi}\,)=&\frac{ \alpha_\varrho+2\vec\alpha_{U}\cdot\vec\xi+\alpha_T\left(\xi^2-\frac{3}{2}\right)+\frac{4}{15}\vec\alpha_{q}\cdot\vec\xi\left(\xi^2-\frac{5}{2}\right) }{ 1+i\tau\vec{\theta}\cdot\vec{\xi} } {f_\text{eq}}.
\end{aligned}
\end{equation}

Finally, multiplying Eq.~\eqref{y0_solution_CIS} with $\phi(\vec{\xi})$ and integrating the resultant equations with respect to $\vec{\xi}$,  we obtain 8 linear algebraic equations for 8 unknown elements in $\alpha$ with the help of Eq.~\eqref{relation}. These algebraic equations can be written in the matrix form as
\begin{equation}
C_8\alpha^\top=e\alpha^\top, 
\end{equation} 
where the superscript $\top$ is the transpose operator. The error decay rate can be obtained by numerically computing the eigenvalues of matrix $C_8$ and taking the maximum absolute value of $e$; the result as a function of the Knudsen number is shown in Fig.~\ref{fig:SR}. If is clear that when the Knudsen number $\tau$ is large, $e$ goes to zero so that the error decays quickly. This means that the CIS is very efficient for highly rarefied gas flows. On the other hand, $e\rightarrow1$ when $\tau\rightarrow0$, which means that it is hard to obtain converged solutions by using CIS in the near-continuum flows.

\section{The general synthetic iteration scheme}\label{sec:gsis}

\begin{figure}
	\centering
	\includegraphics[width=0.9\textwidth]{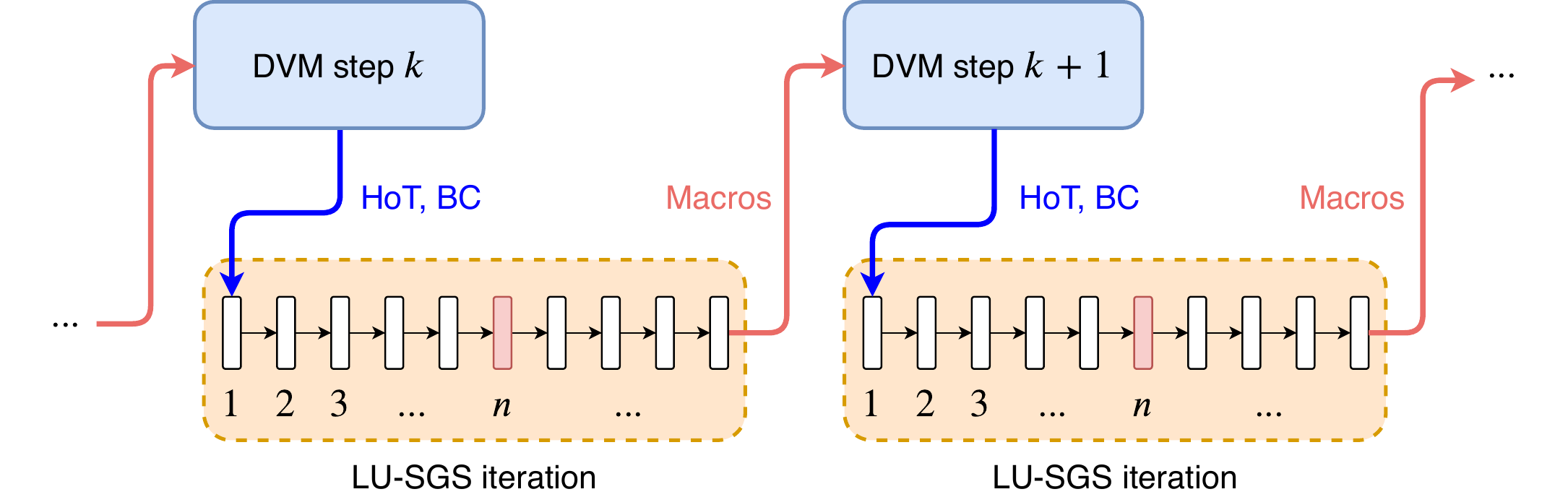}
	\caption{Flow chat of the computational procedures in GSIS. HoT stands for the high-order terms and BC stands for boundary conditions. LU-SGS is the implicit time-stepping procedure solving the macroscopic synthetic equations after each DVM step.}
	\label{fig:gsis_algo}
\end{figure}

The GSIS proposes a strategy to accelerate the iteration of conventional DVM schemes for gas kinetic equations: on top of the  CIS, it adds macroscopic synthetic equations to boost the convergence to steady-state solutions in the near-continuum flow regime. The flowchart of GSIS is visualized in Fig.~\ref{fig:gsis_algo}: after the CIS at the $k$-th step, the nonlinear macroscopic synthetic equations are solved to the converged state by sophisticated CFD techniques, with the boundary conditions and high-order constitutive relations from the CIS. The obtained macroscopic quantities are fed back to the CIS, which provides macroscopic quantities and VDFs for the CIS to execute at the $(k+1)$-th step. Details of GSIS are given below.

\subsection{Macroscopic synthetic equations}

For generality we consider the derivation of macroscopic synthetic equations from the full Boltzmann equation.
By multiplying Eq.~\eqref{be} with 1, $\vec \xi$, and $\xi ^2$, and integrating them with respect to $\mathrm{d}^3\xi$, we have:
\begin{align}\label{conservativeeq}
 \begin{split}
 \pp{\rho}{t} + \vec \nabla \cdot (\rho  \vec U) &= 0,\\
 \pp{\rho \vec U}{t} +  \vec \nabla \cdot (\rho  \vec U \vec U) + \vec \nabla p +  \vec \nabla \cdot \bm \sigma &=0,\\
 \pp{\rho E}{t} +\vec \nabla \cdot \left ( \rho E \vec U +   p \vec U + \vec U \cdot \bm \sigma +  \vec q \right)&= 0, \\ 
 \end{split}
\end{align}
where $E = c_\mathrm v T + U^2/2$ is the total energy with $c_\mathrm v=3R/2$ being the heat capacity at constant volume. This equation is not closed because the shear stress and heat flux are not known. From the Chapman-Enskog expansion to the first-order of Knudsen number~\cite{chapmanMathematicalTheoryNonuniform1970}, they are given by the NSF constitutive relations:
\begin{equation}\label{nsfrelation}
\begin{aligned}
&\sigma_{ij} \approx \sigma_{ij,\text{NSF}} = -2\mu \pp{U_{\la i}}{r_{j \ra}}=\frac{1}{2} \mu \left( \pp{U_i}{r_j} + \pp{U_j}{r_i} \right) - \frac{1}{3}\mu \vec \nabla \cdot \vec U\delta_{ij}, \\
&\vec q \approx \vec q_{\text{NSF}} = -\kappa \vec \nabla {T}, 
 \end{aligned}
\end{equation}
where the heat conductivity $k$ is related to the viscosity and $\Pr$ by $\kappa = \mu c_\mathrm p/\Pr$, with $c_\mathrm p =5R/2$ being the heat capacity at constant pressure.

However, under rarefied condition this approximation is inaccurate, thus the NSF constitutive relation fails. To be consistent with the gas kinetic equation, the shear stress tensor and heat flux have to compute from the VDF itself, without any truncation. In the linear GSIS~\cite{suCanWeFind2020}, the shear stress and heat flux used in the macroscopic synthetic equations are expressed in terms of the first-order NSF constitutive relation and higher-order terms (HoTs),
where the explicit separation of the NSF constitutive relation out of the diffusive fluxes is essential to fast convergence~\cite{suFastConvergenceAsymptotic2020}.
Here we do the same for nonlinear GSIS: 
\begin{align}\label{twoparts}
\begin{split}
&\sigma_{ij} = \sigma_{ij,\text{NSF}} + \HoT_{\sigma_{ij}},\\
& \vec q = \vec q_{\text{NSF}} + \HoT_{\vec q}.
\end{split}
\end{align}

In the linear GSIS~\cite{suCanWeFind2020}, HoTs are calculated from the spatial derivatives of even higher-order VDF moments than the heat flux and stress tensor, which is equivalent to use the governing equations of stress tensor and heat flux in the Grad 13-moment systems but close these moment equations using the VDF from CIS, rather that the one reconstructed using low-order macroscopic quantities. In the nonlinear GSIS here, we can also apply this approach directly. Multiplying the steady-state kinetic equation with $C_{\la i} C_{j \ra}$ and $\vec CC^2/2$, and integrating in the velocity space, we have: 
\begin{equation}\label{hotintdiff}
\begin{aligned}
\sigma^{*}_{ij} = \underline{- \frac{\mu^*}{p^*}\int C^*_{\la i} C^*_{j \ra}  \vec \xi \cdot \vec \nabla f^* \mathrm{d}^3 \vec\xi}
 - \frac{\mu^*}{p^*}\int C^*_{\la i} C^*_{j \ra}  [\mathcal{Q}(f^*)-\mathcal{Q}^s(f^*)] \mathrm{d}^3 \vec\xi, \\
\vec q^{\hspace{0.15em}*} = \underline{-\frac{\mu^*}{ 2p^* \Pr}\int \vec C^{\hspace{0.15em}*}(C^{*})^2 \vec \xi \cdot \vec \nabla f^* \mathrm{d}^3 \vec \xi}
-\frac{\mu^*}{ 2p^* \Pr}\int \vec C^{\hspace{0.15em}*}(C^{*})^2  [\mathcal{Q}(f^*)-\mathcal{Q}^s(f^*)] \mathrm{d}^3 \vec \xi,
\end{aligned}
\end{equation}
where the superscript ``*" means that both the VDF and macroscopic quantities are obtained from CIS. It should be noted that the last terms in each equation are much smaller than the corresponding underlined terms. For example, for the Boltzmann collision operator of Maxwell molecules the last term in each equation vanishes~\cite{chapmanMathematicalTheoryNonuniform1970}.

To obtain the HoTs in Eq.~\eqref{twoparts}, we simply subtract the NSF parts from the complete starred stress and heat flux, yielding 
\begin{align}\label{hotdiffint}
\begin{split}
\HoT_{\sigma_{ij}} = \sigma^{*}_{ij} - \sigma^*_{ij,\text{NSF}},\\
\HoT_{\vec q_{\sstar}} = \vec q^{\sstar} - \vec q_{\text{NSF}}^{\sstar},
\end{split}
\end{align}
with the NSF parts $\sigma_{ij,\text{NSF}}^*$ and $\vec q_{\text{NSF}}^{\sstar}$ calculated using NSF constitutional relations from the starred macroscopic variables. This will be called scheme I in the following paper.


Alternatively, instead of using the derivatives of  higher-order moments to calculate $\sigma^*_{ij}$ and $\vec q^{\sstar}$, we can calculate them directly according to their definitions. Then the HoTs are calculated as
\begin{align}\label{hotdirect}
\begin{split}
\HoT_{\sigma_{ij}} = \int f^* C^*_{\la i}C^*_{j \ra} \dCCC^{\sstar} - \sigma^*_{ij,\text{NSF}},\\
\HoT_{\vec q} = \frac{1}{2} \int f^* \vec C^* (C^*)^2 \dCCC^{\sstar} - \vec q^{\sstar}_{\text{NSF}},
\end{split}
\end{align}
which will be called scheme II in the following paper. 

\begin{remark} It is clear that the scheme I is much more complicated than the scheme II, because (i) it involves the calculation of Boltzmann collision operator in the general case and (ii) the underlines terms contain spatial derivations which may lead to numerical instabilities around sharp solid corners, while the scheme II does not have this problem. Therefore, if both scheme share the similar value of error decay rate, the scheme II will be used in our numerical simulations. What's more, the scheme II can be directly applied to Boltzmann equations involving multi-species and chemical reactions.

\end{remark}

\subsection{Convergence rate of GSIS: the scheme II}
We analyze the error decay rate of the GSIS based on the linearized Shakhov model and the scheme II. In GSIS, when $f^k$ is known, $f^*$ is obtained by solving Eq.~\eqref{bgkfd} with $k+1$ replaced by $*$. Then the macroscopic quantities at the $(k+1)$-th iteration step are obtained by solving the following synthetic equations (note that the time derivative is dropped for steady-state solutions):
\begin{equation}\label{eq123_lin}
\begin{aligned}
\frac{\partial {U^{k+1}_i}}{\partial{r_i}}=0,\\
\frac{\partial {\varrho^{k+1}}}{\partial{r_i}}+\frac{\partial {T^{k+1}}}{\partial{r_i}}+\frac{\partial {{\sigma^{k+1}_{ij}}}}{\partial{r_j}}=0, \\
\frac{\partial {{q^{k+1}_{i}}}}{\partial{r_i}}=0,	
\end{aligned}
\end{equation}
with
\begin{equation}\label{hotdirect_Lin}
\begin{split}
& \sigma^{k+1}_{ij}  = 2\int \left(\xi_i\xi_j-\frac{\xi^2}{3}\delta_{ij}\right) f^* \mathrm{d}^3\vec{\xi} +2\tau \pp{U^*_{\la i}}{r_{j \ra}} -2\tau \pp{U^{k+1}_{\la i}}{r_{j \ra}},\\
&\vec{q}^{\,k+1} = \int  \vec{\xi}\left(\xi^2-\frac{5}{2}\right) f^*
\mathrm{d}^3\vec{\xi} +\frac{15}{8}\tau \vec \nabla {T^*} -\frac{15}{8}\tau \vec \nabla {T^{k+1}},
\end{split}
\end{equation}
which are the linearized version of Eqs.~\eqref{conservativeeq}, \eqref{twoparts} and~\eqref{hotdirect}.  Therefore,
to calculate the convergence rate of GSIS, the error functions in Eqs.~\eqref{Diff_Y}, \eqref{Macro_difference}, and \eqref{ansatz} are redefined as
\begin{equation}\label{Y_ansatz2} 
\begin{aligned}
Y^*(\vec{r},\vec{\xi}\,)=f^*(\vec{r},\vec{\xi}\,)-f^{k}(\vec{r},\vec{\xi}\,)=e^{k}\bar{Y}(\vec{\xi}\,)\exp(i\vec{\theta}\cdot{\vec{r}}\,),\\
\Phi^{k+1}(\vec{r}\,)=M^{k+1}(\vec{r}\,)-M^{k}(\vec{r}\,)=e^{k+1}\alpha\exp(i\vec{\theta}\cdot{\vec{r}}\,),
\end{aligned}
\end{equation}
where the solution of $\bar{Y}(\vec{\xi}\,)$ is still given by Eq.~\eqref{y0_solution_CIS}. Note that the definitions for $\Phi$ remain unchanged, but in GSIS they are calculated from macroscopic synthetic equations, rather than from the VDF $Y^*$.

With Eqs.~\eqref{y0_solution_CIS}, \eqref{eq123_lin} \eqref{hotdirect_Lin} and~\eqref{Y_ansatz2}, we obtain the following 8 linear algebraic equations for 8 unknowns in $\alpha_M$:
\begin{equation}\label{L_lin1}
\begin{aligned}
e(i\theta_x\alpha_{U_x}+i\theta_y\alpha_{U_y}+i\theta_z\alpha_{U_z})=0, \\
e[i\theta_x(\alpha_\varrho+\alpha_T)+\tau\alpha_{u_x}]=S_2,\\
e[i\theta_y(\alpha_\varrho+\alpha_T)+\tau\alpha_{u_y}]=S_3,\\
e[i\theta_z(\alpha_\varrho+\alpha_T)+\tau\alpha_{u_z}]=S_4,\\
e(i\theta_x\alpha_{q_x}+i\theta_y\alpha_{q_y}+i\theta_z\alpha_{q_z})=0,\\
e\left(\frac{15}{8}i\theta_x\tau\alpha_{T}+\alpha_{q_x}\right)=S_6,\\
e\left(\frac{15}{8}i\theta_y\tau\alpha_{T}+\alpha_{q_y}\right)=S_7,\\
e\left(\frac{15}{8}i\theta_z\tau\alpha_{T}+\alpha_{q_z}\right)=S_8,
\end{aligned}
\end{equation}
where the source terms, due to the HoTs  in Eq.~\eqref{hotdirect_Lin}, are also linear functions of $\alpha_M$:
\begin{equation}\label{L_lin2}
\begin{aligned}
S_2=\int\left[\tau{}\xi_x-2i\theta_x\left(\xi_x^2-\frac{\xi^2}{3}\right)-2i\theta_y\xi_x\xi_y-2i\theta_z\xi_x\xi_z\right]\bar{Y}(\vec{\xi}\,)\mathrm{d}^3\vec{\xi}, \\
S_3=\int\left[\tau{}\xi_y-2i\theta_y\left(\xi_y^2-\frac{\xi^2}{3}\right)-2i\theta_x\xi_x\xi_y-2i\theta_z\xi_y\xi_z\right]\bar{Y}(\vec{\xi}\,)\mathrm{d}^3\vec{\xi}, \\
S_4=\int\left[\tau{}\xi_z-2i\theta_z\left(\xi_z^2-\frac{\xi^2}{3}\right)-2i\theta_x\xi_x\xi_z-2i\theta_y\xi_y\xi_z\right]\bar{Y}(\vec{\xi}\,)\mathrm{d}^3\vec{\xi}, \\
S_6=\int\left[\frac{15}{8}i\theta_x\tau{}\left(\frac{2}{3}\xi^2-1\right)+\xi_x\left(\xi^2-\frac{5}{2}\right)\right]\bar{Y}(\vec{\xi}\,)\mathrm{d}^3\vec{\xi},\\
S_7=\int\left[\frac{15}{8}i\theta_y\tau{}\left(\frac{2}{3}\xi^2-1\right)+\xi_y\left(\xi^2-\frac{5}{2}\right)\right]\bar{Y}(\vec{\xi}\,)\mathrm{d}^3\vec{\xi},\\
S_8=\int\left[\frac{15}{8}i\theta_z\tau{}\left(\frac{2}{3}\xi^2-1\right)+\xi_z\left(\xi^2-\frac{5}{2}\right)\right]\bar{Y}(\vec{\xi}\,)\mathrm{d}^3\vec{\xi}.
\end{aligned} 
\end{equation} 

The error decay rate of the scheme II can be obtained by solving Eqs.~\eqref{L_lin1} and~\eqref{L_lin2}. That is, these equations are firstly rewritten in the matrix form as $Le\alpha^\top=R\alpha^\top$, where $L_8$ and $R_8$ are two $8\times8$ matrices. By introducing $G_1=L_8^{-1}R_8$ and numerically computing its eigenvalues we obtain the error decay rate $e$ of GSIS, see the results in Figure~\ref{fig:SR}. It is seen that the value of $|e|$ is much reduced when $\tau\rightarrow0$, which demonstrates that the GSIS is able to boost convergence in near-continuum flows. However, the error decay rate increases to one when $\tau\rightarrow\infty$.

To fix this problem, macroscopic quantities at the (k+1)-th iteration step are not all updated by the solution $M_{syn}$ from macroscopic synthetic equations, when $\tau$ is large. Rather, they are updated in the following manner
\begin{equation}\label{GSIS_K3}
M^{k+1}(\vec{r}\,)=\beta{}M_{syn}+(1-\beta)M^{\ast}(\vec{r}\,),
\end{equation}
where the relaxation parameter $\beta$ is chosen as
\begin{equation}\label{relax_parameter}
\beta=\frac{min(\tau,\tau_{th})}{\tau}.
\end{equation}
with $\tau_{th}$ being the threshold Knudsen number. That is, $\beta=1$ when the Knudsen number is smaller than $\tau_{th}$; when $\tau>\tau_{th}$, $\beta$ gradually decreases to zero as the Knudsen number approaches infinity. The error decay rate of this GSIS can be obtained by computing the eigenvalue of the matrix $G=\beta{L_8^{-1}R_8}+(1-\beta)C_8$, where the results at the threshold Knudsen number of values 1 and 5 are shown in Fig.~\ref{fig:SR}. Clearly, by choosing approximate value of $\beta$, we can make the maximum error decay rate less than 0.5 for all Knudsen numbers; this means that the error can be reduced by at least three orders of magnitude in 10 iterations. Thus, theoretically, GSIS can reach fast convergence in the whole range of Knudsen number.

\subsection{Convergence rate of GSIS: the scheme I}

Schemes I and II differ only in HoTs. In the scheme I, the shear stress and heat flux for the linearized Shakhov model equation are
\begin{equation}\label{hotintdiff_shcemeI}
\begin{aligned}
\sigma^{*}_{ij} =- 2\tau \int \xi_{\la i} \xi_{j \ra}  \vec \xi \cdot \vec \nabla f^* \mathrm{d}^3 \vec\xi, \\
q^{\hspace{0.15em}*}_i = -\frac{3}{2}\tau\int \xi_i\left(\xi^2-\frac{5}{2}\right) \vec \xi \cdot \vec \nabla f^* \mathrm{d}^3 \vec \xi,
\end{aligned}
\end{equation}
hence Eq.~\eqref{L_lin2} is modified as
\begin{equation}\label{L_lin_shcemeI}
\begin{aligned}
S_2=\tau\int\left[\xi_x-2\Theta
(\theta_x\xi_{\la x} \xi_{x \ra}+\theta_y\xi_x \xi_y+\theta_z\xi_x \xi_z)\right]\bar{Y}(\vec{\xi}\,)\mathrm{d}^3\vec{\xi}, \\
S_3=\tau\int\left[\xi_y-2\Theta (\theta_x\xi_x \xi_y+\theta_y\xi_{\la y} \xi_{y \ra}+\theta_z\xi_y \xi_z)\right]\bar{Y}(\vec{\xi}\,)\mathrm{d}^3\vec{\xi}, \\
S_4=\tau\int\left[\xi_z-2\Theta (\theta_x\xi_x \xi_z+\theta_y\xi_y \xi_z+\theta_z\xi_{\la z} \xi_{z \ra})\right]\bar{Y}(\vec{\xi}\,)\mathrm{d}^3\vec{\xi}, \\
S_6=i\tau\int\left[\frac{15}{8}\theta_x\left(\frac{2}{3}\xi^2-1\right)-\frac{3}{2}\Theta\xi_x\left(\xi^2-\frac{5}{2}\right)\right]\bar{Y}(\vec{\xi}\,)\mathrm{d}^3\vec{\xi},\\
S_7=i\tau\int\left[\frac{15}{8}\theta_y{}\left(\frac{2}{3}\xi^2-1\right)-\frac{3}{2}\Theta\xi_y\left(\xi^2-\frac{5}{2}\right)\right]\bar{Y}(\vec{\xi}\,)\mathrm{d}^3\vec{\xi},\\
S_8=i\tau\int\left[\frac{15}{8}\theta_z{}\left(\frac{2}{3}\xi^2-1\right)-\frac{3}{2}\Theta\xi_z\left(\xi^2-\frac{5}{2}\right)\right]\bar{Y}(\vec{\xi}\,)\mathrm{d}^3\vec{\xi},
\end{aligned} 
\end{equation}
where $\Theta=\theta_x\xi_x+\theta_y\xi_y+\theta_z\xi_z$.

With Eqs.~\eqref{L_lin1}, \eqref{L_lin_shcemeI}, \eqref{GSIS_K3} and~\eqref{relax_parameter}, we obtain the error decay rate of the scheme I, which is also shown in Fig.~\ref{fig:SR}. It is seen that when the Knudsen number $\tau\rightarrow0$, the error decay rate goes to zero, which means that the GSIS with scheme I is very efficient in obtaining the steady-state solution of the gas kinetic equations.

\begin{remark} The Fourier stability analysis is conducted in the spatial periodic systems. In reality, however, solid walls are always present, and the Knudsen layer (exists in a region within a few mean free path away from the wall) always make the effective Knudsen number $\tau\sim1$. Therefore, what's important is the maximum error decay rate in the whole range of Kn. In this sense, from Fig.~\ref{fig:SR} we see that schemes I and II have the similar efficacy in boosting the convergence rate to steady-state solutions. We therefore choose the scheme II over scheme I because it is simpler and can be easily applied to other Boltzmann collision operators.
\end{remark}

\section{Numerical schemes}\label{sec4}

\subsection{The DVM scheme on curved structured mesh}\label{sec:dvm2}

For irregular computational domain, general structured body-fitted meshes are preferred. In order to use the sophisticated techniques in computational fluid dynamics on such meshes, we keep the time derivative in the gas kinetic equation. 
On using the forward Euler scheme for the time derivative and applying implicit treatment to the convection term and $f$ in the collision term, we have
\begin{equation}\label{bgkkp2}
  \frac{f^{k+1} - f^k}{\Delta t} +  \vec \xi \cdot \vec \nabla {f^{k+1}} = \frac{1}{\tau^k} [f^{\mathrm{s},k} - f^{k+1}],
 \end{equation}
which, in order to enable a simple matrix-free implicit solving of the semi-discretized equation, is rewritten 
in the so-called ``delta" form,
\begin{equation}\label{eq31}
  \left( \frac{1}{\Delta t} + \frac{1}{\tau^k} \right)\Delta f^k 
  + \vec \xi \cdot \vec \nabla \Delta f^k
  = \frac{1}{\tau^k} [f^{s,k} - f^{k}] 
  - \vec \xi \cdot \vec \nabla f^k,
\end{equation}
by introducing the incremental VDF $\Delta f^k = f^{k+1} - f^{k}$. 

The gradient operators $\vec \nabla$ at the left-hand-side (LHS) and right-hand-side (RHS) of Eq.~\eqref{eq31} will be calculated by the first-order upwind scheme and a second-order scheme, respectively. With such a treatment, the implicit part allows a simple matrix-free solving with the  Lower-Upper Symmetric Gauss–Seidel (LU-SGS) technique, while the converged solution will be second-order accurate. 

We apply the finite volume method to solve the above gas kinetic equation.  After the volume integration and applying the Gauss theorem, for each cell indexed by ($i,j$) on a structured grid, we have
\begin{equation}
  \left( \frac{1}{\Delta t} + \frac{1}{\tau^k_{i,j}} \right)\Omega_{i,j}\Delta f_{i,j}^k + 
  \sum_{m}^{} \vec S_m \cdot \vec \xi \Delta f^k_m
  = \frac{\Omega_{i,j}}{\tau^k_{i,j}} (f^{s,k}_{i,j} - f_{i,j}^{k})
  - \sum_{m}^{} \vec S_m \cdot \vec \xi f^k_m, 
\end{equation}
where $\Omega_{i,j}$ is the cell's volume, $m$ is the index of the faces belonging to the cell, and $\vec S_m$ is the face's normal vectors pointing out of the cell with its magnitude being the face area. Variables with subscript $i,j$ are the cell averaged quantities on the cell center, while $\Delta f_m^k$ and $f_m^k$ are reconstructed variables on cell faces. For the reconstruction of $\Delta f_m^k$, the first-order upwind scheme is applied: at the left face of cell 
$\Delta f_{i-1/2,j}^k = \Delta f_{i-1,j}^k$ if $\vec \xi \cdot \vec S_{i-1/2,j} > 0$,  otherwise it is $\Delta f_{i-1/2,j}^k = \Delta f_{i,j}^k$. For the reconstruction of $f_m^k$, various second-order limited interpolation scheme can be applied. In this study, $f_m$ is calculated from the upwind cell center by first-order Taylor expansion, where the slope is calculated with van Leer slope limiter. 


With the above discretization, the linear equation system for all cells can be written in the following matrix form
\begin{equation}\label{dluf}
 \mbD_{i,j} \Delta f_{i,j}^k 
 +  \mbL^x_{i,j} \Delta f^k_{i-1,j} + \mbU^x_{i,j} \Delta f^k_{i+1,j}
 + \mbL^y_{i,j} \Delta f^k_{i,j-1} + \mbU^y_{i,j} \Delta f^k_{i,j+1}
 = \text{RHS}_{i,j}
\end{equation}
where the matrix elements are
\begin{align}
 & \mbD_{i,j} = \frac{\Omega_{i,j}}{\Delta t} + \frac{\Omega_{i,j}}{\tau^k_{i,j}} + |\vec S_{i}\cdot \vec \xi | + | \vec S_{j}\cdot \vec \xi |, \\
 &  \mbL^x_{i,j} =  \frac{1}{2} \vec S_{i}\cdot \vec \xi \left[1-\text{sign}(\vec \xi \cdot \vec n_{i}) \right],  \quad
   \mbL^y_{i,j} =  \frac{1}{2} \vec S_{j}\cdot \vec \xi \left[1-\text{sign}(\vec \xi \cdot \vec n_{j}) \right], \\
 &  \mbU^x_{i,j} =  -\frac{1}{2}  \vec S_{i}\cdot \vec \xi \left[1+\text{sign}(\vec \xi \cdot \vec n_{i}) \right], \quad
   \mbU^y_{i,j} =  -\frac{1}{2}  \vec S_{j}\cdot \vec \xi \left[1+\text{sign}(\vec \xi \cdot \vec n_{j}) \right],
\end{align}
with $\vec n$ = $\vec S/|\vec S|$ and $\text{sign}(x)$ the sign function that returns 1 if $x>0$ and $-1$ otherwise. The approximation $ \vec S_i = \frac{1}{2}\left(\vec S_{i-1/2} + \vec S_{i+1/2} \right) \approx \vec S_{i-1/2} \approx -\vec S_{i+1/2}$ has been used. By applying the LU-SGS technique to Eq.~\eqref{dluf}, the incremental VDF is solved by a forward sweeping and a backward sweeping:
\begin{align}
  \text{Forward:}& \quad \mbD_{i,j} \Delta f_{ij}^*
  + \mbL^x_{i,j}\Delta f_{i-1,j}^* 
  + \mbL^y_{i,j}\Delta f_{i,j-1}^* 
  = \text{RHS}_{ij}, \\
  \text{Backward:}& \quad \Delta f_{ij}^k = \Delta f_{ij}^* 
  - \mbD_{i,j}^{-1} \mbU^x_{i,j}\Delta f_{i+1,j}^*
  - \mbD_{i,j}^{-1} \mbU^y_{i,j}\Delta f_{i,j+1}^*,
\end{align}
and the VDF is then updated as $f^{k+1}_{i,j} = f^{k}_{i,j} + \Delta f^{k}_{i,j}$. 

\subsection{Numerical scheme for the macroscopic synthetic  equations}\label{sec:lusgs}

The macroscopic synthetic  equations~\eqref{conservativeeq} can be viewed as compressible NSF equation with HoTs as constant source terms, where the steady-state solution  can be obtained by using sophisticated time-implicit schemes and shock capturing schemes. Again, we use the LU-SGS technique to handle the implicit time stepping in a matrix-free manner. 

Integrating Eq.~\eqref{conservativeeq} in a control volume $\Omega$ of the finite volume mesh and applying the Gauss theorem, we have
\begin{equation}\label{macrointegral}
 \frac{\partial}{\partial t} \int_{\Omega} \vec W \mathrm{d} \Omega+\oint_{\partial \Omega}\left[\vec F_{\mathrm{c}}+\vec F_{\mathrm{v}}(\bm \sigma_{\text{NSF}}, \vec q_{\text{NSF}})\right] \mathrm{d} \vec S =- \oint_{\partial \Omega} \vec F_{\mathrm{v}}^{\mathrm{HoT}}
 \mathrm{d} \vec S,
\end{equation}
where $\vec W$ is the vector of conservative variables and $\vec F_\text{c}$ is the vector of convective fluxes:
\begin{equation}
\vec{W}=\left[\begin{array}{c}{\rho} \\ {\rho U_x} \\ {\rho U_y} \\ {\rho E}\end{array}\right], 
\quad \vec{F}_{\mathrm{c}}=\left[\begin{array}{c}{\rho V} \\ {\rho U_x V+ n_{x} p/2} \\ {\rho U_y V+n_{y} p/2} \\ {\rho H V}\end{array}\right].
\end{equation}
Here, $H=E+p/\rho$, $V = \vec U \cdot \vec n$ with $\vec n$ being the unit normal vector of $\mathrm{d}\vec S$, and 
\begin{equation}
\vec F_\mathrm{v}(\bm \sigma, \vec q \,) = \left[\begin{array}{c}{0} \\{n_{x} \sigma_{x x}/2+ n_{y} \sigma_{x y}/2} \\ {n_{x} \sigma_{y x}/2+n_{y} \sigma_{y y}/2} \\ {n_{x} \Theta_{x}(\bm \sigma, \vec q\,)+n_{y} \Theta_{y}(\bm \sigma, \vec q\,)}\end{array}\right], 
\end{equation}
where $\Theta_{x}(\bm \sigma, \vec q\,)=-U_x \sigma_{x x}-U_y \sigma_{x y}+ q_{x}$ and $\Theta_{y}(\bm \sigma, \vec q\,)=-U_x \sigma_{y x}-U_y \sigma_{y y}+ q_{y}$.
In the RHS of Eq.~\eqref{macrointegral}, $\vec{F}^{\text{HoT}}_{\mathrm{v}} \equiv \vec{F}_{\mathrm{v}}(\bm\sigma^*,\vec q^{}\hspace{0.1em}*) - \vec{F}_{\mathrm{v}}(\bm\sigma^*_{\text{NSF}},\vec q^{\hspace{0.1em}*}_{\text{NSF}}) $ is the viscous flux due to the HoTs in shear stress and heat flux. The starred variables have the same meaning as in Eq.~\eqref{hotdiffint}.

Applying implicit scheme for the fluxes at the LHS of Eq.~\eqref{macrointegral}, we have, for each cell,
\begin{equation}\label{macroimplicit}
 \left[ \left(\frac{\Omega}{\Delta t_p}\right)_{i,j} \vec I + \left(\pp{\vec R}{\vec W}\right)_{i,j} \right]\Delta \vec W^{n}_{i,j}= -\vec R^n_{i,j} + \vec R_{i,j}^\mathrm{HoT},
\end{equation}
where $\Delta{t_p}$ is the pseudo time step, $\vec I$ is the identity matrix, and $\vec R$ stands for the residues including the one in NSF equation and the one due to HoTs:
\begin{equation}
\begin{aligned}
 \vec R_{i,j}^n = \sum_{m\in N(i,j)}[\vec F_\mathrm{c}^n + \vec F_\mathrm{v}(\bm \sigma^n_{\mathrm{NSF}},\vec q^{\:n}_{\mathrm{NSF}} )]_m \Delta S_m,\\ 
 \vec R_{i,j}^{\mathrm{HoT}} = \sum_{m\in N(i,j)}(\vec F_\mathrm{v}^{\mathrm{HoT}})_m \Delta S_m,
 \end{aligned}
\end{equation}
with the index $m$ looping through all faces of the current cell, represented by $N(i,j)$. As the iteration converges ($\Delta \vec W^n$ approaches to zero), the RHS of Eq.~\eqref{macroimplicit} also approaches to zero.

The LU-SGS technique employs a factorization of the implicit operator in Eq.~\eqref{macroimplicit} as
\begin{equation}
 (\mathbf{D} + \mathbf L)\mathbf D^{-1}(\mathbf D + \mathbf U) \Delta \vec W^n = -\vec R^n + \vec R^{\mathrm{HoT}}.
 \end{equation}
Note that the symbols $\mathbf{D}$, $\mathbf{L}$ and $\mathbf{U}$ are different from the ones in the DVM~\eqref{dluf}.
The solving of the linear equation system in terms of $\Delta \vec W^n$ can be easily executed as  a forward sweep and a backward sweep procedure on a structured mesh~\cite{blazekComputationalFluidDynamics2015}, as only the lower- or upper-half matrix coefficients are non-zero,
 \begin{align}
  \begin{split}
  \text{Forward:\quad} & (\mathbf{D} + \mathbf{L}) \Delta \vec W^{(1)} = -\vec R^{n}+ \vec R^{\mathrm{HoT}},\\
  \text{Backword:\quad} &(\mathbf{D} + \mathbf{U}) \Delta \vec W^n = \mathbf{D}\Delta \vec W^{(1)},
  \end{split}
 \end{align}
where, for each cell the lower, the upper and diagonal matrix elements are:
\begin{align}
 \begin{split}
\mathbf{L}_{i,j}=&\left(\bar{A}^{+}+\bar{A}_{\mathrm{v}}\right)_{i-1} \Delta S_{i-1 / 2}+\left(\bar{A}^{+}+\bar{A}_{\mathrm{v}}\right)_{j-1} \Delta S_{j-1 / 2},\\
\mathbf{U}_{i,j}=&\left(\bar{A}^{-}-\bar{A}_{\mathrm{v}}\right)_{i+1} \Delta S_{i+1 / 2}+\left(\bar{A}^{-}-\bar{A}_{\mathrm{v}}\right)_{j+1} \Delta S_{j+1/ 2}, \\
 \mathbf{D}_{i,j}=& \frac{\Omega}{\Delta t} \mathbf{I}+\left(\bar{A}^{-}-\bar{A}_{\mathrm{v}}\right) \Delta S_{i-1 / 2}+\left(\bar{A}^{-}-\bar{A}_{\mathrm{v}}\right) \Delta S_{j-1 / 2} \\&+\left(\bar{A}^{+}+\bar{A}_{\mathrm{v}}\right) \Delta S_{i+1 / 2} +\left(\bar{A}^{+}+\bar{A}_{\mathrm{v}}\right) \Delta S_{j+1 / 2},
 \end{split}
\end{align}
with $\bar A^{\pm}$ being the positive and negative convective flux Jocabian due to the flux-vector splitting scheme, and $\bar A_\mathrm{v}$ the viscous flux Jocabian. 
During the forward and backward sweeps, the product of  convective flux Jocabian and change of conservation variables can be approximated as~\cite{blazekComputationalFluidDynamics2015}:
\begin{equation}
 (\bar{A}^{\pm} \Delta S)\Delta \vec W^n \approx \frac{1}{2}\left(\Delta \bar{F}^n_{\mathrm{c}} \Delta S \pm r_{A} \bar{I}\Delta \vec W^n \right), \quad r_{A}=w \hat{\Lambda}_{\mathrm{c}},
 \end{equation}
where $\Delta F$ is the change of convective flux due to the change of conservative variables, $\hat \Lambda_c$ is the convective flux Jocabian's spectral radius, and $w$ is the over-relaxation factor  in the range of $1 < w \leq 2$. Higher $w$ increases the stability but slows down the convergence speed. In this paper, we use $w = 1$. Depending on the orientation of the interface ($I$- or $J$-direction), $\hat \Lambda_c$ is evaluated as
\begin{equation}
 \hat{\Lambda}^I_{\mathrm{c}} = (| \bU_{i,j} \cdot \bn_{I}| + c_{i,j})\Delta S_{I} \quad\text{and}\quad \hat{\Lambda}^J_{\mathrm{c}} = (| \bU_{i,j} \cdot \bn_{J}| + c_{i,j})\Delta S_{J}, 
\end{equation}
where $\bn_{I} =\left( \bn_{i+1/2,j} + \bn_{i-1/2,j} \right)/2$, $\Delta S_{I} = \left( \Delta S_{i+1/2,j} + \Delta S_{i-1/2,j} \right)/2$ with $c_{i,j}$ being the sound of speed and $\bm n_{I\pm1/2,j}$ being the right/left face normal vector.  Similar definition are used for the $J$-oriented face. The viscous flux Jocabian are approximated by its spectral radius, i.e. $\bar{A}_{\mathrm{v}} \Delta S \approx \hat{\Lambda}_{\mathrm{v}}$, and for the $I$- or $J$-oriented  faces,
\begin{equation}
 \hat{\Lambda}^{I}_{\mathrm{v}}=\max \left(\frac{4}{3 \rho_{i,j}}, \frac{\gamma}{\rho_{i,j}}\right)\left(\frac{\mu_{i,j}}{\operatorname{Pr}}\right) \frac{(\Delta S_I)^{2}}{\Omega_{i,j}}, \quad \text{and} \quad 
 \hat{\Lambda}^{J}_{\mathrm{v}}=\max \left(\frac{4}{3 \rho_{i,j}}, \frac{\gamma}{\rho_{i,j}}\right)\left(\frac{\mu_{i,j}}{\operatorname{Pr}}\right) \frac{(\Delta S_J)^{2}}{\Omega_{i,j}},
\end{equation}
where $\gamma = c_\text{p}/c_\text{v}$. 

For the explicit calculation of viscous flux in $R^n$, we use the MUSCL3 reconstruction scheme and 2nd-order Roe flux scheme, while the 2nd-order central scheme is adopted for the viscous flux computation. The time step is determined according to 
\begin{equation}
 \Delta t_p = \alpha \frac{\Omega}{\hat\Lambda_c^I + \hat\Lambda_c^J + \hat\Lambda_v^I + \hat\Lambda_v^J },
\end{equation}
where $\alpha$ is the Courant-Friedrichs-Lewy (CFL) number. 

\subsection{Updating of  macroscopic variables and correction to the VDF}
The converged solution of  macroscopic variables of the  synthetic equations is used in the next DVM step to calculate the equilibrium VDF. A relaxation coefficient $0 \le \beta < 1$ is introduced in the updating processes~\eqref{GSIS_K3} to improve the stability of GSIS for high Kn flows,
\begin{equation}\label{updatemacros}
\vec W^{\,k+1} = \beta \vec W' + (1-\beta) \vec W^{k,\,*}, \quad \bm \sigma^{\hspace{0.1em}k+1} = \beta \bm \sigma' + (1-\beta) \bm \sigma^{k,\hspace{0.1em}*}, \quad \vec q^{\hspace{0.1em}k+1} =  \beta \vec q\,' + (1-\beta) \vec q ^{\hspace{0.14em}k,*},
\end{equation}
where $\vec W'$, $\bm \sigma'$ and $\vec q$ are the converged macroscopic solution of the inner loop between the $k$- and $(k+1)$-th DVM steps, and $\vec W^{k,\hspace{0.1em}*}$, $\bm \sigma^{\hspace{0.1em}k,*}$ , $\vec q^{\hspace{0.1em}k,*}$ are calculated by numerical quadratures after $k$-th DVM step. In the practical numerical simulations, the relaxation coefficient is adapted according to a local NSF breakdown parameter~\cite{mengBreakdownParameterKinetic2014}:
\begin{equation}
\beta  =  1 - \text{min}(1,E_c^\text{NSF}), \quad  E_c^\text{NSF} = \sqrt{ \frac{ \int(f-f^{\text{G13}}) \text{d}^3 \vec \xi}{\int(f^m)^2 \text{d}^3 \vec \xi}},
\end{equation}
where $f^{\text G13}$ is the VDF reconstructed following the one used in Grad 13-moment method. For continuum flows, $E_c^\text{NSF}$ approaches to zero and $\beta$ approaches to 1, which means the macroscopic variables in the DVM are almost entirely replaced by the solution of macroscopic synthetic equations. For high Kn flows, $E_c^\text{NSF}$ may be higher than 1 and $\beta$ becomes zero, so the solution of macroscopic synthetic equationS is not used in the DVM and the GSIS reduces to CIS which is already efficient for these flows.

The VDF is also adjusted to reflect the changes of  leading-order moments $\vec W$. This is achieved by replacing equilibrium part of the VDF with the one computed from the new moments, while keeping the non-equilibrium part unchanged:
\begin{equation}\label{correctvdf}
\begin{aligned}
  f^{k+1} = f^k+ \beta \left[ f^m(\vec W^{k+1}) - f^m(\vec W^{\hspace{0.1em}k,*})\right]. 
  \end{aligned}
\end{equation}


\subsection{Overview of the GSIS algorithm}\label{sec:overview}

Here we summarize the GSIS algorithm for nonlinear gas kinetic equation proposed. The overall computing procedure is a nested loop, with the outer and inner loop indexes as $k$ and $n$, as illustrated in Fig.~\ref{fig:gsis_algo}. The outer loop solves the gas kinetic equation with the iterative or time-stepping DVM method, and the inner loop solves the macroscopic synthetic equations using the LU-SGS technique. Each inner loop starts from the latest macroscopic state, together with HoTs and boundary conditions from the current step in the outer loop. The step-by-step procedures are listed as below,

\begin{enumerate}
 \item Initialize macroscopic variables in both the DVM and macroscopic equation solvers.
 \item Initialize VDF in the DVM solver.
 \item Solve the NSF equations (the macroscopic equation with HoTs as zero) to its converged state.
 \item Execute one iterate/time step in the DVM solver, in which the latest converged macroscopic variables are used to compute equilibrium.
 \item Calculate the HoTs of shear stress and heat flux from VDF via Eq.~\eqref{hotdirect}. Calculate the macroscopic boundary conditions from the VDF on the boundary.
 \item Solve the macroscopic synthetic equation (with the HoTs and boundary conditions from step 5) using the LU-SGS technique to the converged state.
 \item Update the macroscopic variables and VDF in DVM from the solution in step 6.
 \item Repeat steps 4 to 7 until meeting the defined convergence criterion of the outer loop via Eq.~\eqref{updatemacros} and \eqref{correctvdf}.
 \end{enumerate}





\section{Numerical test cases}\label{sec:tests}

Several 1D and 2D flows are simulated to investigate the accuracy and efficiency of the nonlinear GSIS. In the 1D Fourier flow and Couette flow, the macroscopic synthetic equations can be greatly simplified and solved without resorting to the LU-SGS technique in Sec.~\ref{sec:lusgs}. The 2D cases include the lid-driven cavity flow and supersonic flow past a cylinder, where the gas kinetic equation is solved by the upwind finite difference method and the method in Section~\ref{sec:dvm2}, respectively. In all test cases, $\Pr = 2/3$ and the viscosity index is $\omega = 0.81$. All parameters used in CIS and GSIS are the same, thus we can evaluate the GSIS's efficiency compared with CIS. 

\subsection{Reduced Shakhov model equation}

For 2D flows, the VDF can be reduced to save the computational cost. We introduce the following two reduced VDFs as:
\begin{equation}
\begin{aligned}
g(t,x,y,\xi_x,\xi_y) = \int_{\mathbb{R}} f(t,\vec r, \vec \xi\,) \mathrm{d} \xi_z, \\
 h(t,x,y,\xi_x,\xi_y) = \int_{\mathbb{R}} \xi_z^{2} f(t,\vec r, \vec \xi\,) \mathrm{d} \xi_z,
 \end{aligned}
\end{equation}
whose dynamics are described by the following reduced Shakhov model equation:
\begin{equation}\label{reducedsmodel}
\pp{\Phi}{t} + \vec \xi \cdot \vec \nabla \Phi = \frac{1}{\tau} \left[ \Phi^s - \Phi \right], 
\end{equation}
with $\Phi \equiv [g, h]^T$, $\Phi^s \equiv [g^s, h^s]^T$, and 
\begin{align}
g^s &= \frac{\rho}{2\pi RT}\exp \left( -\frac{C^2}{2RT}\right) \left[ 1 + (1-\Pr)\frac{\vec q \cdot \vec C}{5pRT}\left( \frac{C^2}{RT} - 4 \right) \right], \\
h^s &= \frac{\rho}{2\pi RT}\exp \left( -\frac{C^2}{2RT}\right) \left[ 1 +(1-\Pr)\frac{\vec q \cdot \vec C}{5pRT}\left( \frac{C^2}{RT} - 2 \right) \right]RT.
\end{align}

Note that now all vectors have only two components, e.g. $\vec \xi = (\xi_x, \xi_y)$, $\vec C = (C_x, C_y)$, and $C^2 = C_x^2 + C_y^2$. The macroscopic variables are now calculated as:
\begin{align}\label{macrosfromgh}
\begin{split}
& \rho = \int g\dvv,\quad \rho \vec U = \int g \vec \xi \dvv, \quad \sigma_{ij} = \int \left (g C_{i}C_{j} - \frac{gC^2+h}{3} \delta_{ij} \right) \dvv, \\
& p = \frac{1}{3} \int (g C^2 + h) \dvv, \quad \vec q = \frac{1}{2}\int \vec C (g C^2 + h) \dvv. \\
\end{split}
\end{align}

\subsection{Heat transfer between two parallel plates}
Consider the steady heat transfer of gas confined between two vertically placed static parallel plates with a distance $L$, located at $x_\mathrm{L}=0$ and $x_\mathrm{R}=1$. The left and right plates are maintained at constant temperatures of $T_{\mathrm{L}} = 0.75$ and $T_{\mathrm{R}}= 1.25$, respectively (note that the temperature has been normalized by the reference temperature $T_0$). The boundary conditions of the VDF at $x_\mathrm{L} $ and $x_\mathrm{R}$ are 
\begin{subequations}\label{fbcfourier}
 \begin{align}
  g(\vec \xi )\vert_{x_{\mathrm{L}}, \xi_x > 0} &= \frac{\rho_{\mathrm{L}}}{2\pi RT_\mathrm{L}} \exp \left( -\frac{\xi^2}{2RT_\mathrm{L}} \right), \quad  h(\vec \xi )\vert_{x_{\mathrm{L}}, \xi_x > 0} = \frac{\rho_{\mathrm{L}}}{2\pi} \exp \left( -\frac{\xi^2}{2RT_\mathrm{L}} \right), \quad  \\
  g(\vec \xi)\vert_{x_\mathrm R, \xi_x < 0} &= \frac{\rho_\mathrm{R}}{2\pi RT_\mathrm R} \exp \left( -\frac{\xi^2}{2RT_\mathrm{R}} \right), \quad h(\vec \xi)\vert_{x_\mathrm R, \xi_x < 0} = \frac{\rho_\mathrm{R}}{2\pi} \exp \left( -\frac{\xi^2}{2RT_\mathrm{R}} \right),
  \end{align}
\end{subequations}
where
\begin{equation}
\rho_\mathrm{L} = -\frac{4}{\pi \sqrt{2RT_{\mathrm{L}}}} \int_{\mathbb{R}^2,~\xi_x < 0} \xi_x f \dvv, \quad \rho_\mathrm{R} = \frac{4}{\pi \sqrt{2RT_{\mathrm{R}}}}  \int_{\mathbb{R}^2,~\xi_x > 0} \xi_x g \dvv.
\end{equation}

\begin{figure}[t]
	\centering
	\includegraphics[width=0.48\textwidth]{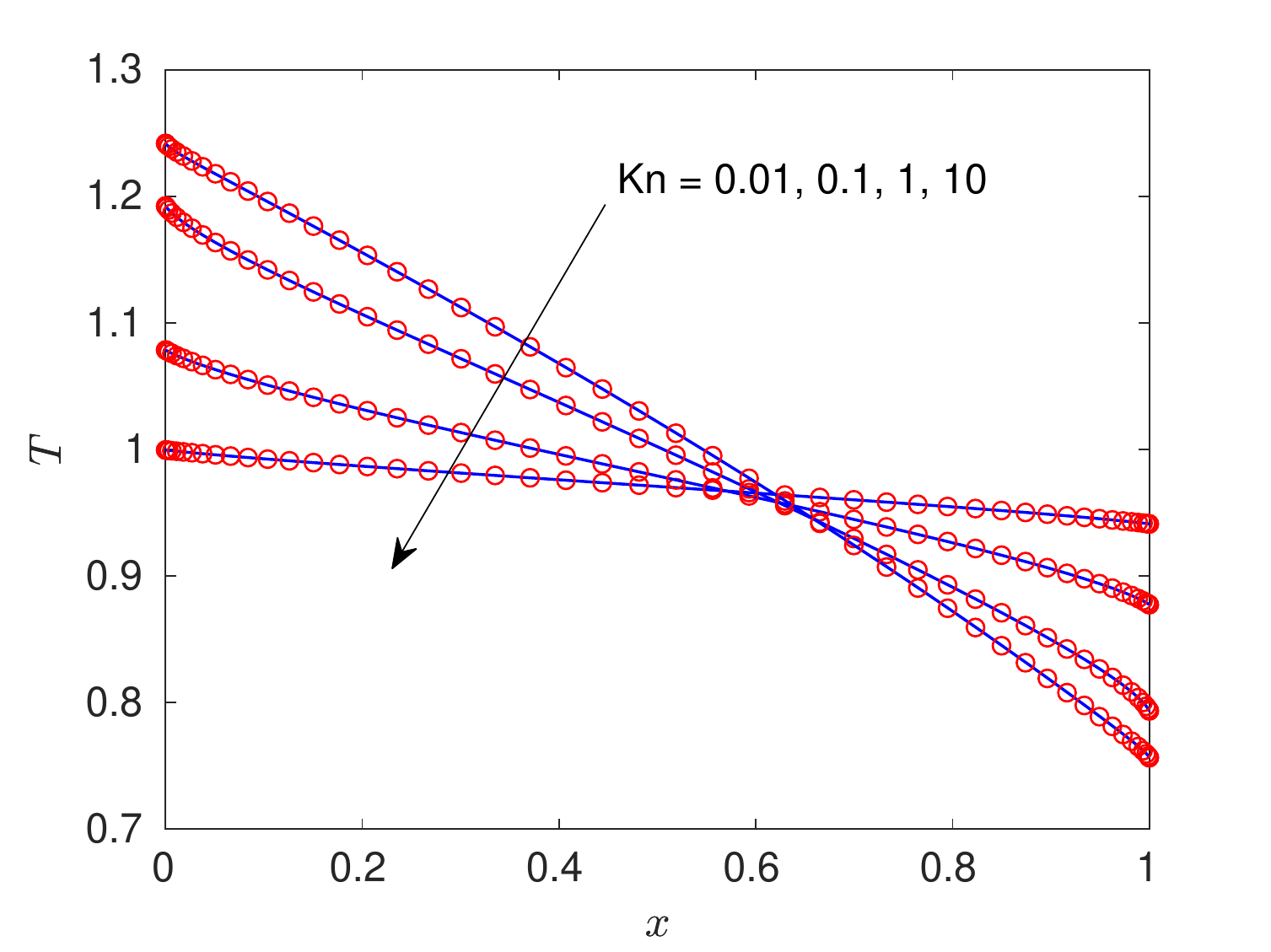}
	\includegraphics[width=0.48\textwidth]{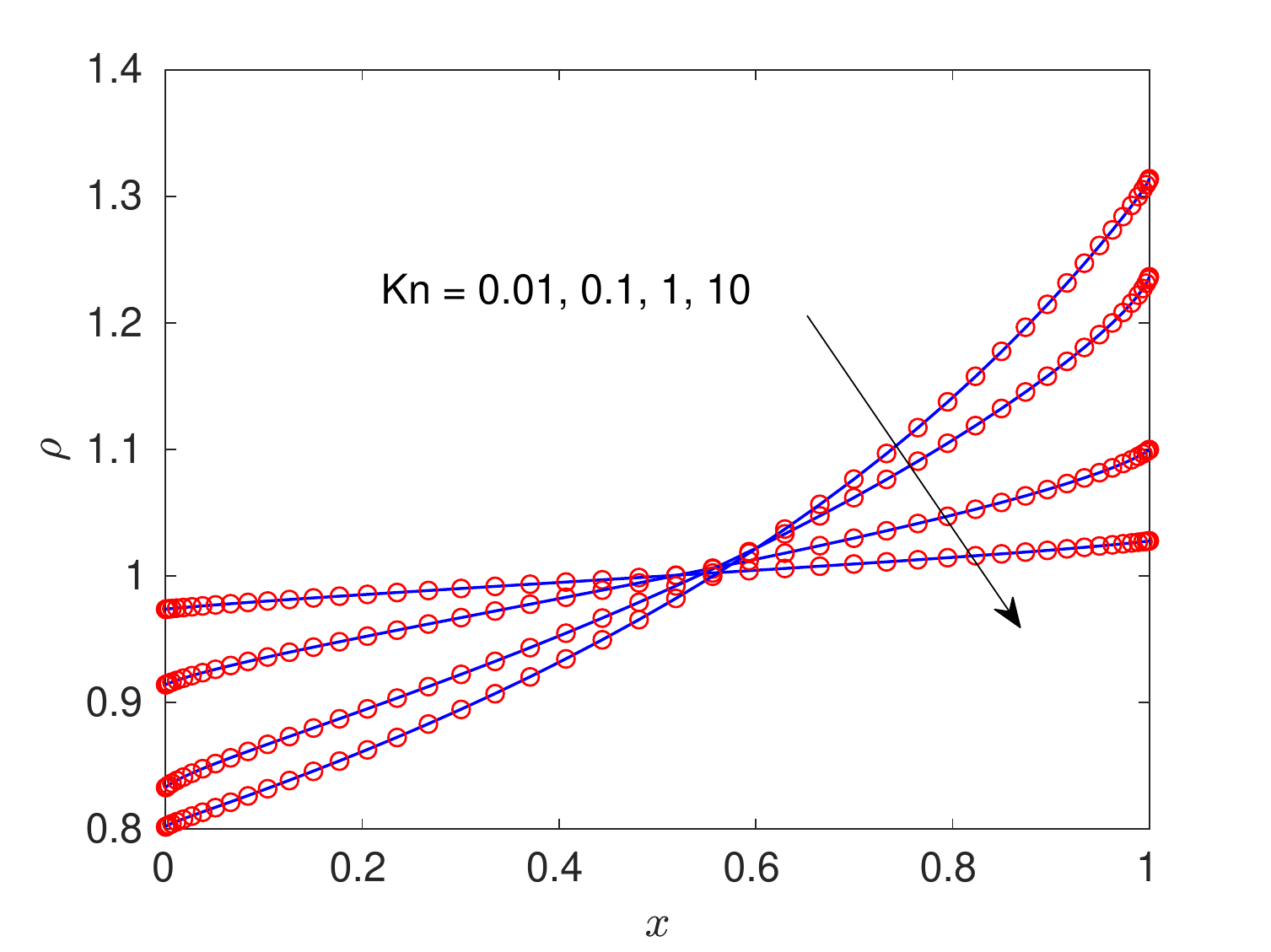}\caption{Temperature (a) and density (b) profiles of the Fourier flow at different Kn numbers. Solid lines and circles represent the CIS and GSIS results, respectively.}
	\label{fig:fourier_accuracy}
\end{figure}

For this 1D problem, we can solve the steady-state synthetic macroscopic equation easily in the following way. We know that $\vec{U}= 0, ~q_y= 0, ~\sigma_{xy} = \sigma_{yx} = 0, ~\partial \phi/\partial y = 0$, where $\phi$ is any quantity. The synthetic equation together with the idea gas equation of state then can be simplified as an equation system of the variables $\rho$, $T$, $p$, $\sigma_{xx}$ and $q_x$:
\begin{subequations}\label{macrofourier}
\begin{align}
   \pp{  p}{  x} + \pp{\sigma_{xx}}{  x} =0, \label{macrofourier:p} \\
   \pp{q_x}{ x} = 0, \label{macrofourier:qx}\\ 
   \sigma_{xx} = \HoT_{\sigma_{xx}} , \label{macrofourier:s11} \\
   q_x = \HoT_{q_x} - k \pp{T}{x}, \label{macrofourier:T} \\
  p = \rho R T \label{macrofourier:rho},
\end{align}
\end{subequations}
together with the boundary values of the variables provided after each DVM step. The HoTs are also computed from the VDF according to Eq.~\eqref{hotdirect} after the last DVM step as
 \begin{subequations}\label{hot_fourier}
  \begin{align}
   & \HoT_{\sigma_{xx}} = \frac{1}{3} \int\left(2g^*C_x^{*2} - g^*C_y^{*,2} - h^*\right) \dvv, \\
   &\HoT_{q_x} = \frac{1}{2}\int C^*_x \left({g^*C^{*,2} + h^* }\right)\dvv + \kappa^* \pp{T^{*}}{x},
   \end{align}
\end{subequations}
where starred variables are from the VDF of the last DVM step, including $T^*$, $k^*$, and $\vec C^{\sstar} \equiv \vec \xi - \vec U^{\sstar}$.
Equation~\eqref{macrofourier} can be directly solved by sequentially solving $\sigma_{xx}$ from Eq.~\eqref{macrofourier:s11}, $q_x$ from Eq.~\eqref{macrofourier:qx}, $T$ from Eq.~\eqref{macrofourier:T}, $p$ from Eq.~\eqref{macrofourier:p}, and $\rho$ from Eq.~\eqref{macrofourier:rho}.

The Knudsen number is defined as $\text{Kn} ={\sqrt{\pi}\mu C_0}/({2p_0 L_0})$ with $\mu = \text{Pr} k/c_\text{p}$, where the reference pressure is $p_0 = \rho_0 RT_0 = 0.5$,  the reference density is $\rho_0 = 1$, the specific gas constant is $R = 0.5$, the reference temperature is $T_0 = 1$, the reference length $L_0 = x_{\mathrm{R}} - x_{\mathrm{L}}=1$,  and the reference velocity $C_0 = \sqrt{2RT_0} = 1$.

\begin{figure}[t]
	\centering
	\includegraphics[width=0.6\textwidth]{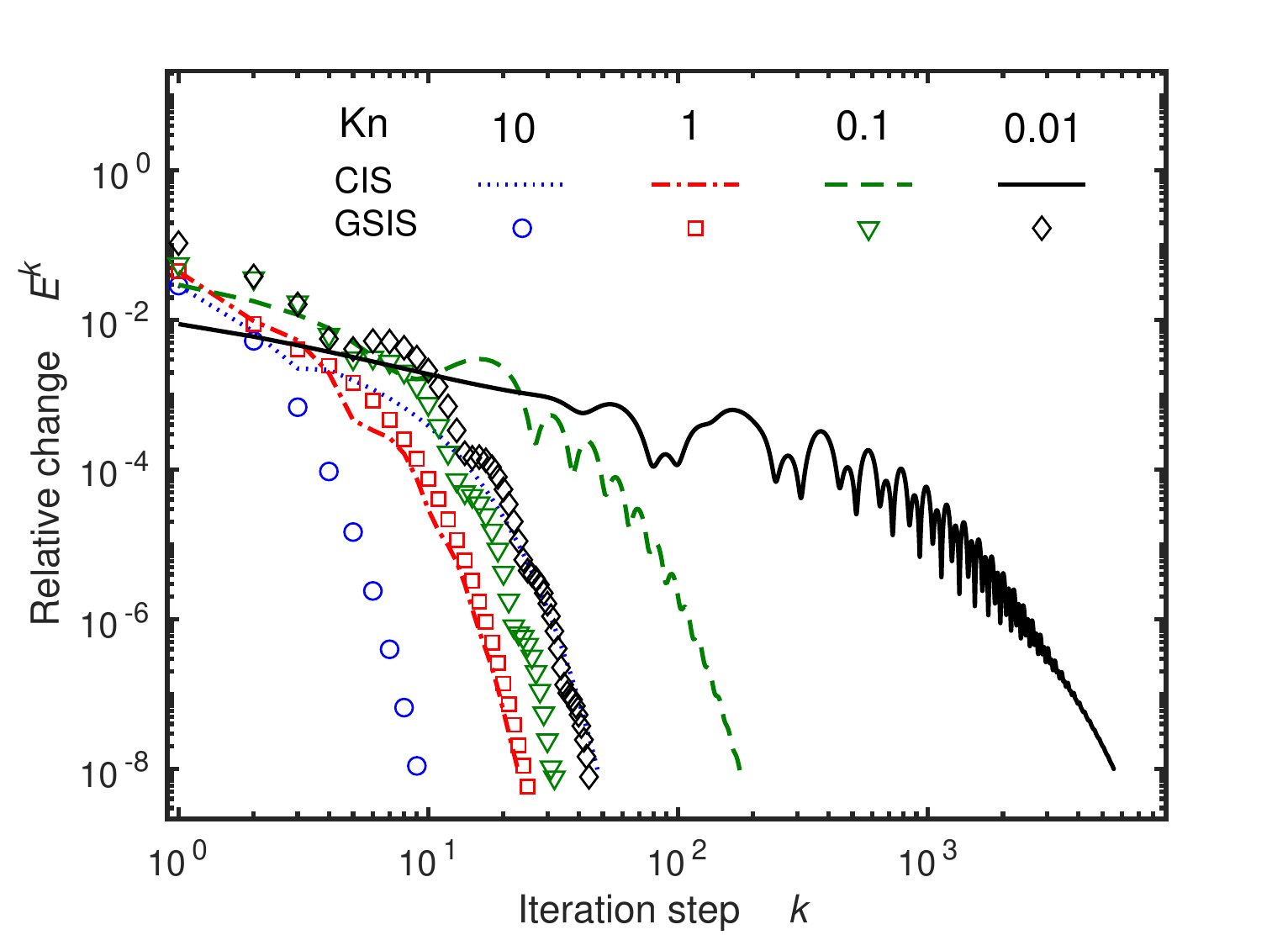}
	\caption{Convergence history of both GSIS and CIS in the simulation of nonlinear Fourier flow.}
	\label{fig:fourier_convergence}
\end{figure}

The velocity space in both $\xi_x$ and $\xi_y$ directions are truncated to $[-6, 6]$, and the discrete velocities are distributed on a non-uniform Cartesian grid with size of $N_V^2$, and the grid line positions determined by the following rule~\cite{wuSolvingBoltzmannEquation2014}:
\begin{equation}\label{wuleidv}
 \xi_x, \xi_y = \frac{6}{\left(N_{\mathrm{v}}-1\right)^{3}}\left[\left(-N_{\mathrm{v}}+1\right)^{3},\left(-N_{\mathrm{v}}+3\right)^{3}, \cdots,\left(N_{\mathrm{v}}-1\right)^{3}\right],
\end{equation}
where $N_V = 32$. Such a non-uniform grid can accurately capture the discontinuity of VDF near the origin of velocity space, which appears in the vicinity of solid walls when the Knudsen number is large. The spatial grid points are distributed in the x-direction non-uniformly according to the following rules:
\begin{equation}
 x_i = s_i^3(10-15s_i+6s_i^2), \quad \text{with} \quad s_i =i/2(N_x - 1), \quad i = 0,1,\ldots,N_x-1,
\end{equation}
where $N_x = 50$ is the number of grid points.
The convergence criterion for the DVM iteration is that the volume-weighted relative change of temperature between two successive iteration steps satisfies
\begin{equation}
 E^k = \frac{\sqrt{\sum_i (T_i^{k} - T_i^{k-1})^2 \Delta x_i}}{\sqrt{\sum_i (T_i^{k-1})^2 \Delta x_i}} < \num{1e-8}.
\end{equation}
The temperature field is chosen is because it converges slower than other macroscopic fields.

Figure~\ref{fig:fourier_accuracy} shows the converged temperature and density profiles calculated by GSIS and CIS. We can see that the results obtained from GSIS agree well with those from CIS. The convergence history of DVM iterations in both GSIS and CIS is shown in Fig.~\ref{fig:fourier_convergence}. It is clear that the CIS is efficient in high Kn cases ($\text{Kn} > 1$), where converged solution can be found within 20 iterations. However, it  becomes very inefficient as $\text{Kn} < 0.1$, for example, it takes 5,000 iterations to produce the converged solution. On the contrary, GSIS converges in less than 60 steps for all Kn cases.

\begin{figure}[t]
	\centering
	\includegraphics[width=0.45\textwidth]{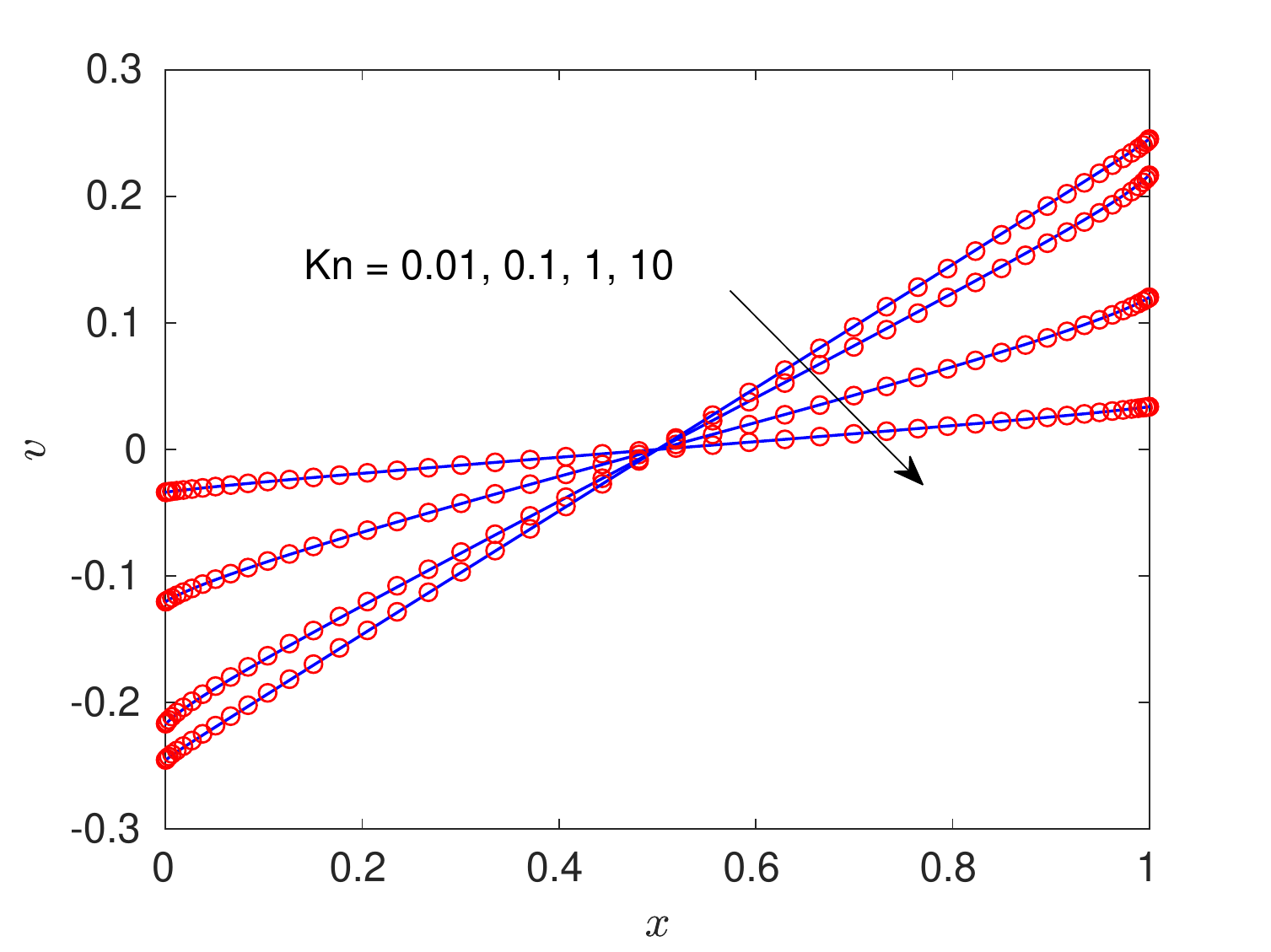}
	\includegraphics[width=0.45\textwidth]{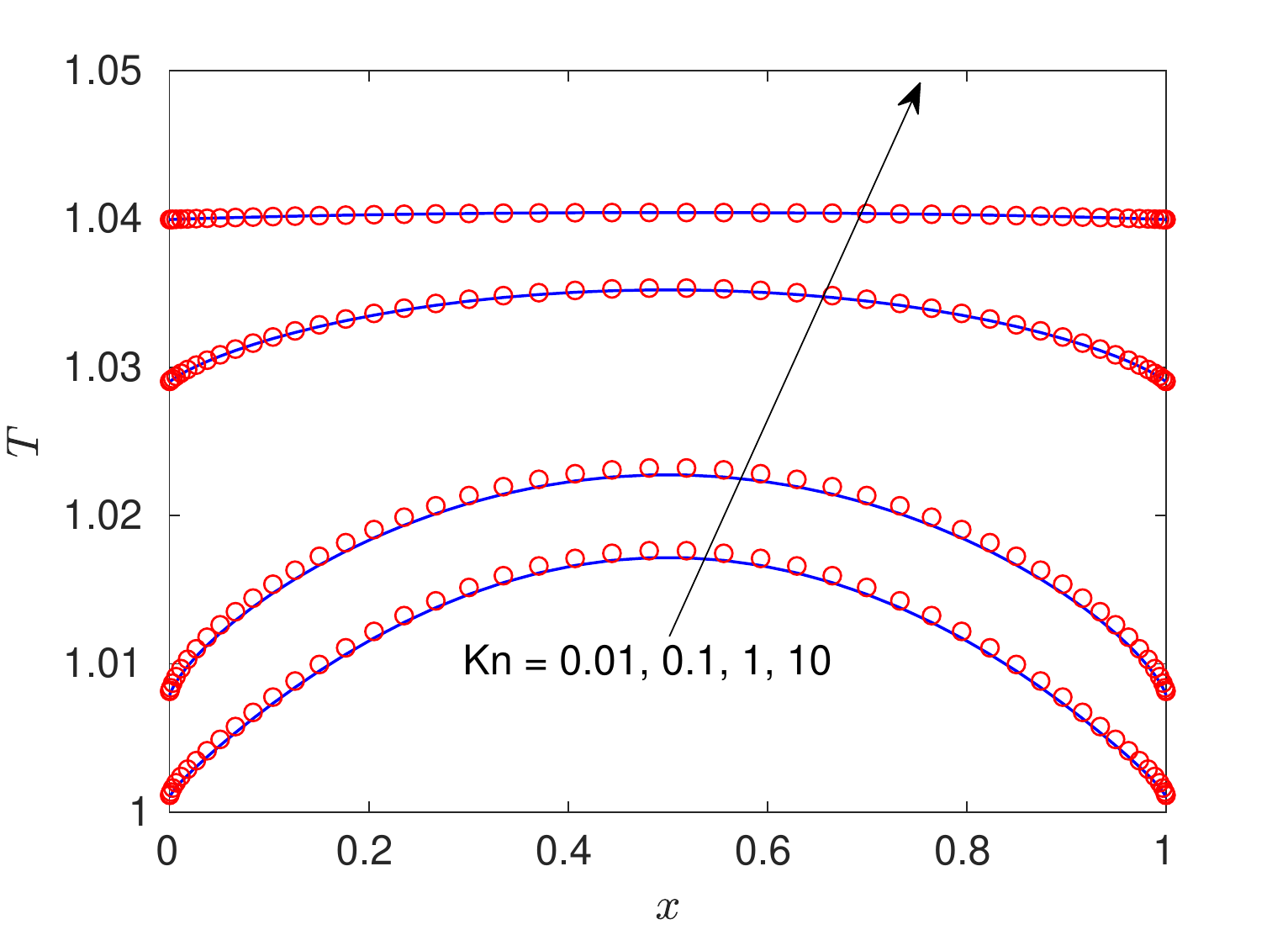} \caption{Velocity (a) and temperature (b) profiles in the nonlinear Couette flow. Solid lines and circles represent the CIS and GSIS results, respectively.}
	\label{fig:couette_accuracy}
\end{figure}

\subsection{Couette flow}
The Couette flow has the same geometry configuration with the Fourier flow, but now the two plates have the same temperature of $T_0 = 1$ and different vertical velocities: $U_{y,\mathrm{L}}= -0.25$ and $U_{y,\mathrm{R}} = 0.25$, respectively. The boundary conditions of the VDF at $x_\mathrm{L}$ and $x_\mathrm{R}$ are 
\begin{subequations}\label{fbccouette}
 \begin{align}
  &g(\vec \xi\,)\vert_{x_\mathrm{L}, \xi_x > 0} = \frac{\rho_\mathrm{L}}{2 \pi R T_{0} } \exp\left[-\frac{ \xi_x^2 + (\xi_y - U_{y,\mathrm{L}})^2 + \xi_z^2 }{2RT_0} \right], \\
    &  h(\vec \xi\,)\vert_{x_\mathrm{L}, \xi_x > 0} = RT_0 g(\vec \xi\,)\vert_{x_\mathrm{L}, \xi_x > 0}, \\
  &g(\vec \xi\,)\vert_{x_\mathrm{R}, \xi_x < 0} = \frac{\rho_\mathrm{R}}{2 \pi R T_{0} } \exp \left[-\frac{ \xi_x^2 + (\xi_y - U_{y,\mathrm{R}})^2 + \xi_z^2 }{2RT_0} \right], \\
  &  h(\vec \xi\,)\vert_{x_\mathrm{R}, \xi_x < 0} = RT_0 g(\vec \xi\,)\vert_{x_\mathrm{R}, \xi_x < 0},
  \end{align}
  \end{subequations}
 with
\begin{equation}
\rho_\mathrm{L} = -\frac{4}{\pi \sqrt{2RT_{\mathrm{L}}}} \int_{\mathbb{R}^2,~\xi_x < 0} \xi_x f \dvv,\quad \text{and} \quad \rho_\mathrm{R} = \frac{4}{\pi \sqrt{2RT_{\mathrm{R}}}} \int_{\mathbb{R}^2,~\xi_x > 0} \xi_x f \dvv.
 \end{equation}

\begin{figure}[t]
	\centering
	\includegraphics[width=0.6\textwidth]{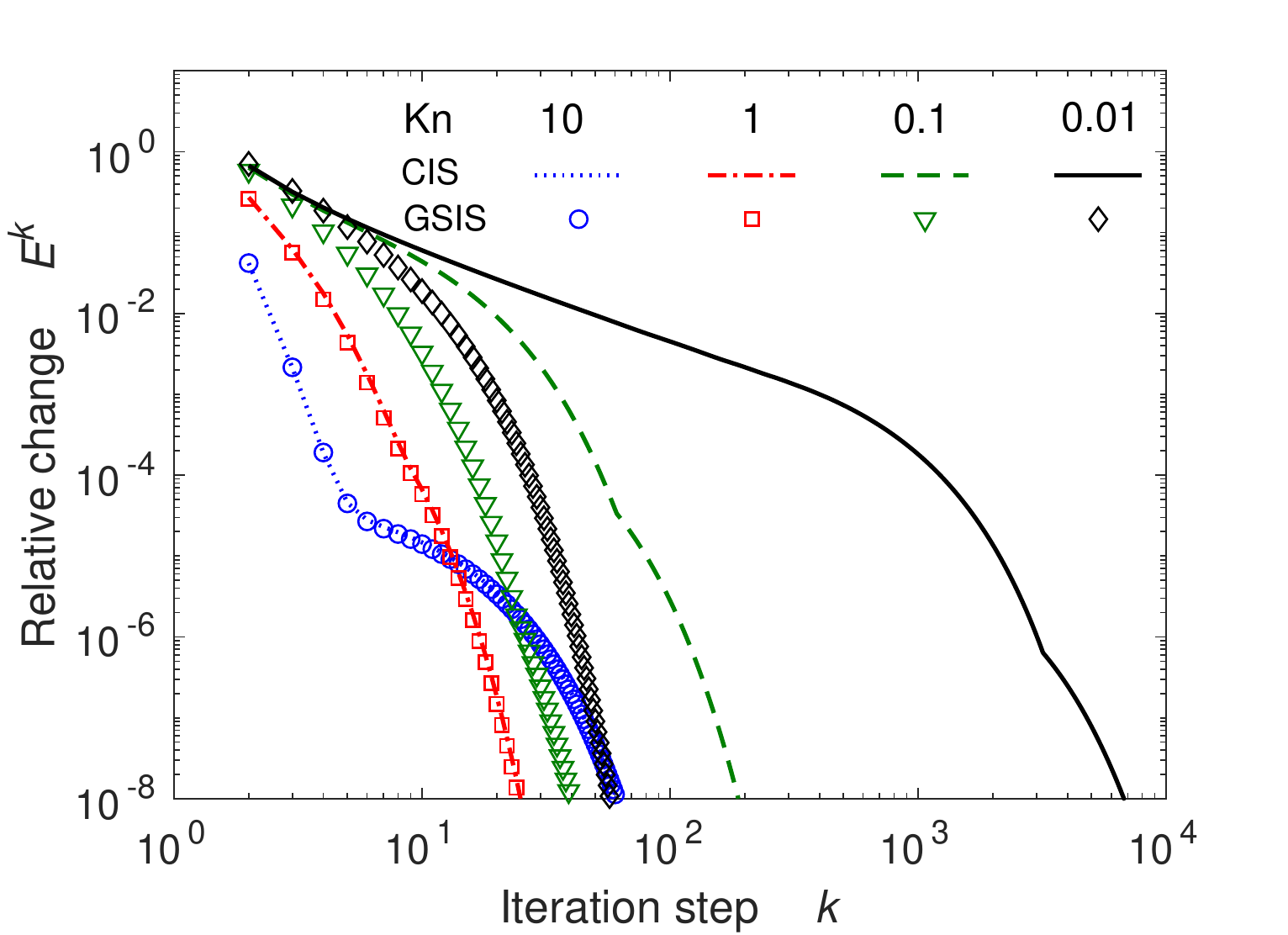}
	\caption{Convergence history of both GSIS and CIS in the simulation of nonlinear Couette flow. }
	\label{fig:couette_convergence}
\end{figure}
 
Similar to the Fourier flow, we solve the macroscopic synthetic equations in the following simplified way.
We know that $U_x = 0$, $q_y = 0$, $\partial \phi /\partial y = 0$ where $\phi$ can be any flow variable. Thus the synthetic equations can be simplified as an equation system of the variables $\rho$, $T$, $p$, $\sigma_{xx}$, $\sigma_{xy}$ and $q_x$ as: 
\begin{subequations}\label{macrocouette}
\begin{align}
  \pp{p}{x} + \pp{\sigma_{xx}}{x} = 0, \label{macrocouette:p} \\
  \pp{\sigma_{xy}}{x} = 0, \label{macrocouette:sxy}\\
  \pp{\sigma_{xy} U_y}{x} + \pp{q_x}{x}= 0, \label{macrocouette:qx} \\
  p = \rho R T, \label{macrocouette:rho}\\
  \sigma_{xy} = \HoT_{\sigma_{xy}} -\mu \pp{U_y}{x}, \label{macrocouette:v}\\
  \sigma_{xx} = \HoT_{\sigma_{xx}}, \label{macrocouette:sxx}\\
  q_x = \HoT_{q_x} -\kappa \pp{T}{x}, \label{macrocouette:t}
  \end{align}
\end{subequations}
where the HoTs are calculated explicitly according to Eq.~\eqref{hotdirect} as 
\begin{subequations}\label{hot_couette}
\begin{align}
 & \HoT_{\sigma_{xx }} = \frac{1}{3}\int \left(2g^*C_x^{*,2} - g^*C_y^{*,2} - h^*\right) \dvv, \\
 & \HoT_{\sigma_{xy}} = \int g^*C_x^* C_y^* \dvv + \mu^* \pp{ v^{*}}{x}, \\
&  \HoT_{q_x} = \int C_x^*(g^*C^{*,2} + h^*)\dvv  + \kappa^*\pp{ T^* }{x}.
 \end{align}
\end{subequations}
The unknown variables in Eq.~\eqref{macrocouette} can be solved in a sequential manner: $\sigma_{xx}$ from Eq.~\eqref{macrocouette:sxx}, $\sigma_{xy}$ from Eq.~\eqref{macrocouette:sxy}, $p$ from Eq.~\eqref{macrocouette:p}, $U_y$ from Eq.~\eqref{macrocouette:v}, $q_x$ from Eq.~\eqref{macrocouette:qx}, $T$ from Eq.~\eqref{macrocouette:t}, and $\rho$ from Eq.~\eqref{macrocouette:rho}.

The velocity-space grid, spatial space grid and the reference variables are set the same as in the Fourier flow cases. The Kn is defined as $\text{Kn} ={\sqrt{\pi}\mu_0 C_0}/({2p_0 L_0 })$, and we consider the cases of $\text{Kn} = 0.01$, 0.1, 1 and 10. The convergence criterion for the DVM iteration is that the volume-weighted relative change of temperature, density and velocity between two iteration steps are all less than $10^{-8}$,
\begin{equation}
 E^k = \left. \frac{\sqrt{\sum_i (\phi_i^{k} - \phi_i^{k-1})^2 \Delta x_i}}{\sqrt{\sum_i (\phi_i^{k-1})^2 \Delta x_i}} \right|_{\mathrm{max}} < \num{1e-8}, \quad \text{for} \quad \phi \in \{\rho, T, v\}.
\end{equation}

The converged velocity and temperature profiles predicted by GSIS and CIS are shown in Fig.~\ref{fig:couette_accuracy}, while the convergence history  is shown in Fig.~\ref{fig:couette_convergence}. It is seen that the GSIS converges in less than 70 steps for all cases, while the CIS needs much more steps when Kn decrease to 0.1. For example, when Kn=0.01, GSIS is faster than CIS by more than two orders of magnitude.

\subsection{Lid-driven cavity flow}

This problem has been simulated extensively for the validation of numerical schemes for gas kinetic equations. The flow domain is a square cavity with a size of $L_0\times L_0=1\times 1$. The top boundary (the lid) of the cavity moves horizontally in the $x$ direction with a velocity of $u_\mathrm w = 0.14828$, while the other walls are fixed. All solid walls are maintained at a uniform reference temperature of $T_\mathrm w = 1$, and are handled as Maxwellian diffusive boundaries in a similar way as in Eq.~\eqref{fbccouette}.

\begin{figure}[htbp]
 \centering
 \includegraphics[width=0.45\textwidth]{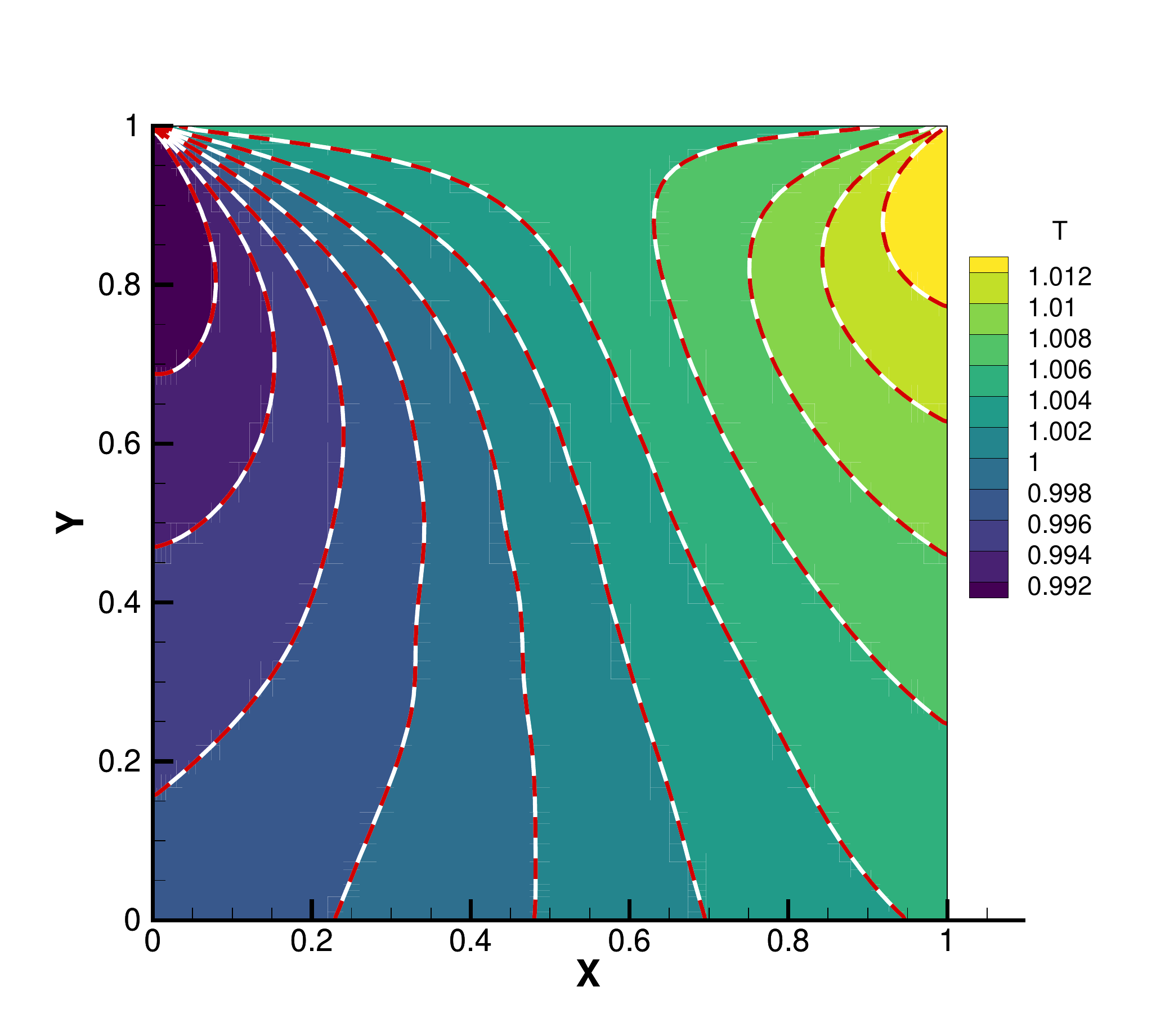}
 \includegraphics[width=0.45\textwidth]{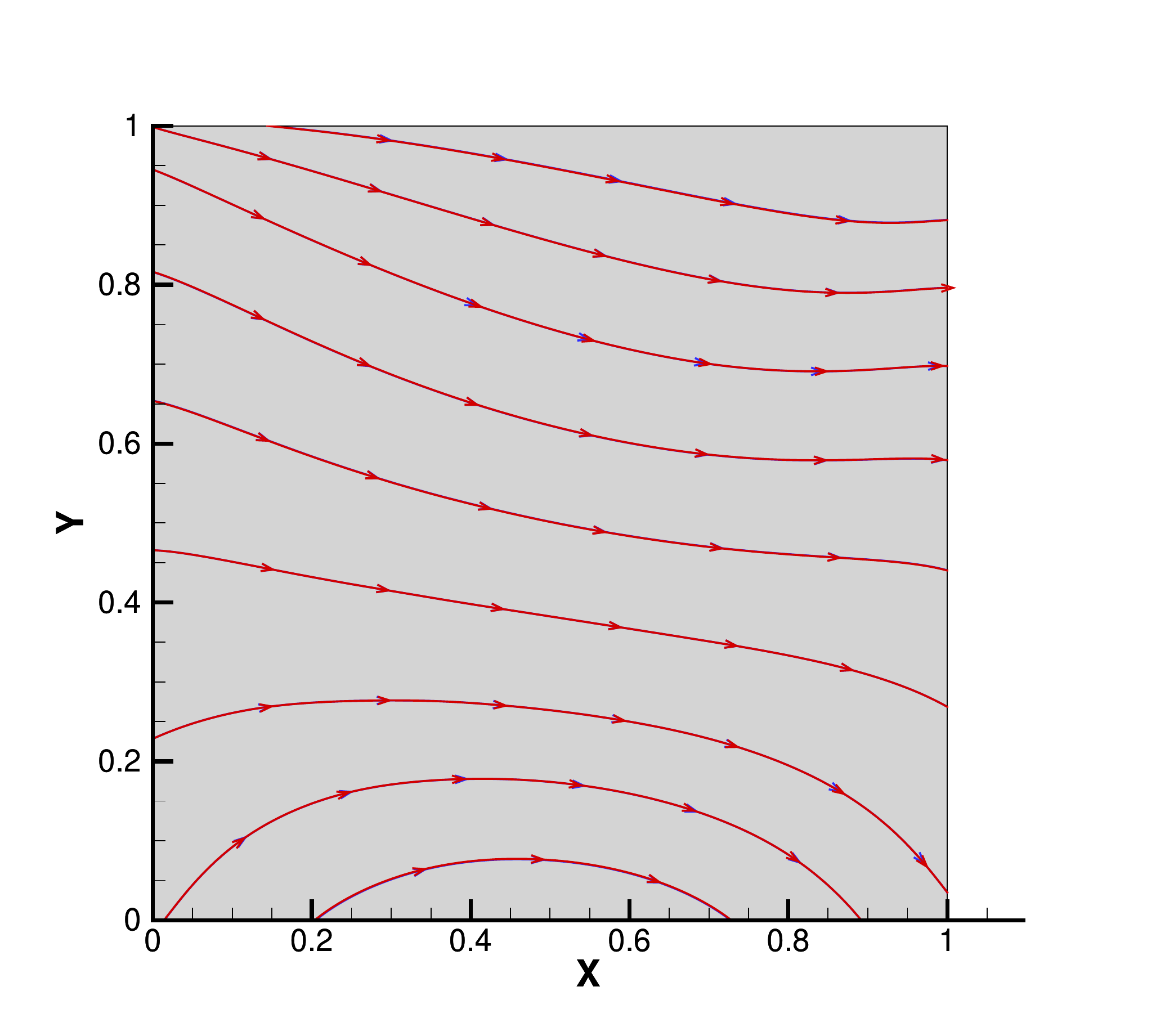}
 \includegraphics[width=0.45\textwidth]{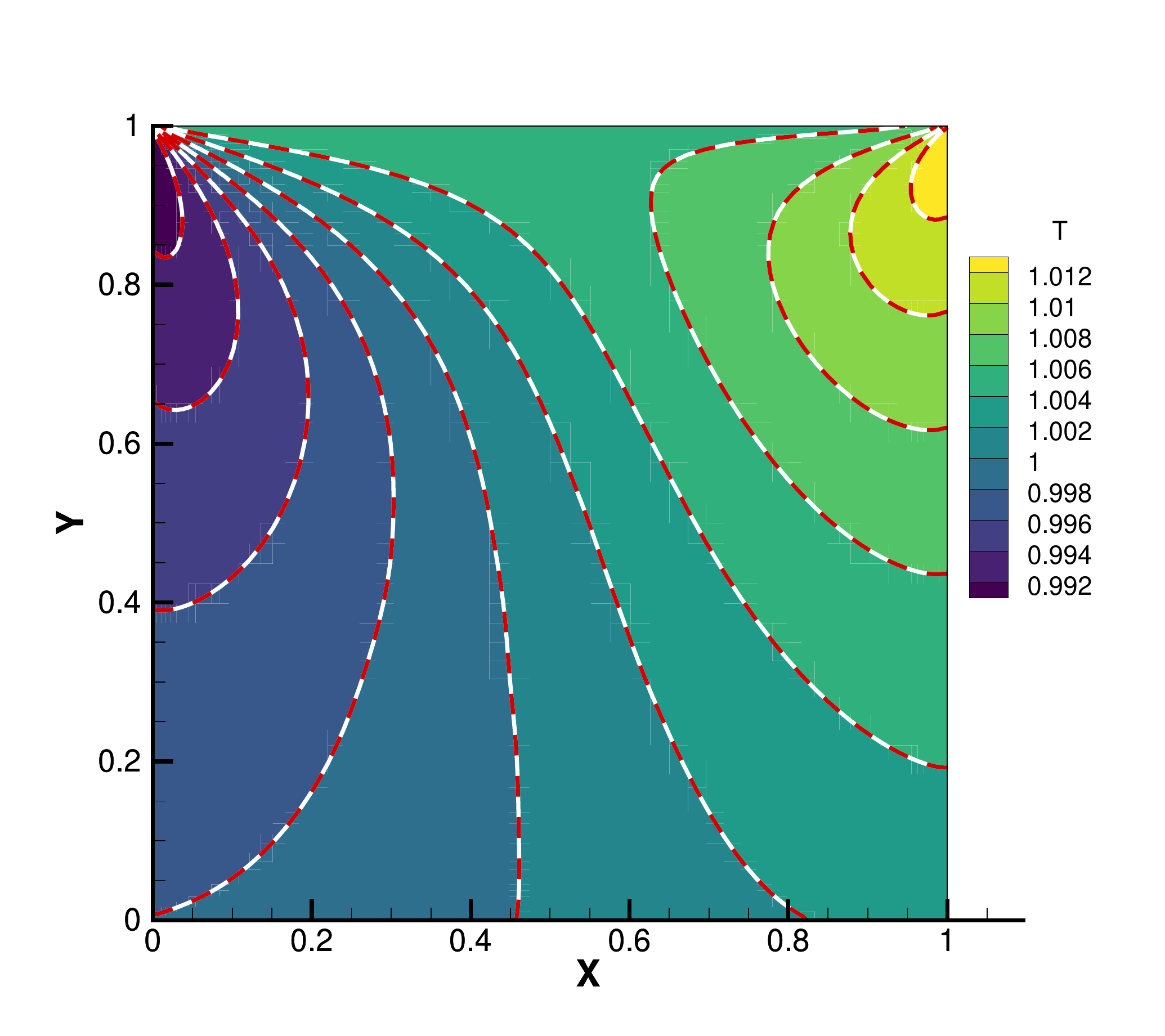}
 \includegraphics[width=0.45\textwidth]{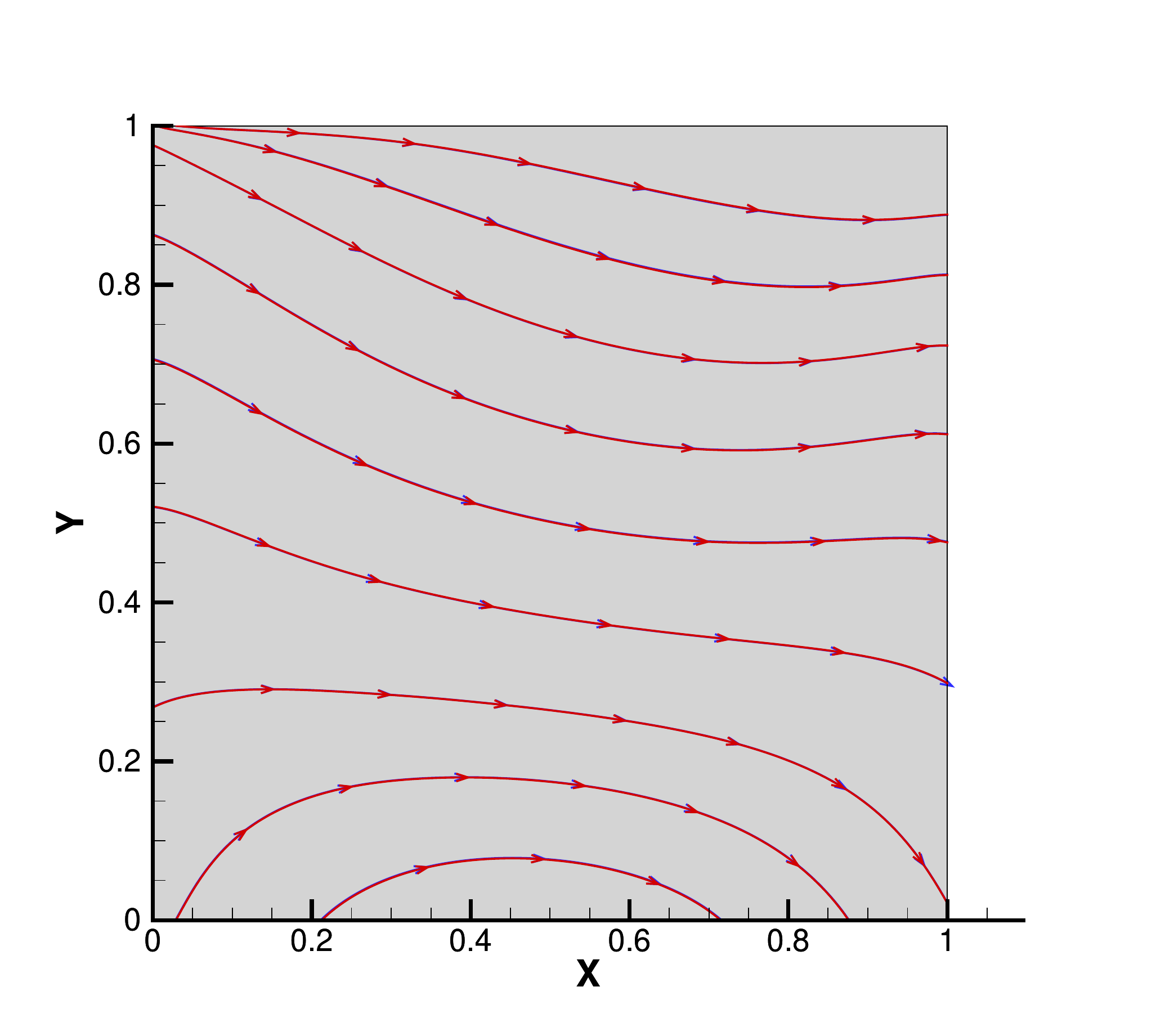}
 \includegraphics[width=0.45\textwidth]{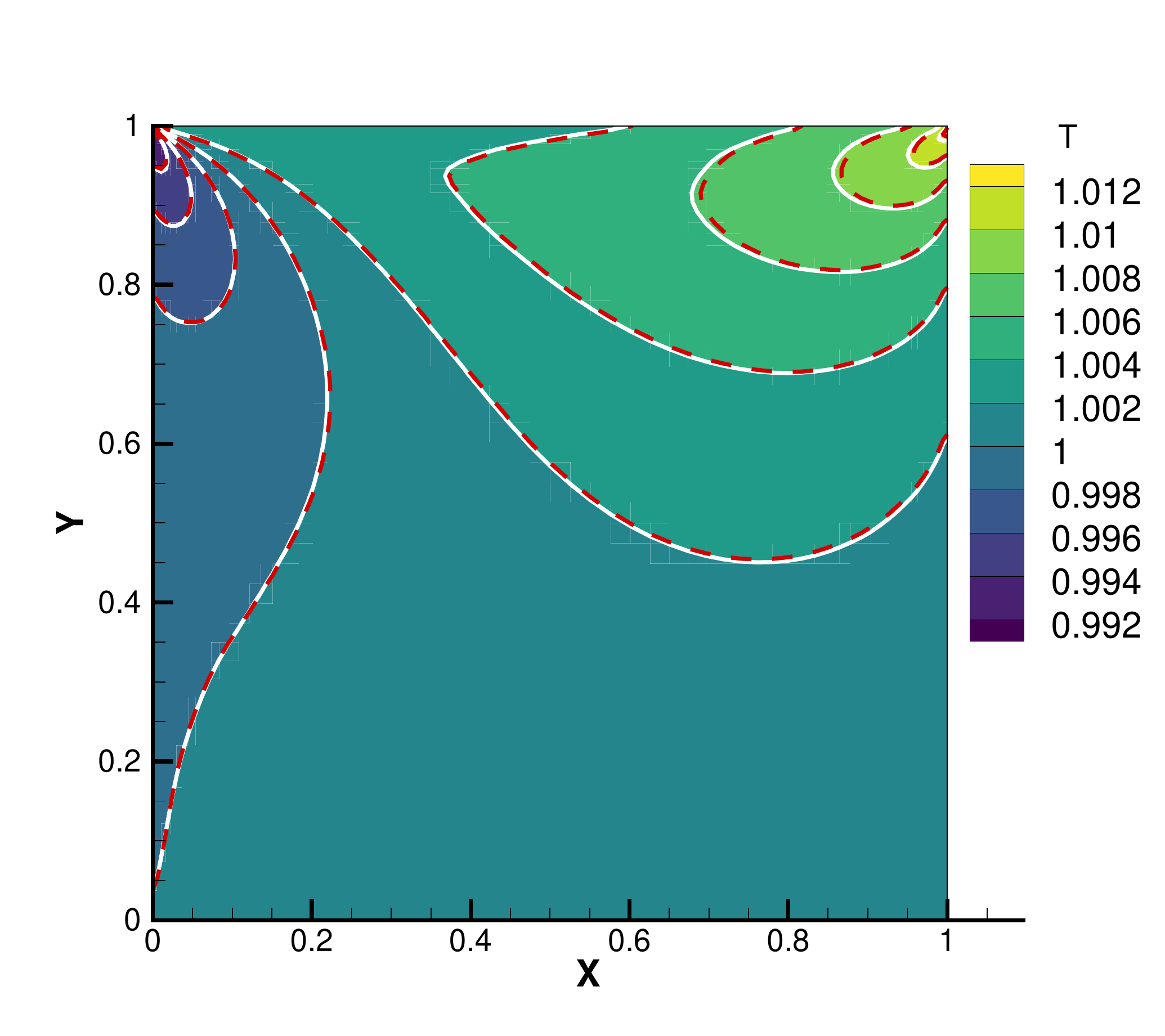}
 \includegraphics[width=0.45\textwidth]{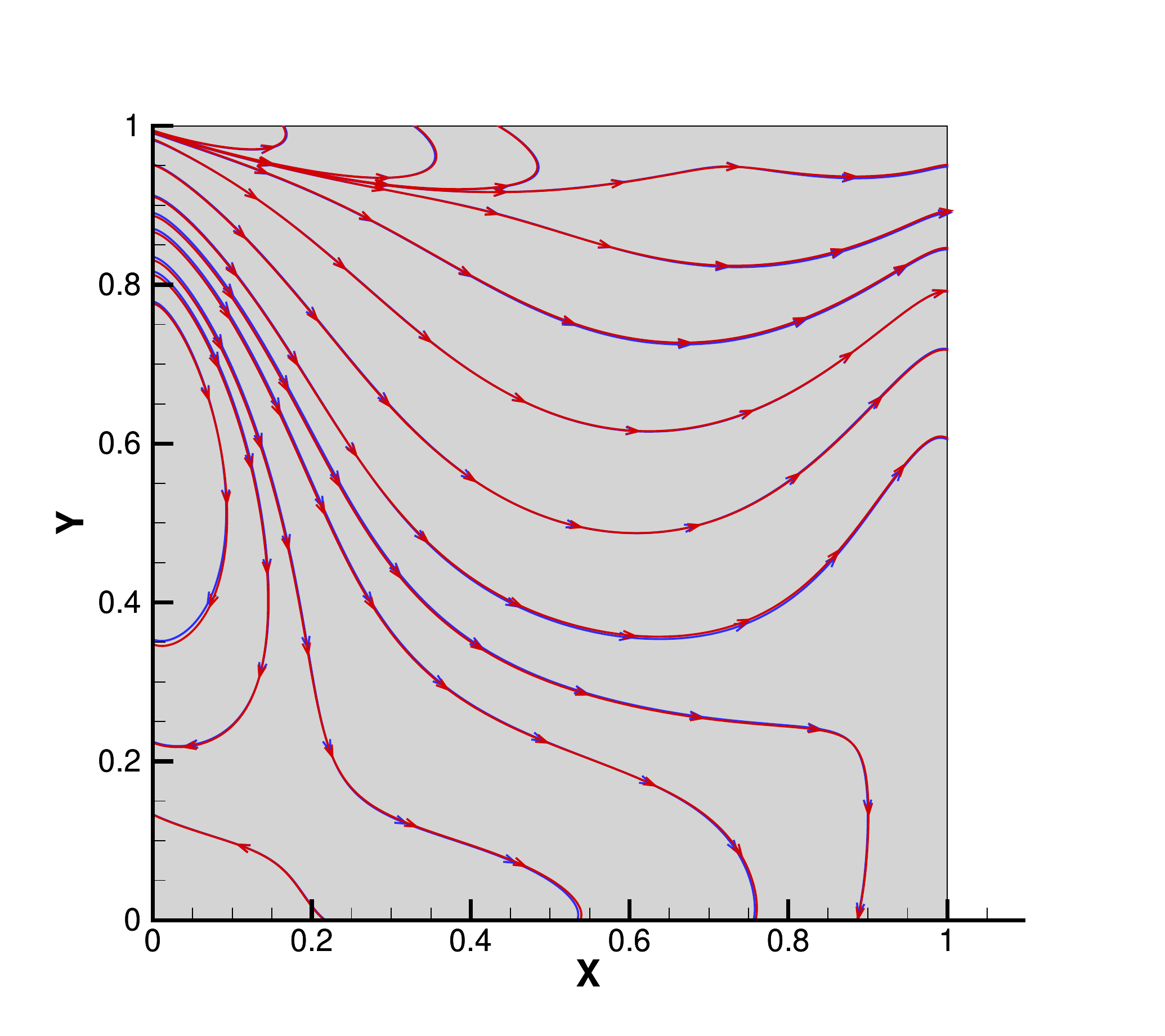}
 \caption{Contour of the temperature field (left) and streamlines of the heat flux field (right) for the cavity flow with Kn = 10 (top), 1 (middle) and 0.075 (bottom). In the temperature contours, CIS results are shown in colored background with white lines, while GSIS results are shown as dashed red lines. In the streamline plots, CIS and GIS results are in blue and red color, respectively. }
 \label{fig:cavity_t_q}
\end{figure}

The spatial space is discretized with Cartesian structured meshes and the gas kinetic equation is solved by the upwind finite difference scheme, while macroscopic synthetic equations are solved using the implicit finite volume method described in Sec.~\ref{sec:lusgs}. The cell centers are the finite difference nodes in the DVM discretization. The finite volume grid line positions are distributed according to  
\begin{equation}
 x_{i},y_{i} =\frac{1}{2}+\frac{\tanh [a(i / N-0.5)]}{2 \tanh (a / 2)}, \quad i=0,1, \ldots, N,
\end{equation}
where $N$ is the mesh size. The parameter $a$ is adjusted such that the height of the first layer of cells adjacent to the wall is the desired value $\Delta x_\mathrm{min} \equiv x_1 = y_1$.
 The convergence criterion for the DVM iteration is that the cell-volume averaged relative change of all conservative variables between two successive steps is less than $\num{1e-8}$, i.e.,
\begin{equation}\label{convtol}
 E^k = \sqrt{ \frac{\sum_{i,j} (\phi^k_{i,j} -\phi^{k-1}_{i,j} )^2\Omega_{i,j} }{ \sum_{i,j} (\phi^{k-1}_{i,j})^2\Omega_{i,j}} } <\epsilon_\text{out}, ~~\text{for} \quad \phi \quad \text{in} \quad \vec W.
\end{equation}
The convergence criterion of solving the macroscopic synthetic equations (the inner loop) is defined exactly the same as the outer loop, but with the superscript $k$ changed to $n$, and $\epsilon_{\text{out}}$ changed to $\epsilon_{\text{in}}$. For this flow problem, $\epsilon_{\text{out}} = \num{1e-8}$ and $\epsilon_{\text{in}} = \num{1e-6}$.

\begin{figure}[t]
 \centering
 \includegraphics[width=0.44\textwidth]{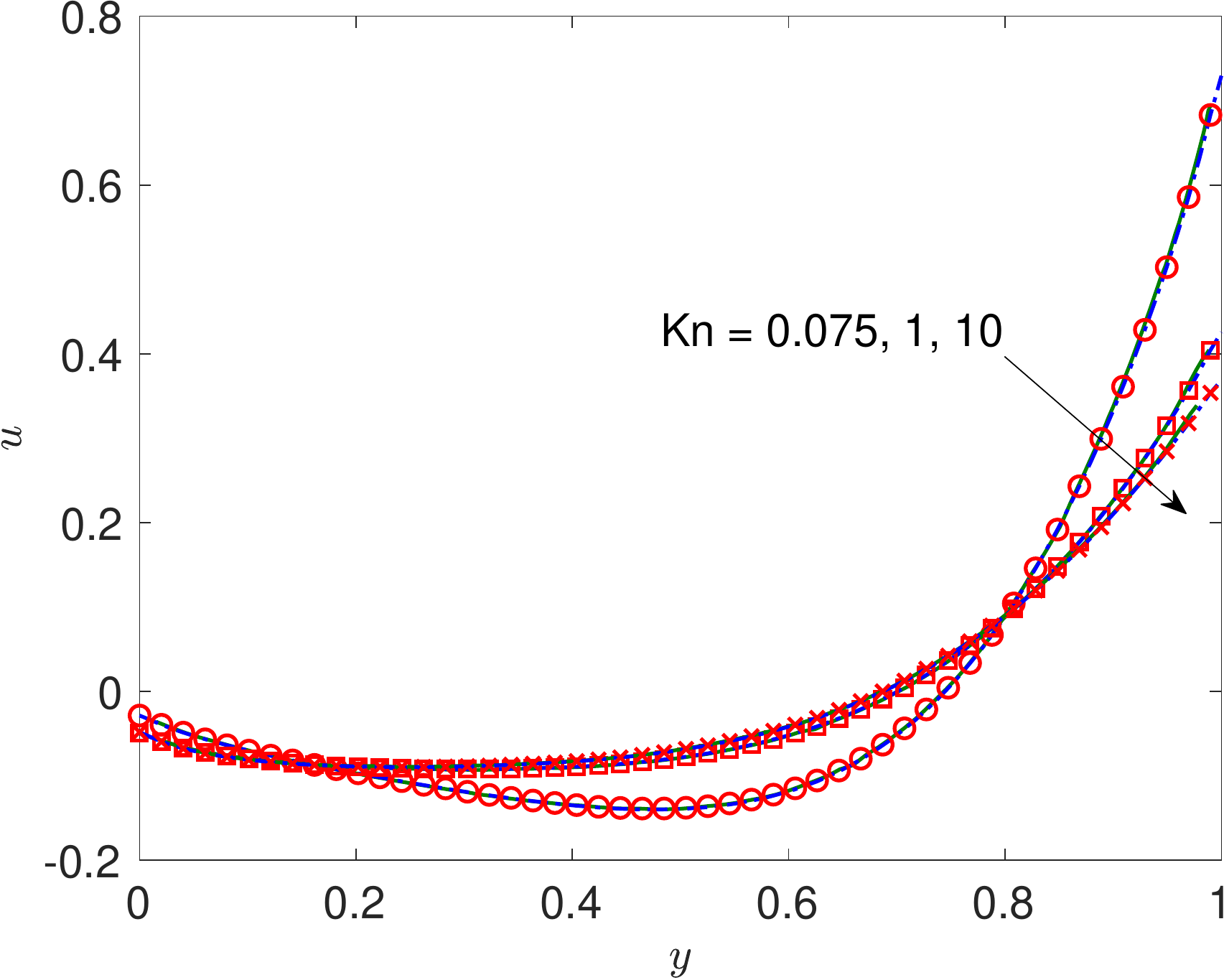}~~~~
 \includegraphics[width=0.45\textwidth]{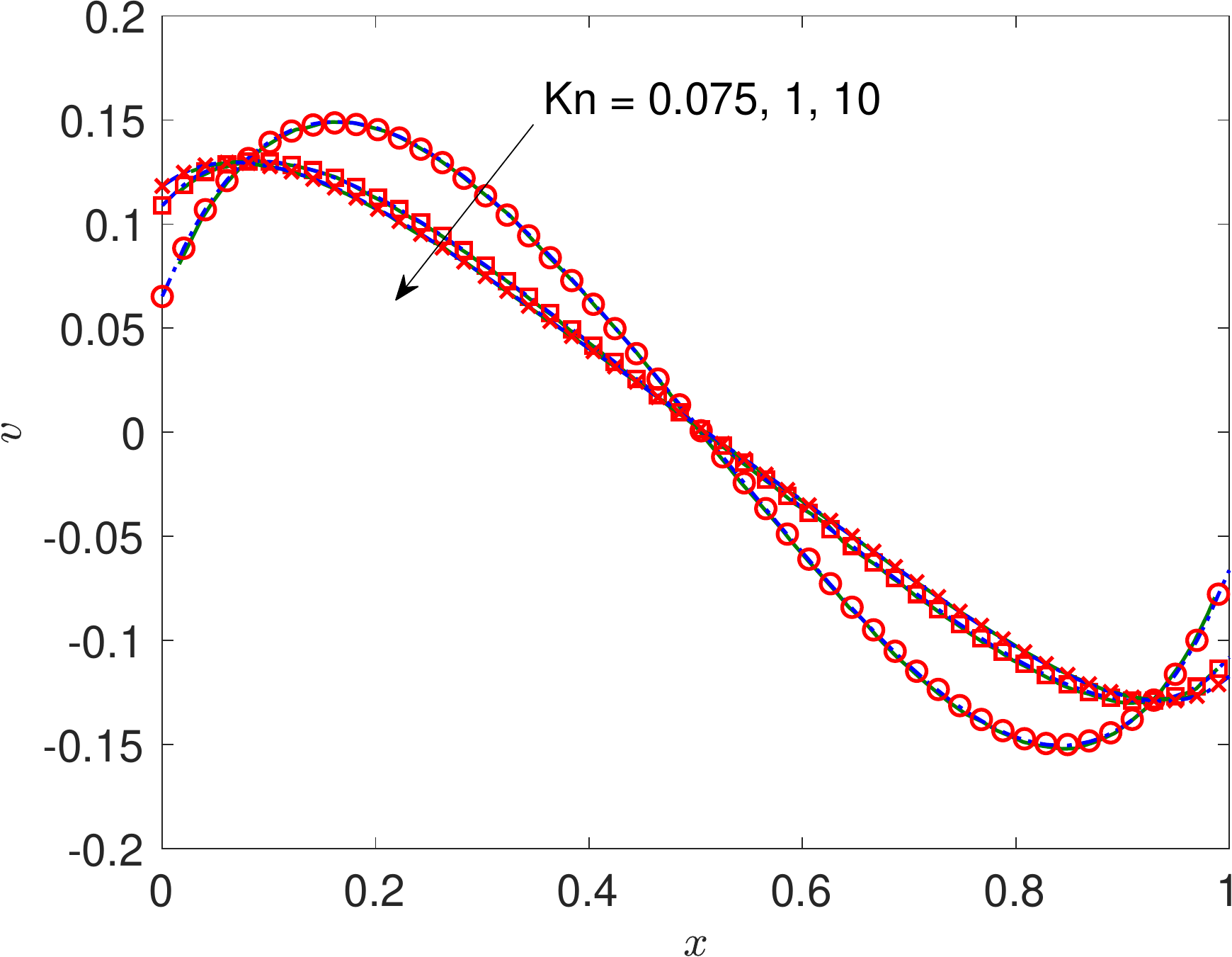}
 \caption{Profiles of the horizontal (left) and vertical (right) velocity component along the vertical/horizontal center lines of the cavity. Following the arrow, the Knudsen numbers corresponding to the lines are 0.075, 1 and 10, respectively. Solid green lines are the results extracted from Ref.~\cite{zhuDiscreteUnifiedGas2016} which is computed with the explicit discrete-UGKS. Blue lines and red markers represent results with the CIS and GSIS, respectively.}
\label{fig:cavity_rarefied_uv}
\end{figure}

We first consider the rarefied gas flows, with Kn = 0.075, 1 and 10. The physical space grid is set as $N = 64$ and $\Delta x_\mathrm{min} = \num{5e-3}$. For Kn = 1 and 10, the velocity-space grid is set according to Eq.~\eqref{wuleidv}, with $N_V = 48$. While for the case of Kn = 0.075, we use a 28-by-28-point velocity grid with the half-range Gauss-Hermite quadrature. The macroscopic synthetic equations are solved in the domain excluding four layers of cells adjacent to the solid walls. Comparisons of temperature and heat flux streamlines are shown in Fig.~\ref{fig:cavity_t_q}, and the velocity profiles across the center lines of the cavity are shown in Fig.~\ref{fig:cavity_rarefied_uv}. These figures show that in the rarefied regime, GSIS and CIS results match well with each other, and both schemes capture the anti-Fourier heat transfer phenomenon (from cold to hot) around the top right corner even Kn is as small as 0.075.

\begin{figure}[t]
	\centering
	\includegraphics[width=0.45\textwidth]{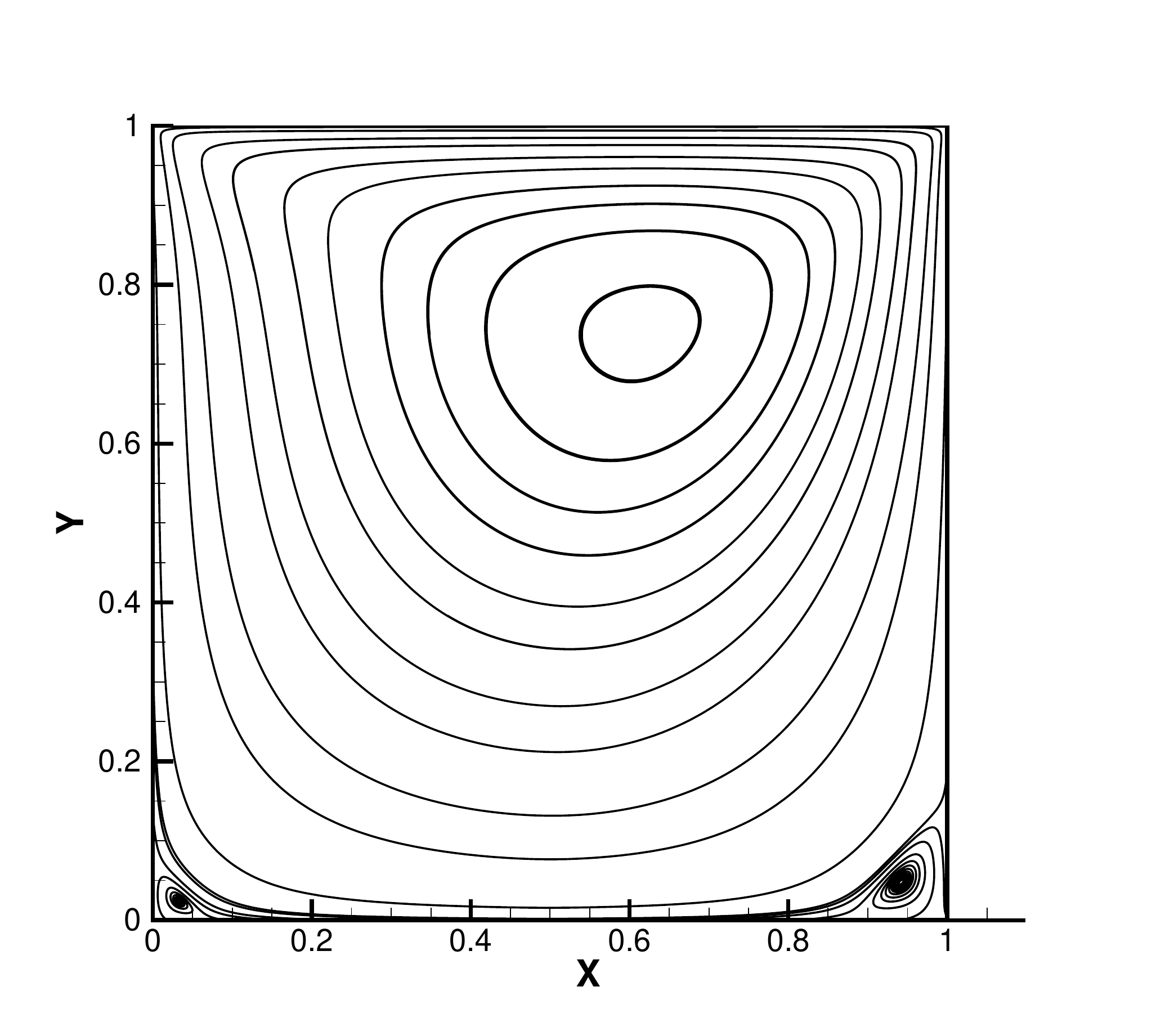}
	\includegraphics[width=0.45\textwidth]{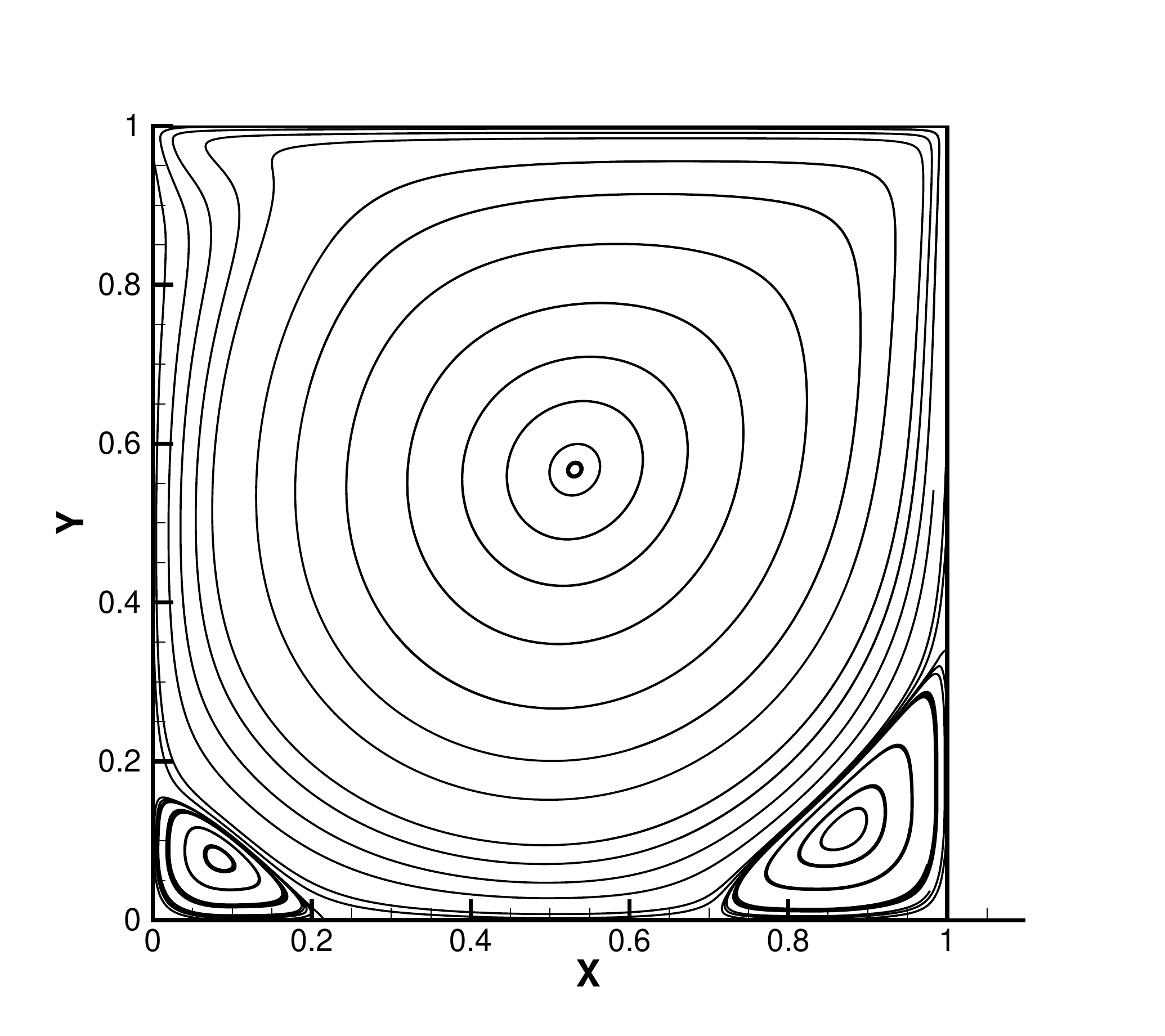}
	\caption{The streamlines of velocity for the lid-driven cavity flow at $\text{Re} = 100$ (left) and 1000 (right)}
	\label{fig:cavity_streamlines}
\end{figure}

\begin{figure}[th]
 \centering
 \includegraphics[width=0.45\textwidth]{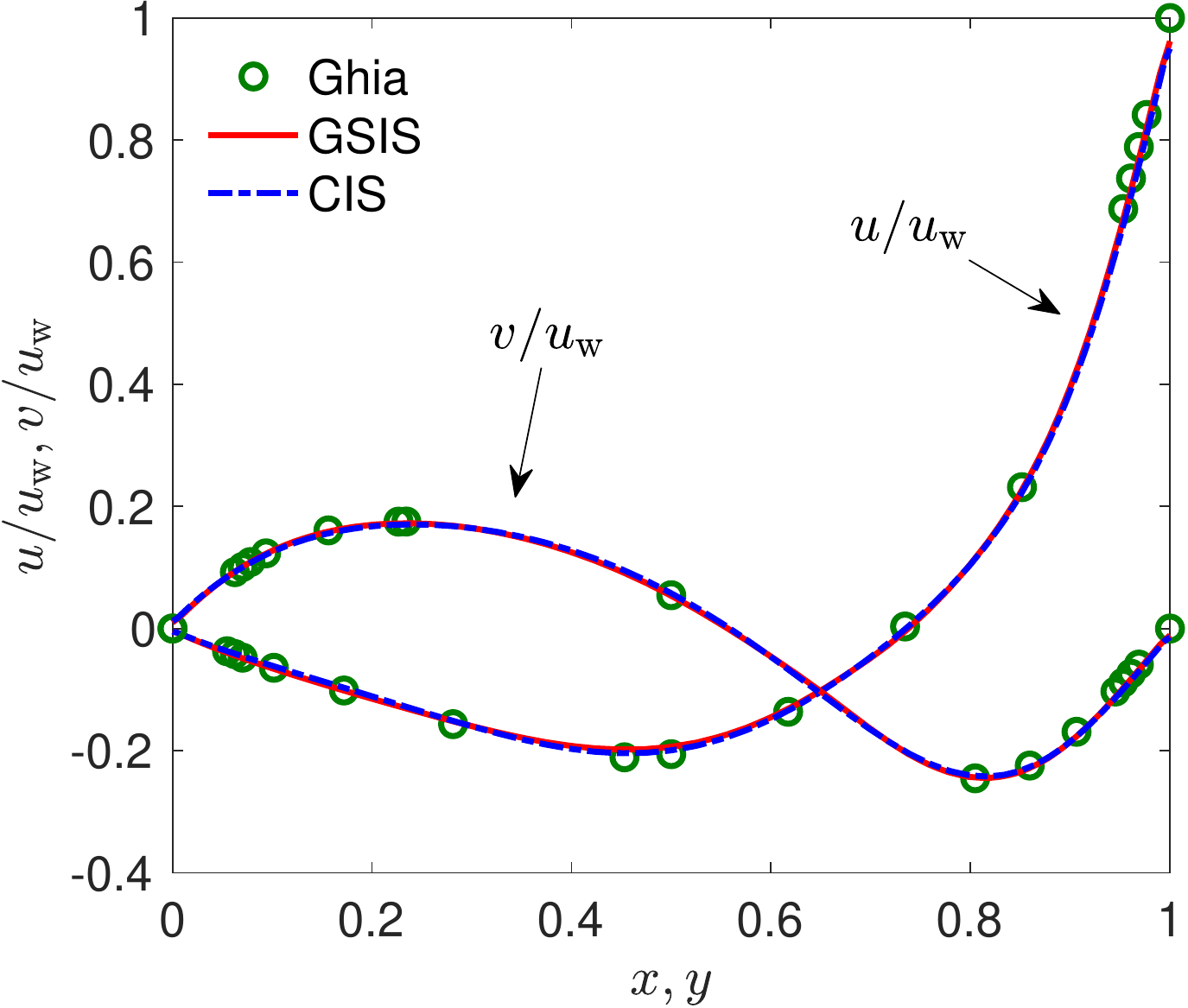}~~~
 \includegraphics[width=0.45\textwidth]{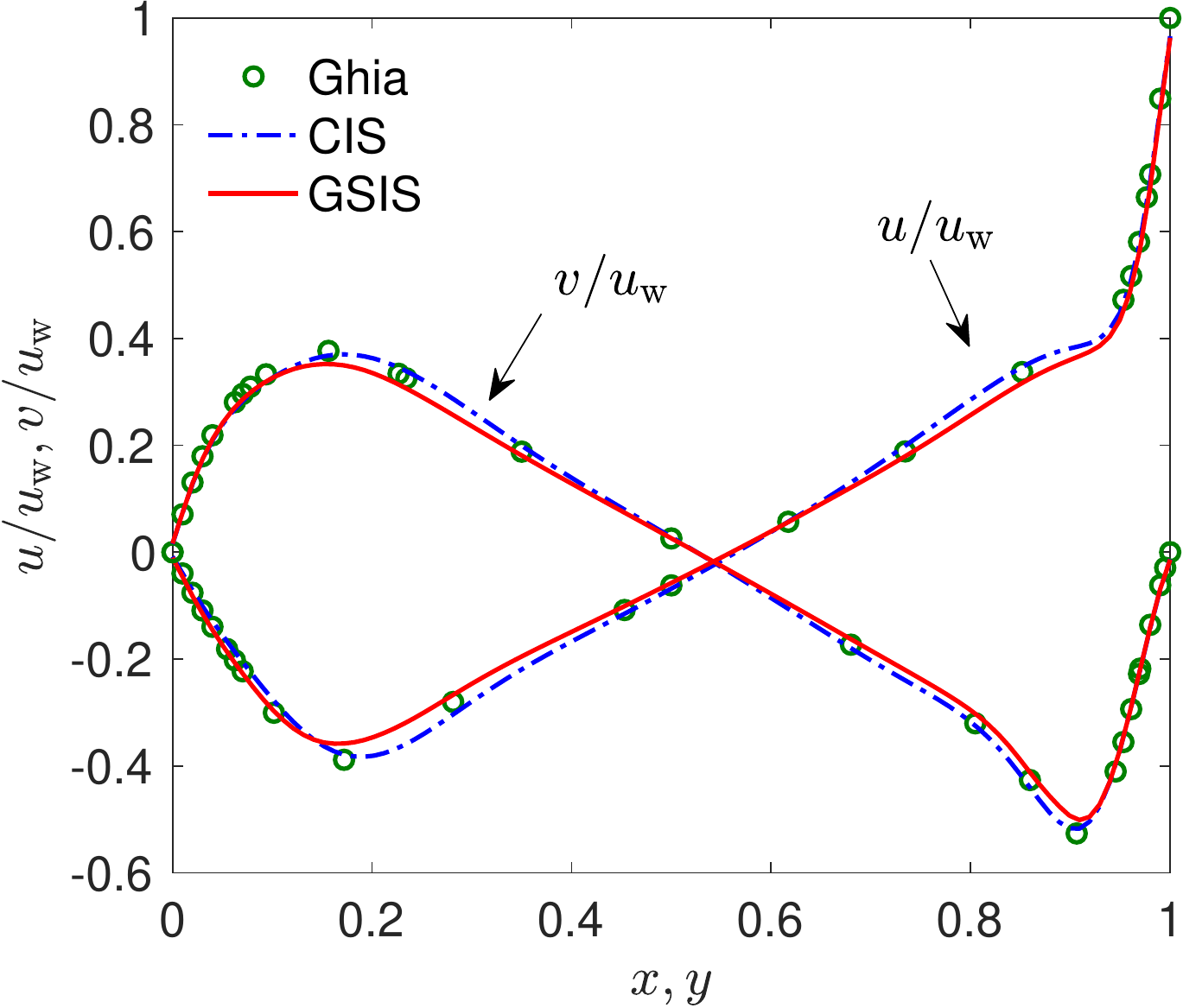}
 \caption{The profiles of normalized horizontal (vertical) velocity components $u$ ($v$) on the vertical (horizontal) central lines of the cavity. Left: Re = 100; Right Re = 1000.}
 \label{fig:cavity_continuum_uv}
\end{figure}

For flows in the near-continuum regime, we consider the case of Re = 100 and Re = 1000, corresponding to Kn = $\num{2.628E-3}$ and $\num{2.628e-4}$, respectively. The spatial grids are set as $N = 64$, $\Delta x_\mathrm{min} = \num{5e-3}$ when Re = 100 and $N = 128$, $\Delta x_\mathrm{min} = \num{2e-3}$ when Re = 1000. The velocity grids are set  the same as the case of Kn = 0.075.
Figure~\ref{fig:cavity_streamlines} shows the velocity streamlines  predicted by GSIS. The vortex patterns, including size and vortex center positions, agree with various literature results. To get a more quantitative comparison, in Fig.~\ref{fig:cavity_continuum_uv} we plot the velocity profiles on the vertical and horizontal centerlines of the cavity, predicted by both CIS and GSIS, together with Ghia's benchmark solution~\cite{ghiaHighReSolutionsIncompressible1982}. We can see that, when Re = 100, CIS and GSIS predicted almost the same solution, and both agree well with Ghia's benchmark solution. When Re = 1000, there is a slight difference between the GSIS and CIS results.

\begin{figure}[t]
 \centering
 \includegraphics[width=0.6\textwidth]{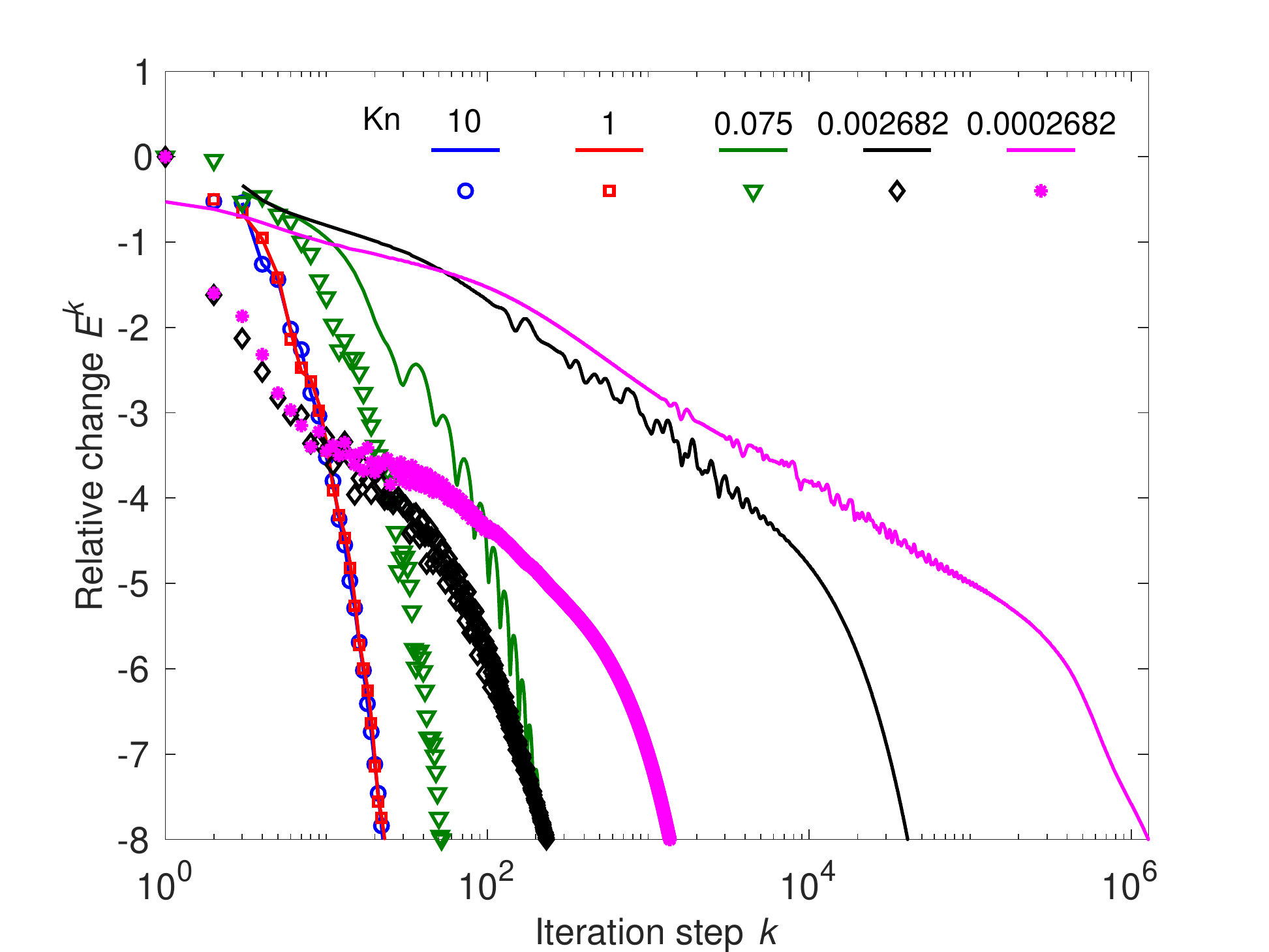}
 \caption{Convergence history of the cavity flow at different Knudsen numbers. Solid lines and  markers represent results of CIS and GSIS, respectively.}
 \label{fig:cavity_convergence}
\end{figure}

The comparison of the convergence history of the DVM iteration is shown in Fig.~\ref{fig:cavity_convergence} for both the rarefied and continuum flow cases. The corresponding CPU time and the number of DVM steps in both CIS and GSIS are listed in Table~\ref{tab:cavity_time}. The serial Fortran program is compiled using the Intel Fortran compiler (version 19.1.1) with the ``-xHost" option, and runs on the Intel\textsuperscript{\textcopyright} Xeon\textsuperscript{\textcopyright} Gold 5118 CPU@2.3GHz. We can see that for the cases of Kn = 1 and 10, the convergence history of CIS and GSIS is very similar, both converge in around 23 DVM steps. Due to the additional cost in solving the macroscopic synthetic equations, with the same number of DVM steps the overall computing cost of GSIS is higher for these highly rarefied cases. However, when Kn $\le 0.075$, GSIS needs much fewer DVM steps than CIS, e.g., when Re = 100, GSIS achieves the convergence criterion in 234 DVM steps, while CIS becomes extremely expensive. For these low Kn cases, because GSIS can reduce the number of DVM steps by several orders, the additional cost for solving the macroscopic synthetic equations is negligible. 

\begin{table}[t]
 \centering
 \caption{Number of DVM steps in CIS and GSIS and the overall CPU time for lid-driven cavity flow at different Knudsen numbers.}
  \begin{tabular}{rrrrrrr}
  \toprule
  \multicolumn{1}{l}{Kn} & \multicolumn{1}{r}{Physical} & \multicolumn{1}{r}{Velocity} &\multicolumn{1}{r}{CIS: DVM} & CIS: CPU & \multicolumn{1}{r}{GSIS: DVM} & GSIS: CPU \\
   \multicolumn{1}{r}{} & \multicolumn{1}{r}{grid size} & \multicolumn{1}{r}{grid size} &\multicolumn{1}{r}{steps} &\multicolumn{1}{r}{time} & \multicolumn{1}{r}{steps}& \multicolumn{1}{r}{time} \\
  \midrule
  $   10$       & $64\times 64$  & $48\times 48$ & $  24$    & $14$ s & 24 & $16$ s \\
  $    1$       & $64\times 64$  & $48\times 48$  & $24$    & $14$ s  & 23  & $ 29   $ s \\
  $ 0.075$       & $64\times 64$   & $28\times 28$ & $226$    & $40$ s & 52  & $2.2$ min \\
 $0.002682$     & $128\times 128$ & $28\times 28$ & $40,917$  & 123 min & 234 & $10.5$ min \\
 $0.0002682$    & $128\times 128$ & $28\times 28$ & $1,283,068$ & $ 64.2$ h & 1410 & $49$ min \\
\bottomrule
  \end{tabular}%
 \label{tab:cavity_time}%
\end{table}%

\begin{figure}[thbp]
	\centering
	\includegraphics[width=0.45\textwidth]{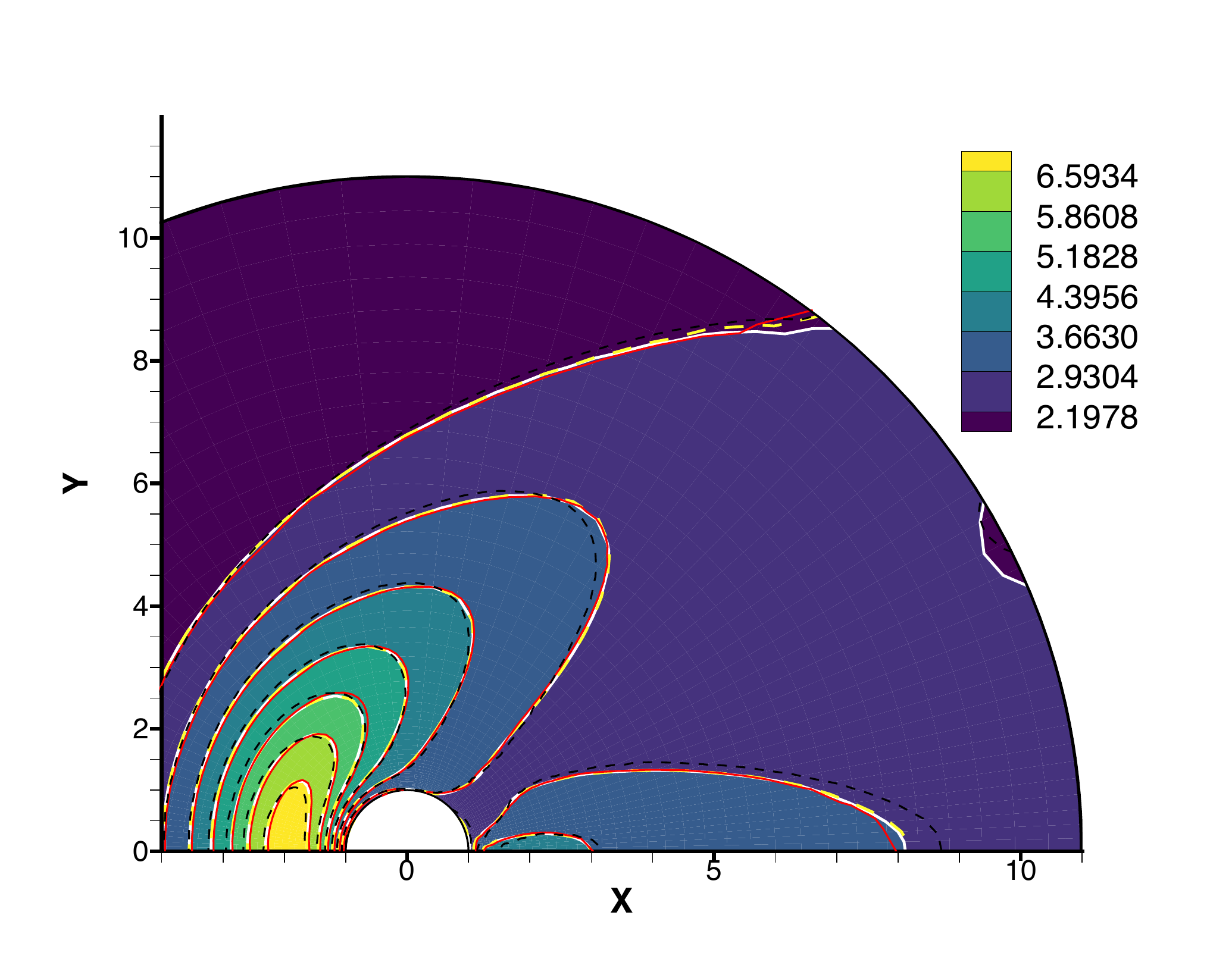}
	\includegraphics[width=0.45\textwidth]{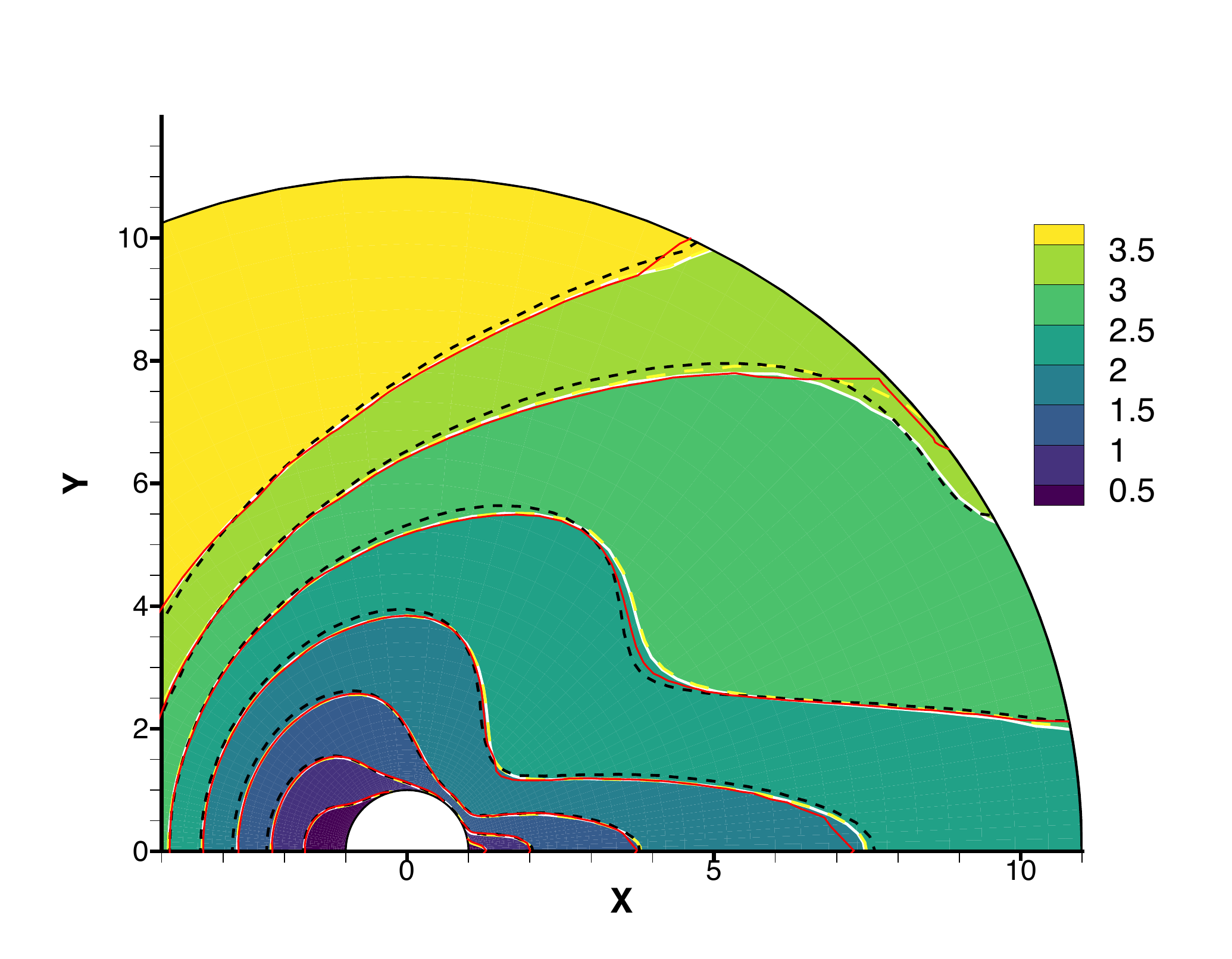} \\
	\includegraphics[width=0.45\textwidth]{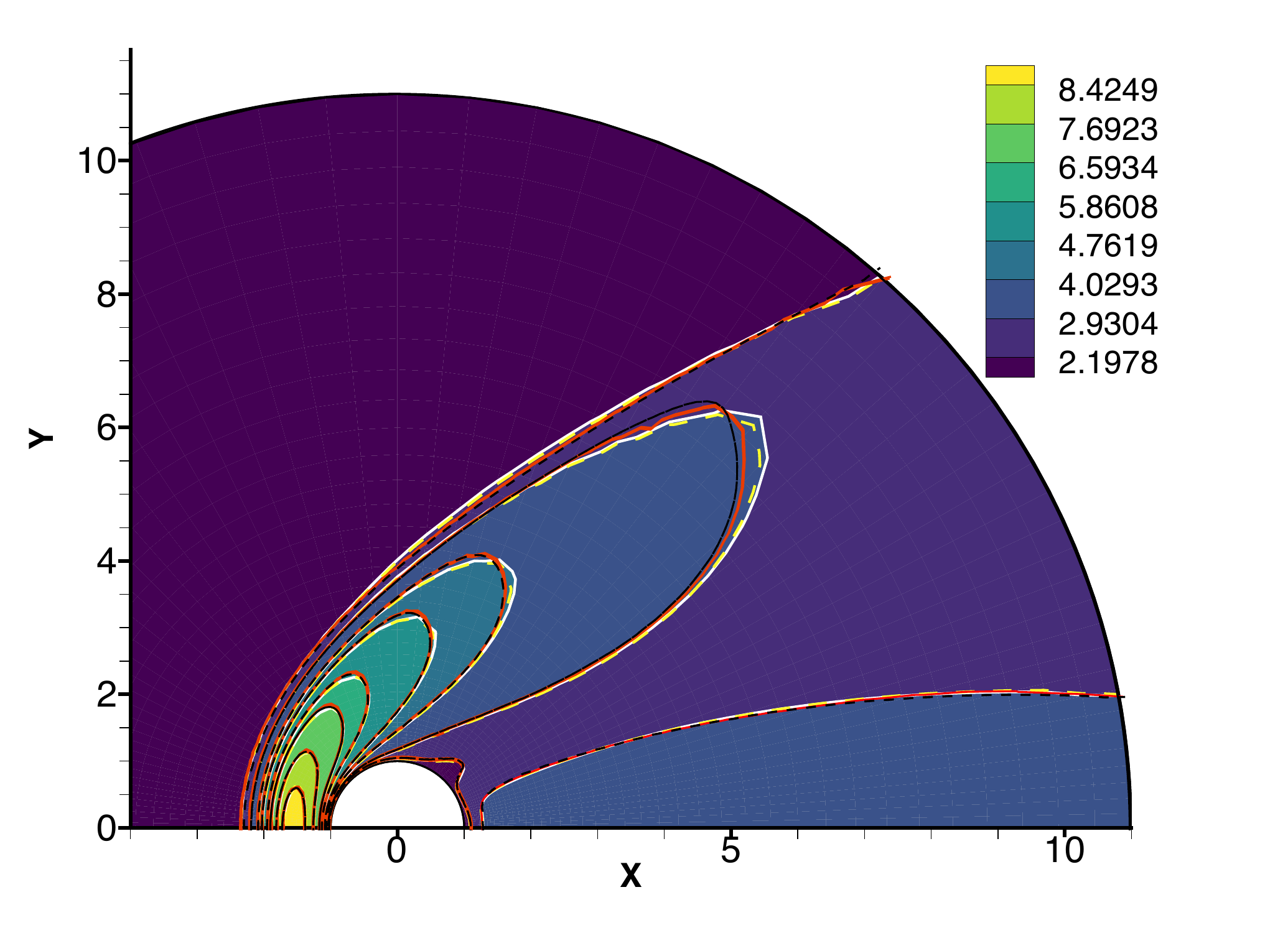}
	\includegraphics[width=0.45\textwidth]{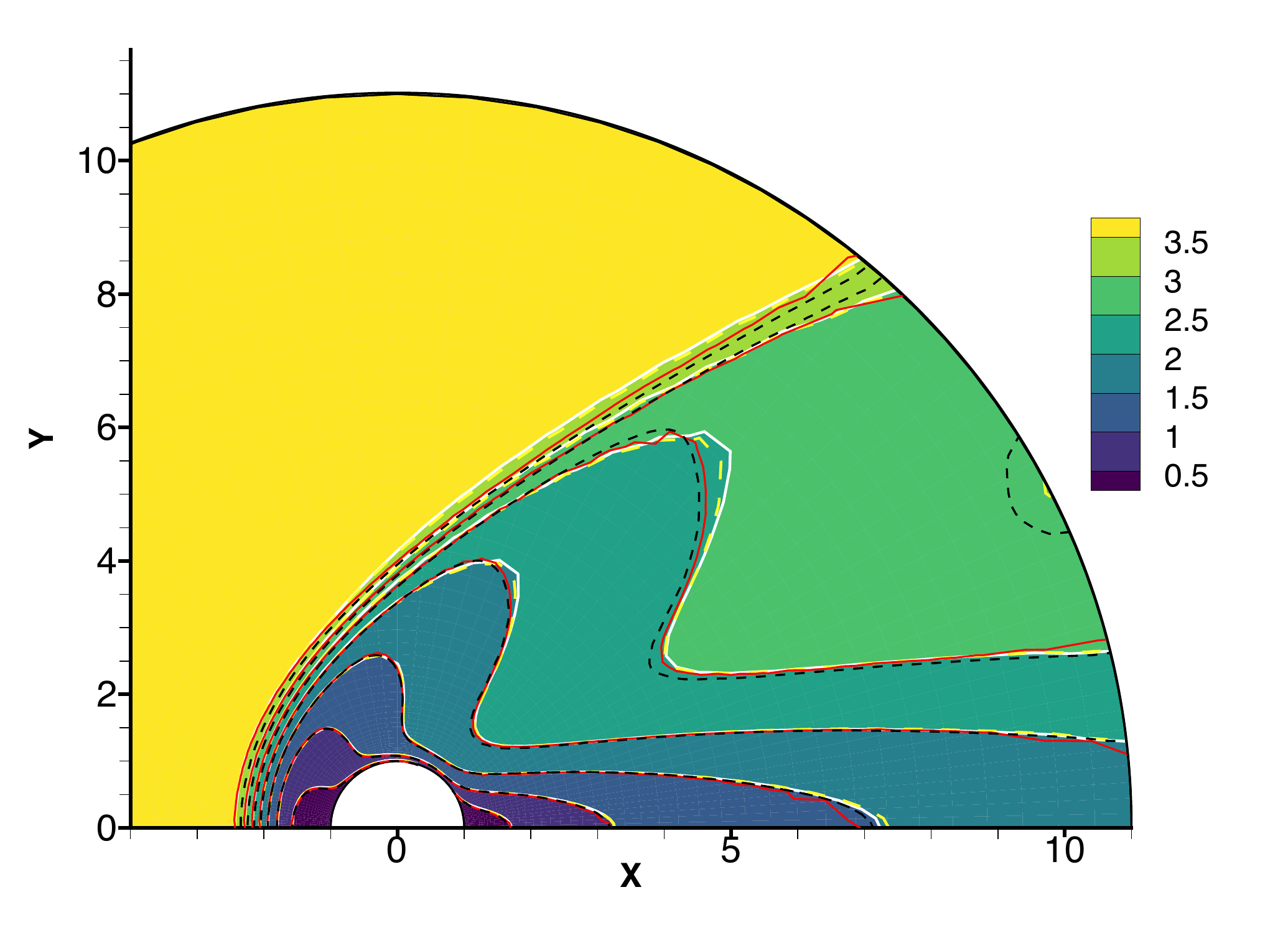}\\
	\includegraphics[width=0.45\textwidth]{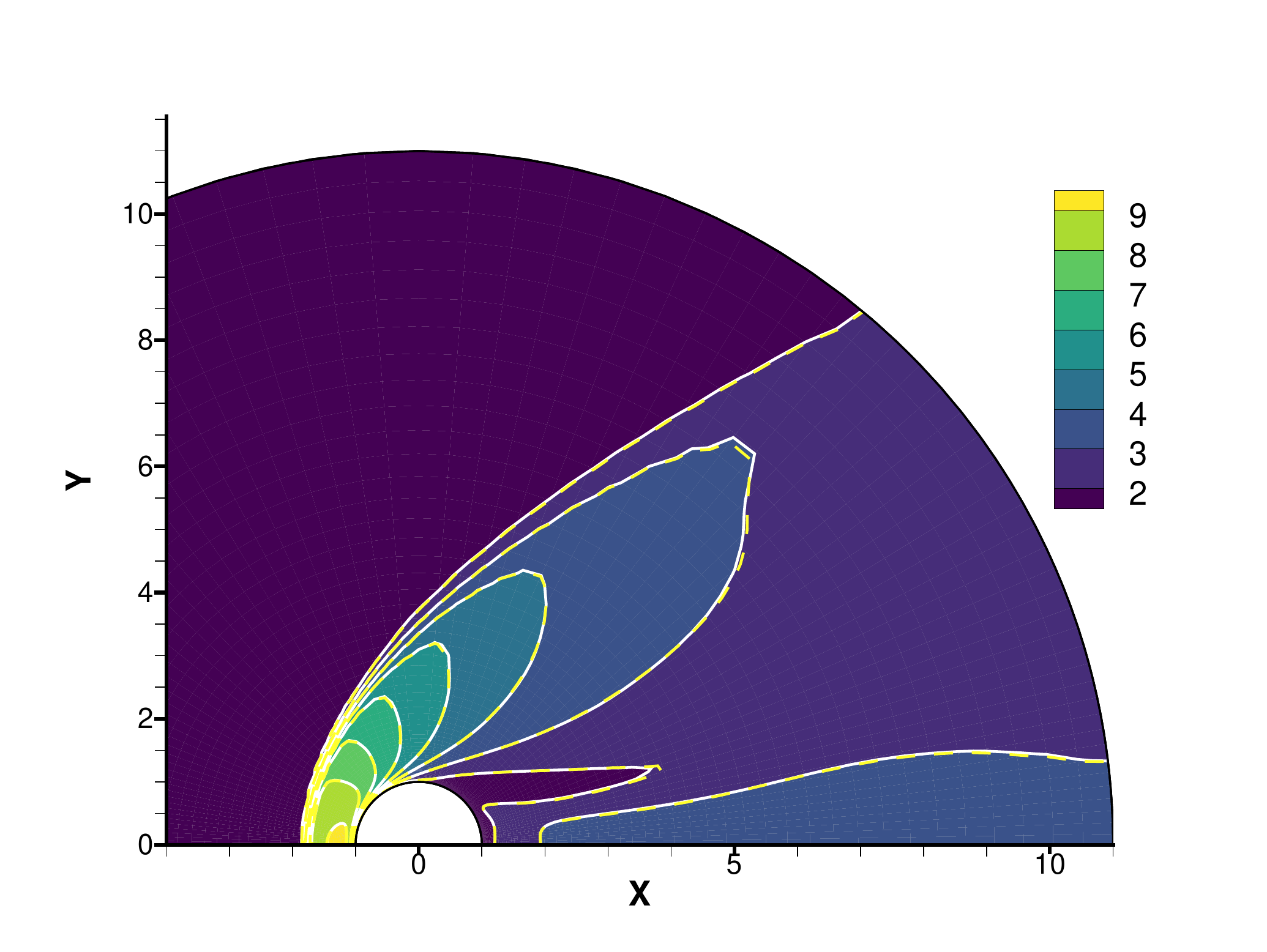}
	\includegraphics[width=0.45\textwidth]{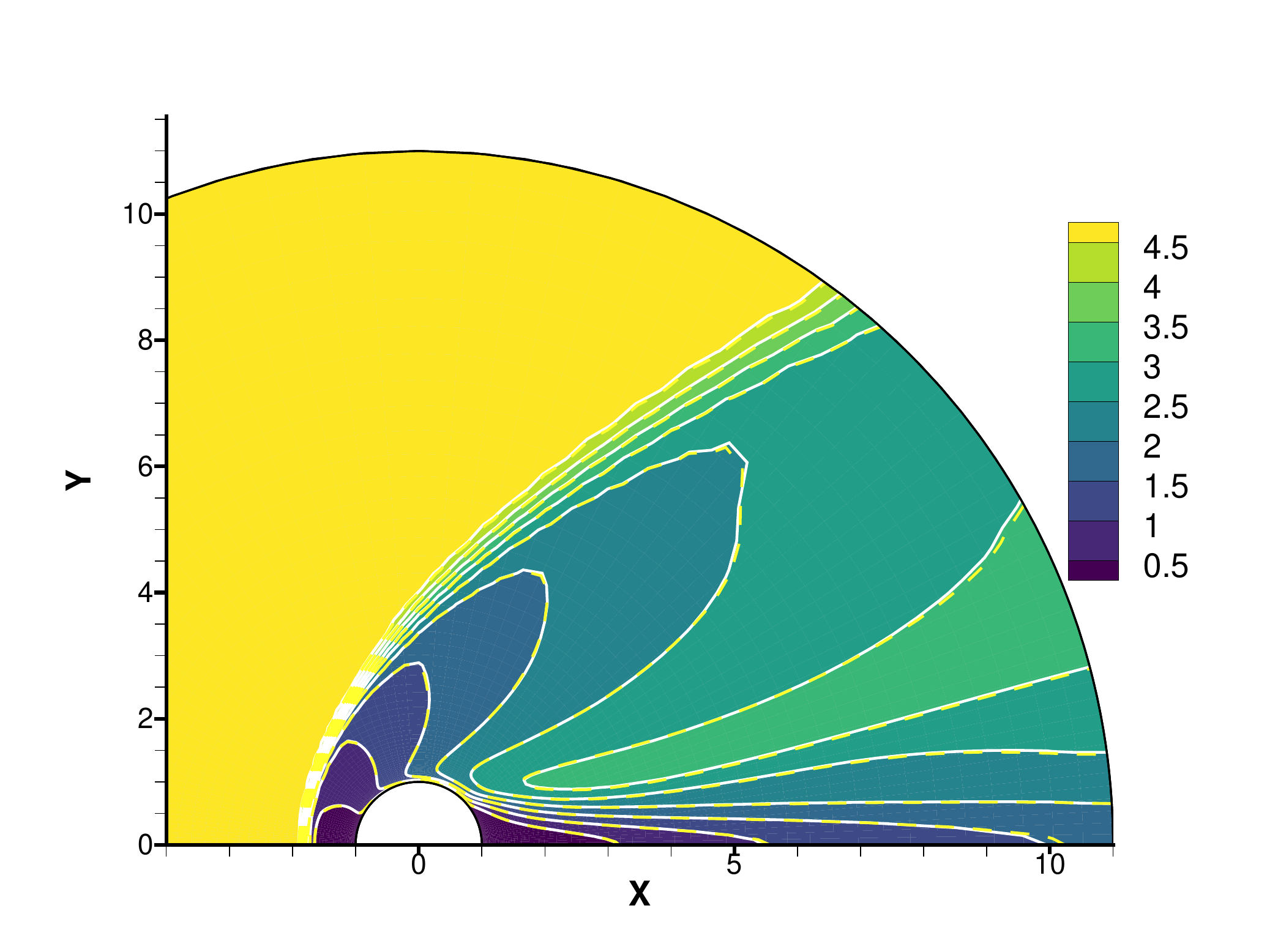}
	\caption{Cylinder flow at $\text{Ma}_\infty = 5$ and (top) $\text{Kn} = 1$, (middle) $\text{Kn=0.1}$, (bottom) $\text{Kn=0.01}$: comparison of the non-dimensional temperature (left) and local Mach number (right) fields obtained by the CIS and GSIS, together with the reference discrete-UGKS and DSMC solutions extracted from Ref.~\cite{zhuDiscreteUnifiedGas2016}. GSIS results are indicated by the colored background with white solid lines, CIS solutions are presented by the dashed yellow lines. The discrete-UGKS and DSMC solutions are represented by the solid red lines and dashed black lines, respectively.}
	\label{fig:cylinder_Kn1}
\end{figure}

\subsection{Supersonic flow past a circular cylinder}

The last testing case is the supersonic rarefied gas flow past by a circular cylinder. The 2D flow domain is an annulus with the inner circle with radius $r_\text{in}$ being the cylinder surface, and the outer circle with radius of $r_\text{out} = 11r_\text{in} $ being the far-field boundary. The free-stream Mach number is $\text{Ma}_{\infty}$. The cylinder surface temperature is set as the same as the free-stream temperature $T_\text{w} = T_\infty$. To properly compare with the literature results~\cite{zhuDiscreteUnifiedGas2016,zhuImplicitUnifiedGaskinetic2016} , the Knudsen number is defined as
\begin{equation}
\text{Kn} = \frac{(5-2\omega)(7-2\omega)\mu_\infty C_\infty}{15\sqrt{\pi}p_\infty r_\text{in}}
\end{equation}
where $\mu_\infty$, $C_\infty$ and $p_\infty$ are the viscosity, most probable molecular velocity and pressure at the free-stream condition, respectively. 

\begin{figure}[t]
	\centering
	\includegraphics[width=0.3\textwidth]{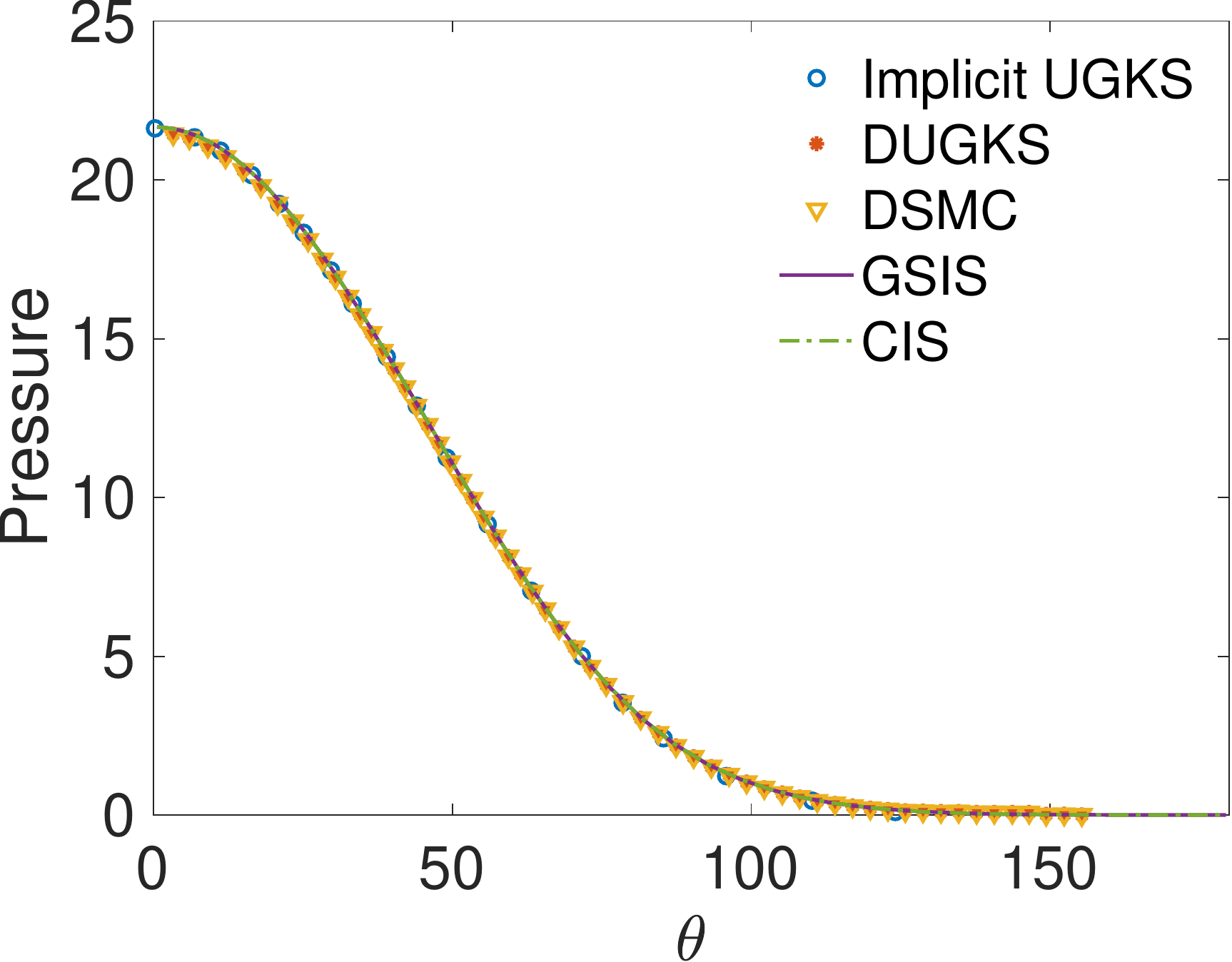}~
	\includegraphics[width=0.3\textwidth]{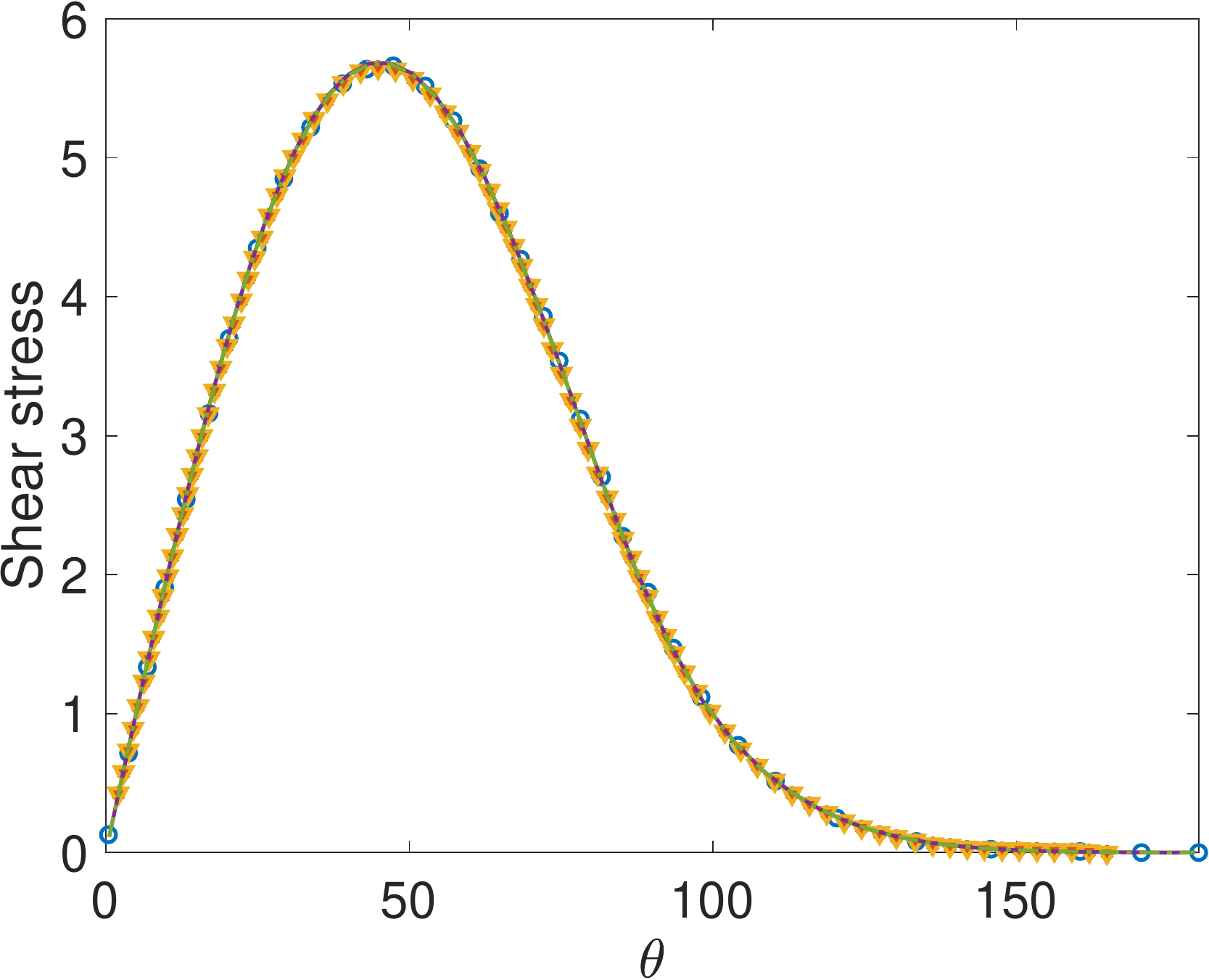}~
	\includegraphics[width=0.3\textwidth]{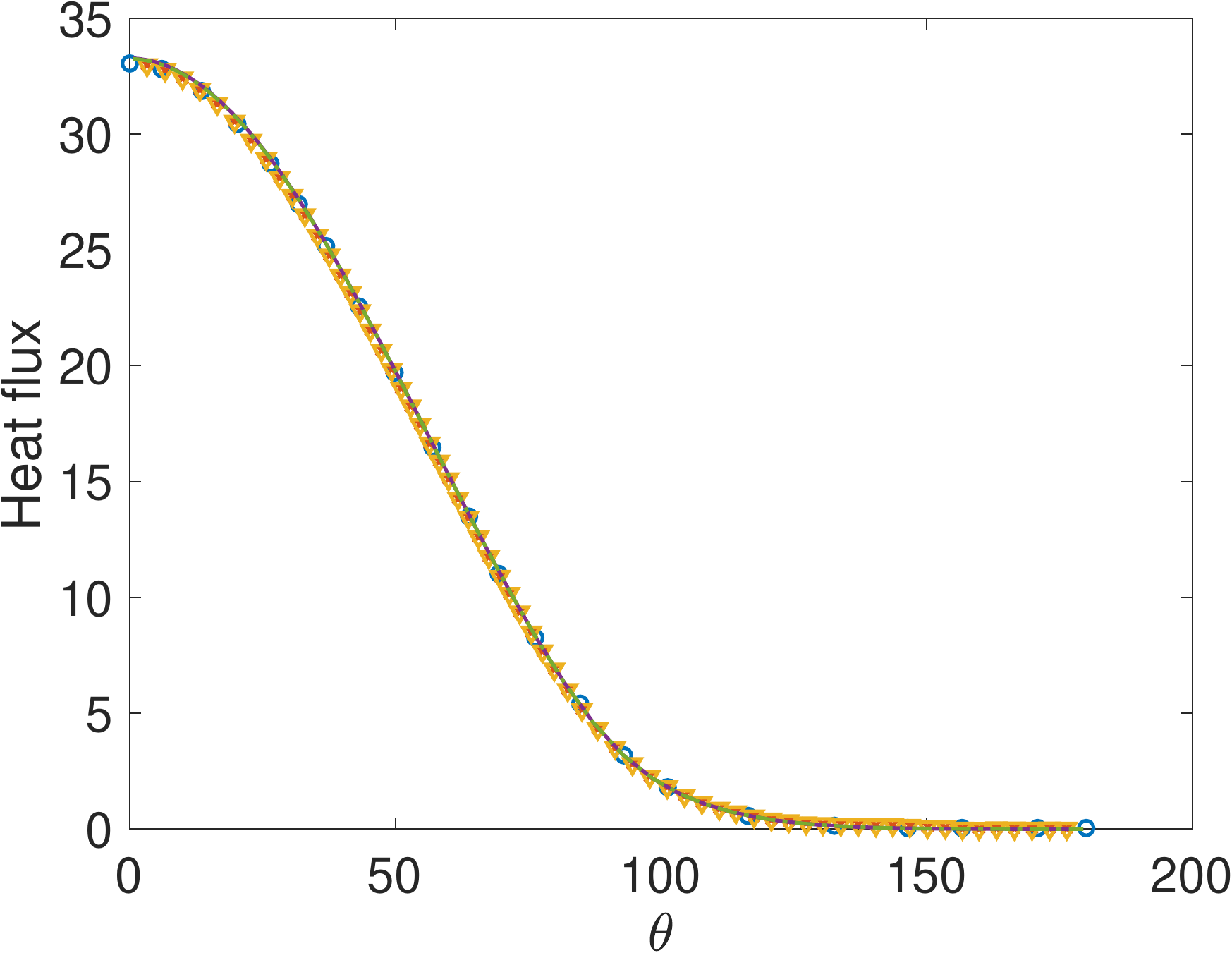}\\ 
	\vskip 0.5cm
	\includegraphics[width=0.3\textwidth]{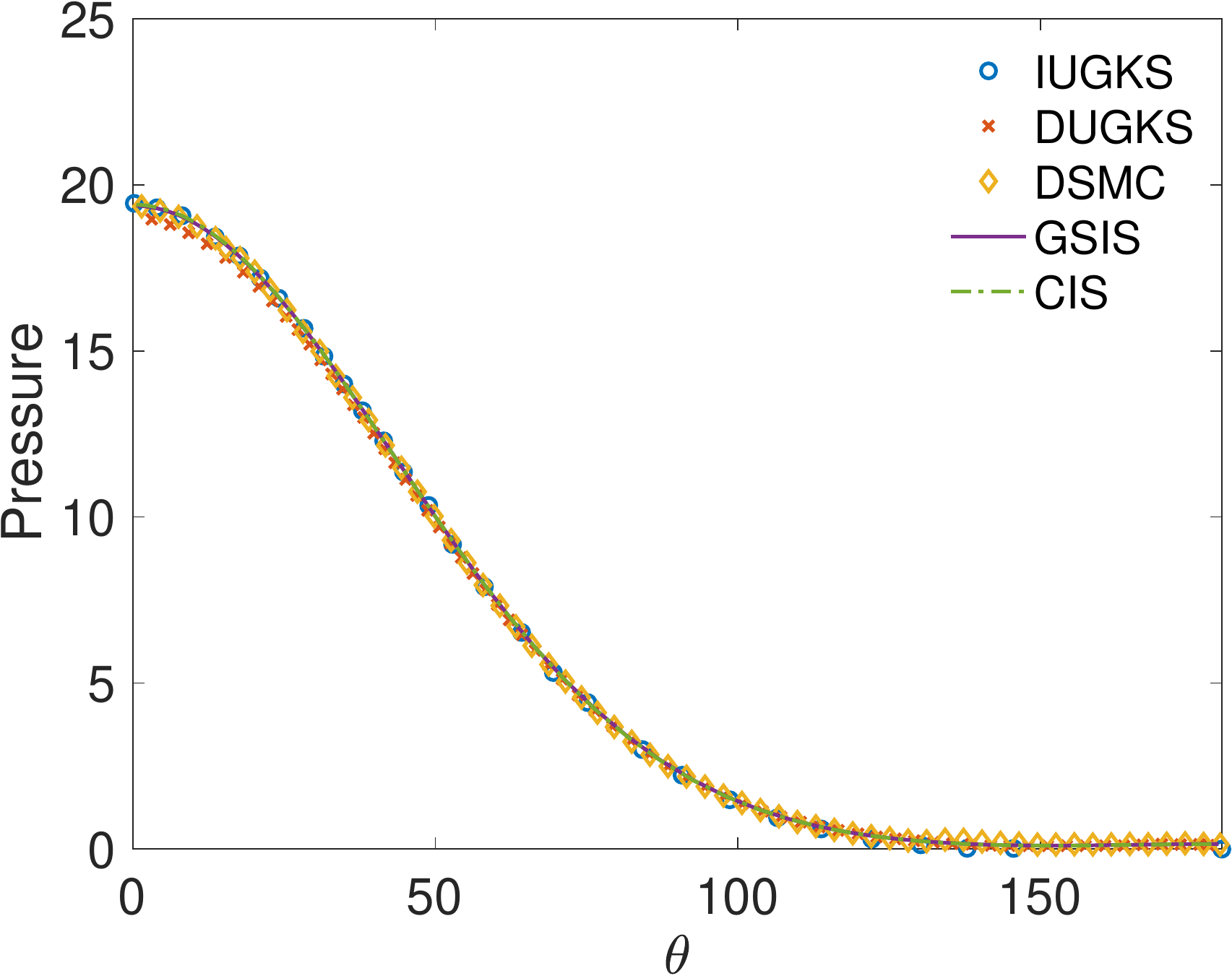}~
	\includegraphics[width=0.3\textwidth]{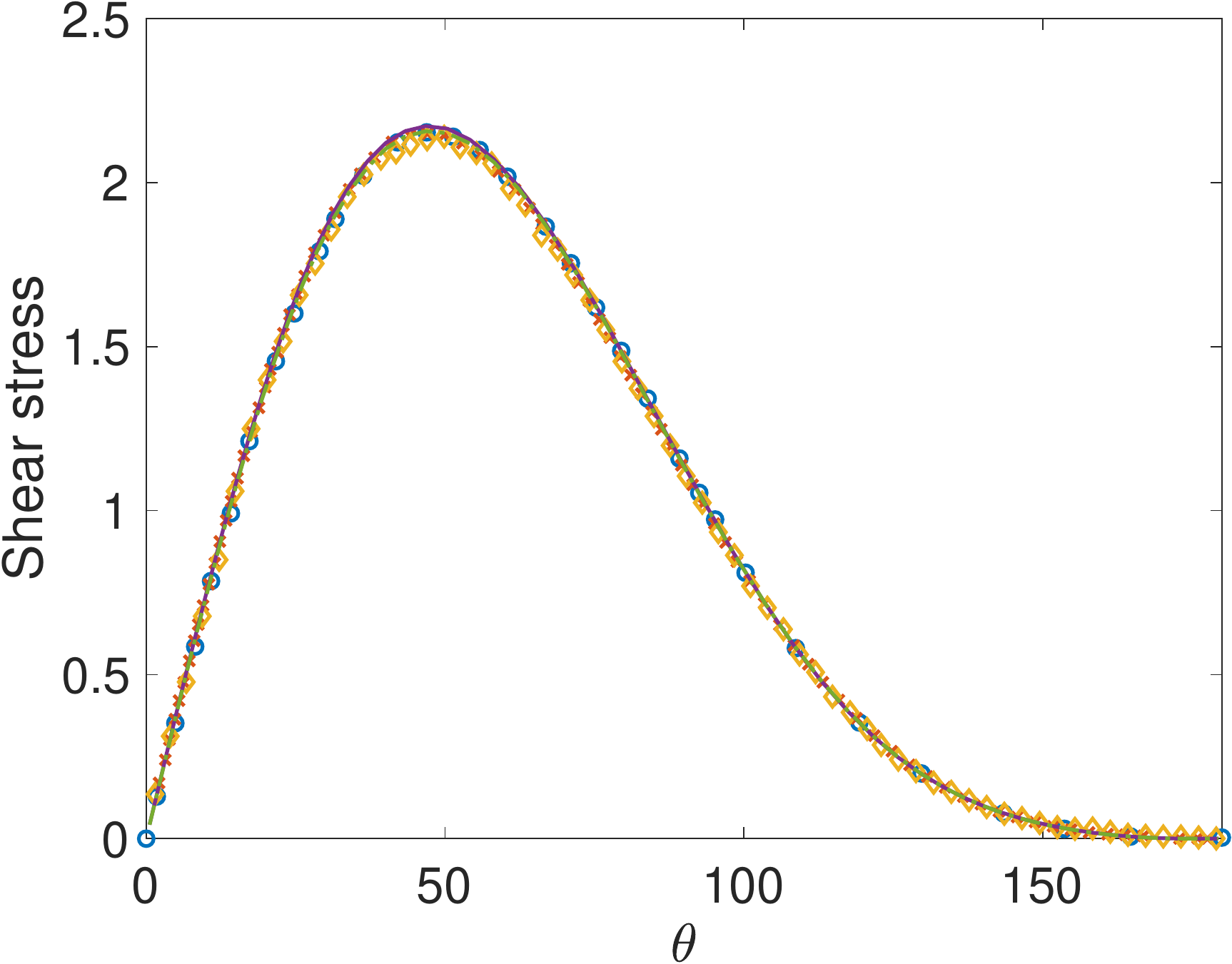}~
	\includegraphics[width=0.3\textwidth]{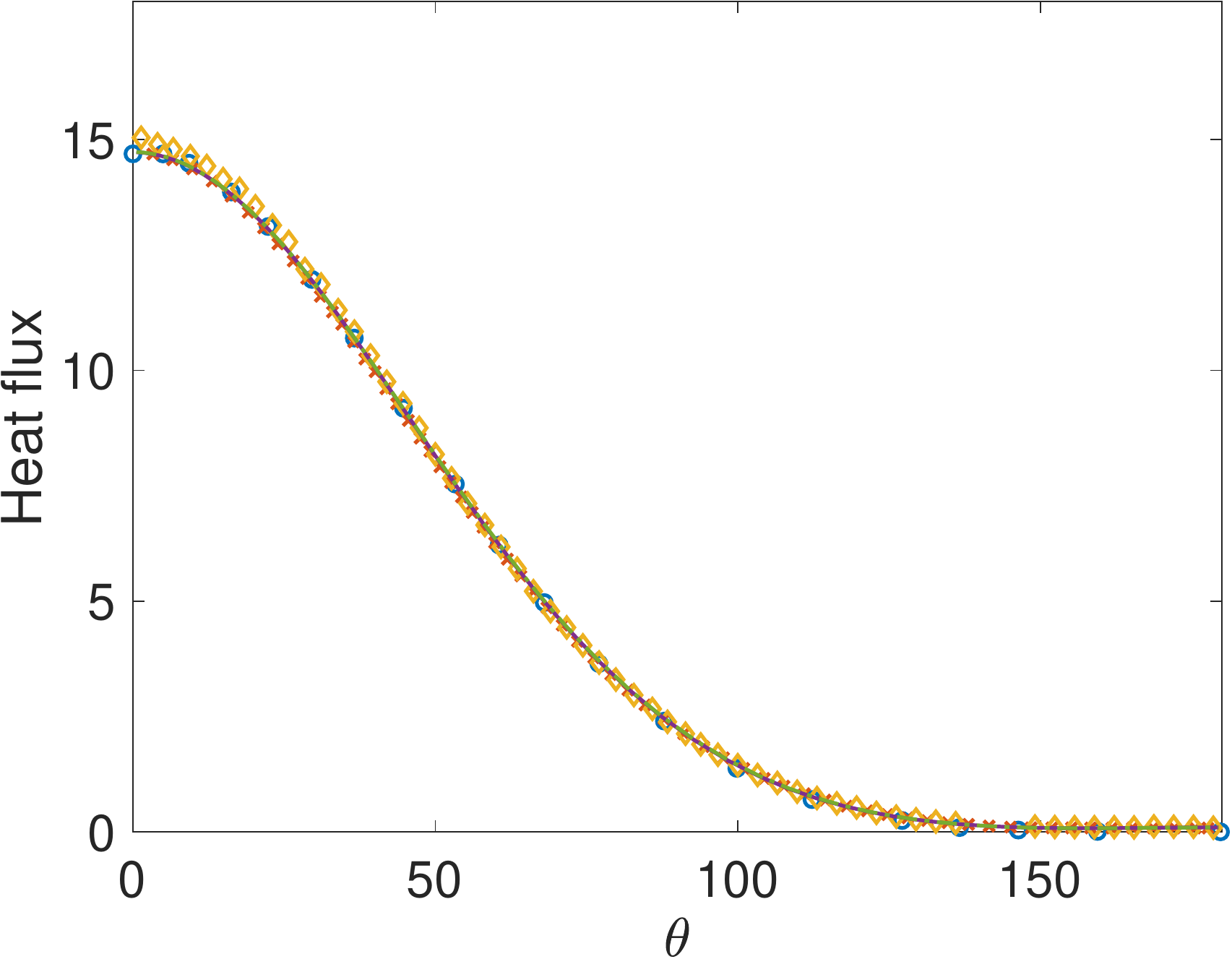} \\ 
	\vskip 0.5cm
	\includegraphics[width=0.3\textwidth]{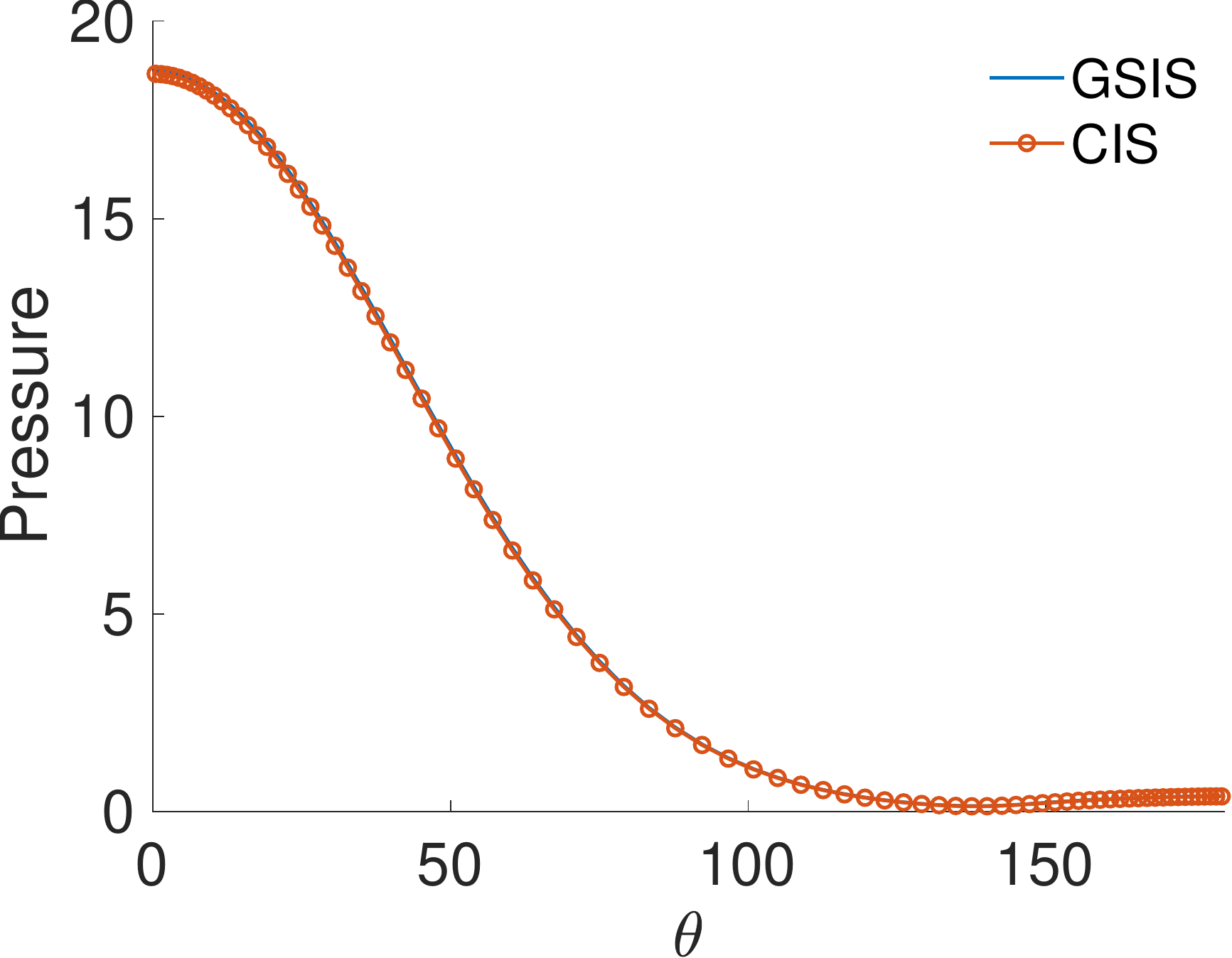}~
	\includegraphics[width=0.3\textwidth]{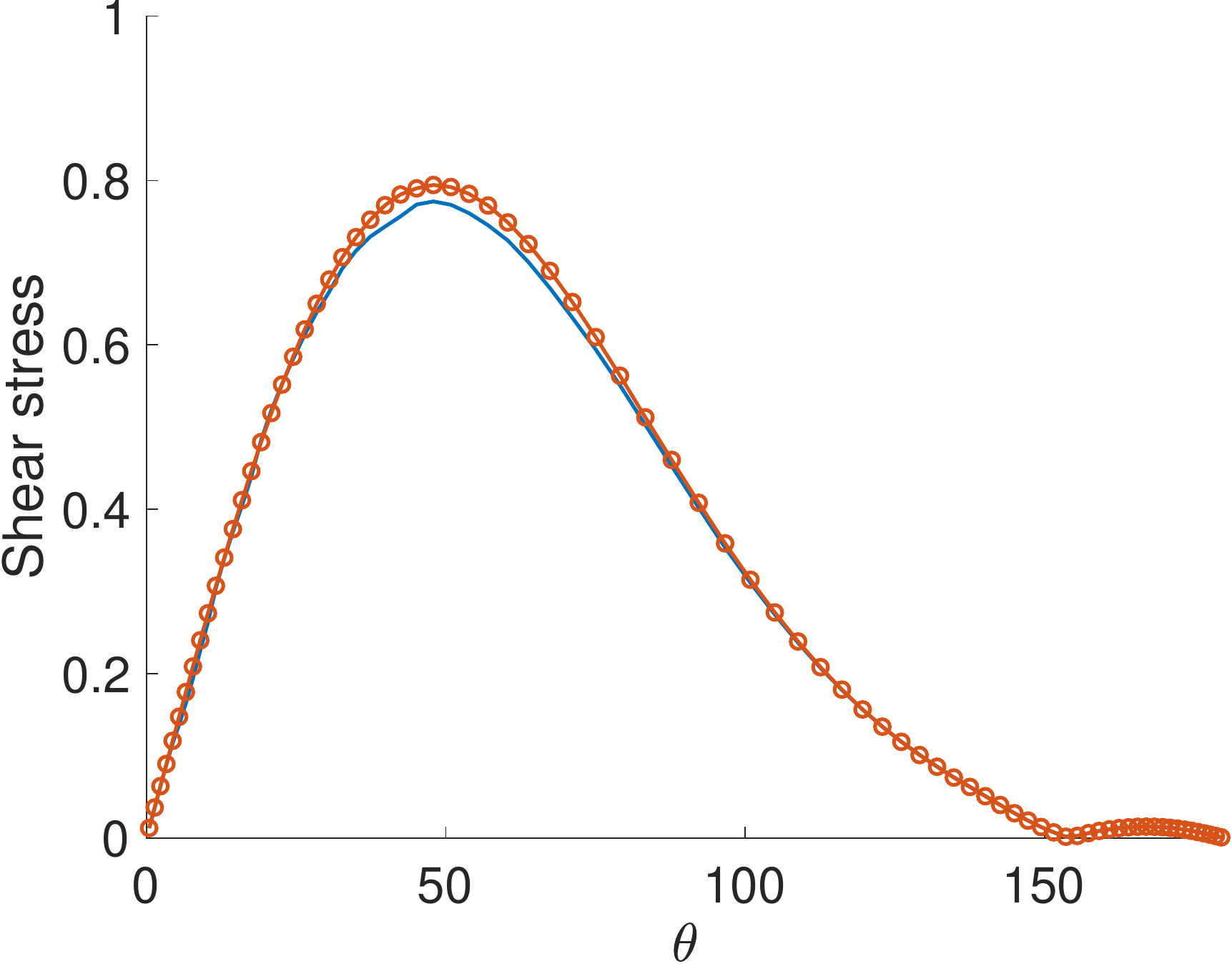}~
	\includegraphics[width=0.3\textwidth]{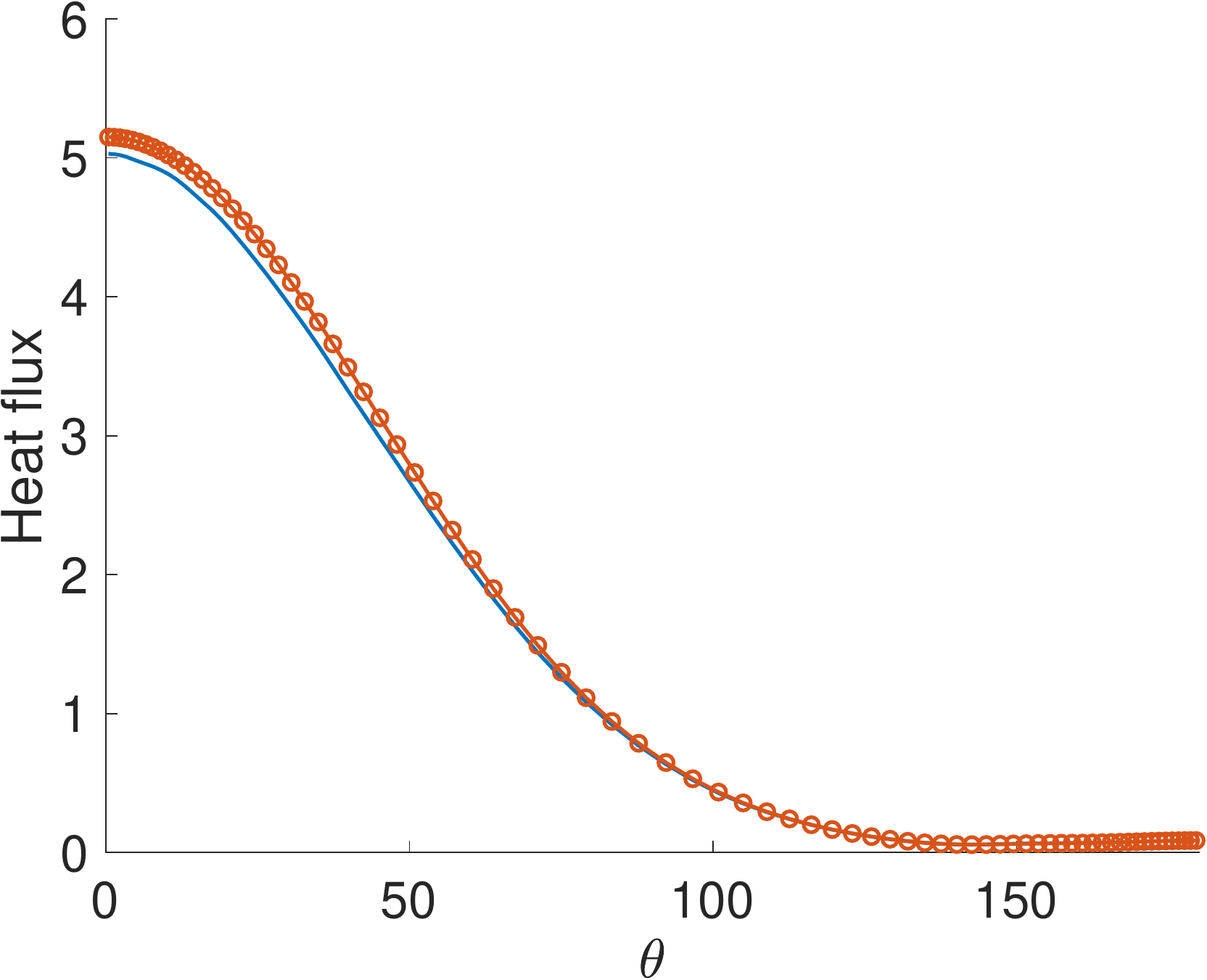} 
	\caption{Comparison of the surface quantities on the cylinder, when (top) $\text{Kn} = 1$, (middle) $\text{Kn=0.1}$, and (bottom) $\text{Kn=0.01}$. The discrete-UGKS and DSMC data are extracted from Ref.~\cite{zhuDiscreteUnifiedGas2016}. The implicit UGKS data are extracted from Ref.~\cite{zhuImplicitUnifiedGaskinetic2016}. The X-axis is the angle ($^{\circ}$) from leading edge of the cylinder.} 
	\label{fig:cylinder_Kn1_surface} 
\end{figure}

Due to symmetry, only the upper-half domain is computed and the symmetric boundary condition is applied. The physical grid size is $M\times N$, where $M$ is the number of cells along the upper surface of the cylinder and $N$ is the number of cells along the radial direction. The cell height along the radial direction grows with a constant expansion ratio from the first layer's height ($\Delta r_\text{min}$). The cell width along the cylinder surface grows from leading and trailing edges of the cylinder toward the upper position with a constant expansion ratio, such that the largest cell's width is five times of the smallest one. For the cases of Kn = 1 and 0.1, the physical grid is set as $N = 50$, $M = 64$ and $\Delta r_\text{min} = 0.01$, while for the case of Kn = 0.01, $N=80$, $M=80$ and $\Delta r_\text{min} = 0.001$. The discrete velocity set is a uniform Cartesian grid with $90^2$ points in the range of $[-15,15]^2$. The DVM method is implemented using the implicit time-stepping scheme as described in Sec.~\ref{sec:dvm2}. The CFL number in the DVM and NS solvers are 1000 and 100, respectively. The convergence criterion of the outer loop and inner loop in Eq.~\eqref{convtol} are set as $\epsilon_{\text{out}} = \num{1e-6}$ and $\epsilon_{\text{in}} = \num{1e-6}$.

Figure~\ref{fig:cylinder_Kn1} presents the temperature and local Mach number contour of the results predicted by GSIS and CIS, which are overlapped with literature results wherever available, in particular the DSMC and discrete-UGKS solution in Ref.~\cite{zhuDiscreteUnifiedGas2016}. We can observe good agreement between the CIS and GSIS solutions, and overall good matches with the literature results. Figure~\ref{fig:cylinder_Kn1_surface}  shows the pressure (normal stress), shear stress and heat flux along the upper surface of the cylinder. Comparison are made with the results from literature including Refs.~\cite{zhuDiscreteUnifiedGas2016} and \cite{zhuImplicitUnifiedGaskinetic2016}. Again, it is shown that current GSIS results match well with the literature results. In the bottom of Fig.~\ref{fig:cylinder_Kn1_surface}, it is seen that both GSIS and CIS capture the flow separation from the surface precisely at the same location around $153^{\circ}$.

\begin{figure}[t]
 \centering
 \includegraphics[width=0.5\textwidth]{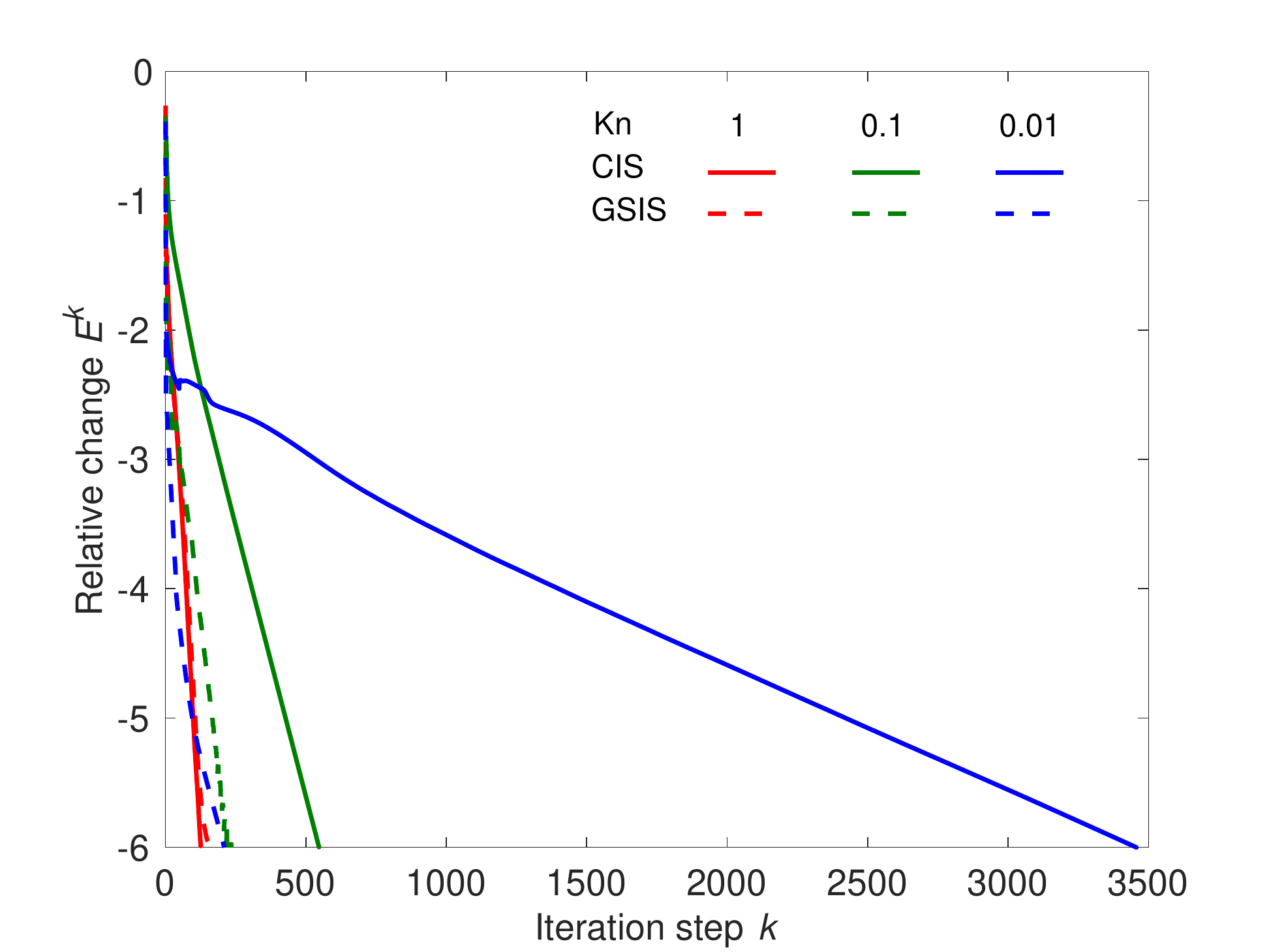}
 \caption{Convergence history of the DVM time stepping in the supersonic cylinder flow at different Knudsen number. The red, green and blue lines are for Kn = 1, 0.1 and 0.01, respectively. The solid and dashed lines represent the CIS and GSIS results, respectively.}
 \label{fig:cylinder_convergence}
\end{figure}

To assess the efficiency of GSIS, we plot the convergence history of the DVM in both GSIS and CIS in Fig.~\ref{fig:cylinder_convergence}. In addition, Table~\ref{tab:cyliner_time} lists the number of DVM steps and overall computing time in the same environment as in the lid-driven cavity flow. Obviously, for highly rarefied flows,  CIS is very efficient: when Kn = 1, the solution converges in 186 steps and the total computing time is around 12 minutes. For this case, GSIS takes more DVM steps than the CIS, and the overall computing time is about twice of CIS. As Kn decreases to 0.1, GSIS becomes slightly more efficient than CIS. When Kn = 0.01,  CIS takes as much as 4925 DVM steps and needs around 8 hours to reach the converge criterion, while GSIS takes only 42 minutes and converges in 210 DVM steps. We note that for small Kn cases, the inner loop solving the macroscopic synthetic equations also takes much fewer time steps to converge, because in this case the Reynolds number is much higher, favoring a fast convergence of NSF solver.

\begin{table}[t]
 \centering
 \caption{Number of DVM steps in CIS and GSIS and the overall CPU time for the supersonic cylinder flow.}
  \begin{tabular}{rrrrrrr}
  \toprule
  \multicolumn{1}{l}{Kn} & \multicolumn{1}{r}{Physical} & \multicolumn{1}{r}{Velocity} &\multicolumn{1}{r}{CIS: DVM} & CIS: CPU & \multicolumn{1}{r}{GSIS: DVM} & GSIS: CPU \\
   \multicolumn{1}{r}{} & \multicolumn{1}{r}{grid size} & \multicolumn{1}{r}{grid size} &\multicolumn{1}{r}{steps} &\multicolumn{1}{r}{time} & \multicolumn{1}{r}{steps}& \multicolumn{1}{r}{time} \\
  \midrule
  
  $1$   & $64\times 50$ &  $90\times 90$ & $186$    & $ 12$ min & 264  & $27 $ min \\
  $0.1$  & $64\times 50$ &  $90\times 90$  & $552$    & $ 36$ min & 232  & $ 24 $ min \\
  $0.01$ & $80\times 80$ & $90\times 90$ & $4,925$  & $508$ min & 210  & $42$ min \\
  \bottomrule
  \end{tabular}%
 \label{tab:cyliner_time}%
\end{table}%

%

\section{Conclusions}\label{Conclusion&outlook}

In summary, we have developed a GSIS to find steady-state solutions of the nonlinear gas kinetic equation, which couples a simple iterative scheme to solve the gas kinetic equation with an implicit scheme to solve the macroscopic synthetic equations. Unlike the pure DVM schemes, GSIS enables the DVM to converge very quickly in the near-continuum flow regime, which is realized by solving the macroscopic synthetic equations to the steady state after each DVM iteration. The viscous fluxes of the macroscopic synthetic equations explicitly include the NSF constitutive relation, while the higher-order terms are calculated from the velocity distribution function in DVM. Such a treatment guarantees the accuracy of GSIS in both continuum and rarefied regimes. In addition, the construction of  higher-order terms is further simplified in this paper, compared with the one in the linear GSIS~\cite{suCanWeFind2020}. Several classical cases have been used to test the accuracy and efficiency of the nonlinear GSIS, based on the Shakhov kinetic model. Numerical results demonstrated that our scheme is able to obtain steady-state solutions of the gas kinetic equation in relatively smaller number of iteration. For high-speed flows, GSIS also shows a significant speed up over the conventional iteration scheme for flows with low Knudsen numbers.

Compared to the implicit unified gas kinetic scheme and its improved versions~\cite{zhuImplicitUnifiedGaskinetic2016,yangImprovedDiscreteVelocity2018}, GSIS does not rely on the relaxation time approximation of the Boltzmann collision operator, thus like the linear GSIS~\cite{suCanWeFind2020} it can be extended to the full Boltzmann equation. Actually, the simple construction of  high-order terms proposed in this work further enhances the potential of GSIS to account for multi-species and vibrational non-equilibrium phenomenon, which are critical in high-speed rarefied gas flows. In the future work, we will also investigate the possibility of coupling the new macroscopic synthetic equations with DSMC, i.e. to produce a DSMC-GSIS algorithm to remove the limitation on cell size and reduce the computing cost for low-Kn flows. 

\section*{Acknowledgements}
L.~Zhu acknowledges the financial support of European Union’s Horizon 2020 Research and Innovation Programme under the Marie Skłodowska-Curie grant agreement number 793007. Financial support in the UK by the Engineering and Physical Sciences Research Council under grant EP/R041938/1, EP/M021475/1, and EP/R029581/1 are greatly acknowledged.

\section*{Reference}

\biboptions{sort,square,comma,compress}
\bibliographystyle{elsarticle-num}
\bibliography{P_GSIS}
\end{document}